\documentclass[fleqn,usenatbib]{mnras}

\usepackage{newtxtext,newtxmath}

\usepackage[T1]{fontenc}

\DeclareRobustCommand{\VAN}[3]{#2}
\let\VANthebibliography\thebibliography
\def\thebibliography{\DeclareRobustCommand{\VAN}[3]{##3}\VANthebibliography}

\usepackage{graphicx}	% Including figure files
\usepackage{amsmath}	% Advanced maths commands
\usepackage{amssymb}	% Extra maths symbols
\usepackage[normalem]{ulem}
\usepackage{url}
\def\msun{\mbox{M$_{\odot}$}}%Solar mass
\def\ms{\mbox{$M_{\ast}$}}%Stellar mass
\def\re{\mbox{$R_{e}$}}% half-light radius
\usepackage[dvipsnames]{xcolor}

\title[Star Formation-Size-Structure Interplay]{Non-Monotonic Relations of Galaxy Star Formation, Radius, and Structure at Fixed Stellar Mass}

\author[Jimena Stephenson, et al.]{
Jimena Stephenson,$^{1}$\thanks{E-mail: xtllez89@gmail.com}
Aldo Rodriguez-Puebla,$^{2}\thanks{E-mail: apuebla@astro.unam.mx }$
S. M. Faber,$^{3}$
Joel R. Primack,$^{4}$ 
Vladimir Avila-Reese,$^{2}$
\newauthor{A. R. Calette,$^{5}$ 
Carlo Cannarozzo,$^{2,6,7}$ 
James Kakos,$^{4}$
Mariana Cano-Díaz,$^{8}$
 David C. Koo$^{3}$, }
\newauthor{Francesco Shankar$^{9}$ and D. F. Morell$^{2}$}
\\
$^{1}$University of Arizona, Tucson, Arizona 85721\\
$^{2}$Universidad Nacional Aut\'onoma de M\'exico, Instituto de Astronom\'ia, A. P. 70-264, 04510, Ciudad de M\'exico, M\'exico\\
$^{3}$UCO/Lick Observatory, Department of Astronomy and Astrophysics, University of California, Santa Cruz, CA 95064, USA\\
$^{4}$Department of Physics, University of California, Santa Cruz, CA 95064, USA\\
$^{5}$Center for Astronomy and Astrophysics, Shanghai Jiao Tong University, Shanghai 200240, China\\
$^{6}$New York University Abu Dhabi, PO Box 129188, Abu Dhabi, United Arab Emirates\\
$^{7}$Center for Astrophysics and Space Science (CASS), New York University Abu Dhabi, Abu Dhabi, PO Box 129188, Abu Dhabi, United Arab Emirates\\
$^{8}$CONACYT Research Fellow—Universidad Nacional Autónoma de México, Instituto de Astronomía, A.P. 70-264, 04510, México, D.F., Mexico\\
$^{9}$Department of Physics and Astronomy, University of Southampton, High- field SO17 1BJ, UK\\
}

\date{Accepted XXX. Received YYY; in original form ZZZ}

\pubyear{2021}

\begin{document}
\label{firstpage}
\pagerange{\pageref{firstpage}--\pageref{lastpage}}
\maketitle

\begin{abstract}
We investigate the relation between galaxy structure and star formation rate (SFR) in a sample of $\sim2.9\times10^{4}$ central galaxies with $z<0.0674$ and axial ratios $b/a>0.5$. The star-forming main sequence (SFMS) shows a bend around the stellar mass of $M_\ast\leq{}M_c=2\times10^{10}{}M_{\odot}$. At $M_\ast\leq{}M_c$ the SFMS follows a power-law $\text{SFR}\propto{}M_\ast^{0.85}$, while at higher masses it flattens. $M_c$ corresponds to a dark matter halo mass of $M_\text{vir}\sim{}10^{11.8}M_{\odot}$ where virial shocks occurs. Some galaxy structure (e.g., half-light radius, $R_e$) exhibits a non-monotonic dependence across the SFMS at a fixed $M_\ast$. We find $\text{SFR}\propto{R_e^{-0.28}}$ at fixed $M_\ast$, consistent with the global Kennicutt-Schmidt (KS) law. This finding suggests that galaxy sizes contribute to the scatter of the SFMS. However, at $M_\ast>M_c$ the relationship between SFR and $R_e$ diminishes. Low-mass galaxies above the mean of the SFMS have smaller radii, exhibit compact and centrally concentrated profiles resembling green valley (GV) and quiescent galaxies at the same mass, and have higher $M_{\text{H}_2}/M_\text{HI}$. Conversely, those below the SFMS exhibit larger radii, lower densities, have no GV or quiescent counterparts at their mass and have lower $M_{\text{H}_2}/M_\text{HI}$. The above data suggest two pathways for quenching low-mass galaxies, $M_\ast\leq{}M_c$: a fast one that changes the morphology on the SFMS and a slow one that does not. Above $M_c$, galaxies below the SFMS resemble GV and quiescent galaxies structurally, implying that they undergo a structural transformation already within the SFMS. For these massive galaxies, CG are strongly bimodal, with SFMS galaxies exhibiting negative color gradients, suggesting most star formation occurs in their outskirts, maintaining them within the SFMS.
\end{abstract}

%%%%%%%%%%%%%%%%%%%%%%%%%%%%%%%%%%%%%%%%%%%%%%%%%%

%%%%%%%%%%%%%%%%% BODY OF PAPER %%%%%%%%%%%%%%%%%%

\section{Introduction}

Most star-forming galaxies are found to have a tight relationship between the rate at which stars are forming, SFR, and the total stellar mass, \ms. This relationship is called the star-forming main sequence (SFMS), and it has been observed at redshifts out to $z\sim3$ \citep{Daddi+2007, Elbaz+2007, Elbaz+2011, Noeske+2007, Whitaker+2012,  Whitaker+2014, Speagle+2014,Leja+2022}. Although the SFMS relationship evolves in time, mainly in its normalization, it maintains its tightness \citep{Speagle+2014} and high number density \citep{Ilbert+2013,Muzzin+2013}. This suggests that galaxies spend a significant fraction of their star-forming lifetimes within it, just like main-sequence stars in the Hertzsprung-Russell diagram. The galaxy SFMS has become a primary tool for understanding the processes that govern the formation and evolution of galaxies as it provides insights into the mechanisms that regulate star formation within galaxies and the factors that influence the growth of galaxies over cosmic time. On the other hand, an important fraction of galaxies deviate downwards from the SMFS in a more sparse region of low values of SFRs for their masses. This quiescent galaxy fraction increases at higher masses and low redshifts, implying mass-dependent changes in their evolutionary stages. What physical mechanism(s) define the position of a galaxy {\it around} the SFMS or whether is {\it in} or {\it out} of it? The details about this are uncertain, but previous research has shown that it is likely related to structural changes of the galaxies and their ability to replenish their gas to keep the galaxy within the SFMS \citep[see, e.g.,][]{Schawinski+2014,Tacchella+2016}. 

%\var{This paragraph need revision. Radius as an extensive property is NOT the same as intensive properties related to it, such as surface density, Sigma1 or compactness...}
%\sout{Proposed as the next significant parameter after \ms, and possibly SFR,} 
%Proposed as another relevant scale parameter, the half-light radius, \re, is beleived to play an
Another significant parameter proposed to play an essential role in understanding galaxy evolution is the half-light radius, \re, of the galaxy \citep[e.g.,][]{Kauffmann+2003a, Omand+2014,vanDokkum+2015, Chen+2020, Lin+2020}. For instance, \citet{Kauffmann+2003a} showed that the effective surface mass density, which combines mass and  half-light radius, $\Sigma_{\rm eff} = \ms/R_e^2 $, is larger for quiescent  than for SF galaxies and that $\Sigma_{\rm eff} = {\rm const.}$  separates SF from quiescent galaxies \citep[see also][]{Krajnovic+2018}. More fundamentally, previous studies have found that SFMS galaxies are, on average, larger than quiescent galaxies for a given \ms\ not only locally \citep{Shen+2003} but up to higher redshifts $z \lesssim 2.5$ \citep{vanderWel+2014,Barro+2017}. 

Galaxy sizes may be correlated with the way they quenched. \citet{vanDokkum+2015} envisioned a star formation quenching model by postulating the ``parallel track'' hypothesis to explain the difference of SFMS and quiescent galaxies in the size-mass plane \citep[see also][]{vanderWel+2009,Cappellari+2016,Lilly+2016}. In this model, the {\it population} of galaxies follows parallel tracks in the size-mass plane, with a of slope $\Delta\log R_e / \Delta \log M_\ast \sim 0.3$, and they form stars at a rate dictated by the SFMS. They experience an increase in size proportional to an increase in mass and quench their SF once they cross a velocity dispersion, $\sigma_e$, threshold.
%\var{, where $\sigma_e$ is roughly proportional to $\sqrt{\ms/\re}$}. 
This model has the profound implication that galaxies that start out with large radii will remain large in a way that \re\ and SFR are nearly independent of each other. \citet{Rodriguez-Puebla+2017} arrived at the conclusion that the track hypothesis was consistent with the stellar mass growth history of galaxies based on semi-empirical modeling of the galaxy-halo connection (\citealp[see, e.g.,][]{Behroozi+2019,Wechsler+2018}) and the evolution of the size-mass relation from \citet{vanderWel+2014}. 

\citet{Chen+2020} used the parallel track hypothesis arguing that the radius of SF galaxies is the next significant parameter after mass in shaping their central SMBH masses:  for a given \ms, large galaxies host lower-mass SMBHs due to their lower central 1 kpc surface densities $\Sigma_1$. Moreover, they showed that assuming SFRs to be independent of \re\ on the SFMS gives the observed sloped quenching boundaries in the size-mass plane, $\Sigma_1-\ms$, and BH-stellar mass relations, simply because larger galaxies will reach a larger stellar mass before quenching. This is consistent with \citet{vandenBosch2016} who noted that the half-light radius, stellar masses and super massive black hole (SMBH) masses form a fundamental plane, $M_{\rm BH}\propto (\ms/\re)^{2.9}$. 

 However, there are still many unknowns. Particularly, the exact shape of the relation between \re\ and SFR for a given \ms\ is still a subject of active research, but it is generally accepted that they are weakly or almost not related, consistent with the track hypothesis. For example, \cite{Omand+2014} analyzed SDSS galaxies by SFR in the $\re-\ms$ plane and found that galaxies with low SFR show smaller radii but found no significant correlation between the SFRs of SFMS galaxies and radius in normal to high mass galaxies \citep[see also][for similar results to higher-$z$]{Whitaker+2017,Brennan+2017,Fang+2018,Lin+2020}. Nevertheless, for lower-mass galaxies with $M\sim 10^{9.5} M_{\odot}$, the authors noted a non-trivial dependence of SFRs, at a fixed stellar mass, on $R_e$. In contrast, \citet{Wuyts+2011} found a strong negative correlation between SFR and \re\ for highly star forming nearby galaxies: star-forming galaxies above the SFMS are a factor of $\sim2$ smaller in the SDSS, with no similar trend at high redshifts.

Previous research has shown that other structural parameters, besides \re, correlate to some degree with the SFRs when analyzing both star-forming and quiescent galaxies at the same stellar mass. Quiescent galaxies tend to have higher bulge mass \citep{Bluck+2014}, effective surface brightness \citep{Kauffmann+2003a,Kauffmann+2012}, S\'ersic index \citep{Bell+2008,Bell+2012,Wuyts+2011}, and surface density within 1 kpc \citep{Cheung+2012,Fang+2013,Woo+2019,Luo+2020}, and tend to be of earlier morphological types \citep[e.g.,][]{Schawinski+2014}.
%\sout{\citep{Kauffmann+2003a, Brinchmann+2004, Franx+2008, Cheung+2012, Fang+2013, Yesuf+2021}.} 

Which structural parameter has the most influence on star formation remains unclear \citep[but see][and below]{Yesuf+2021}. 
Note, however, that some of these structural properties are in some ways mirror quantities of \re\ simply because, for a given mass, quiescent galaxies are smaller (more compact) than star-forming galaxies. For example, \citet{Chen+2020} showed that for a given mass, \re\ maps into $\Sigma_1$ and, at the same time, $\Sigma_1$ into $\sigma_{e}$ \citep[%velocity dispersion, 
see also][]{Fang+2013,Yesuf+2021}. If there is only a weak correlation with \re, then it is logical to expect that these structural properties are weakly correlated as well.

\citet{Yesuf+2021} used the statistical method of mutual information to quantify the relevance of several galaxy structural properties for predicting the sSFRs to a sample of face-on galaxies from the SDSS Stripe 82 deep $i$-band. These authors found that galaxies are a multiparameter family, and that morphological asymmetries (irregular spiral arms, lopsidedness, and perturbations due to mergers) are the best predictors of sSFR $\equiv$ SFR$/M*$ followed by bulge/concentration. SFMS galaxies with higher sSFR are more asymmetric and have stronger bulges. Nevertheless, it is unclear whether there are some variations with mass because the authors did not perform their analysis for different mass bins but for all SFMS galaxies.

The Kennicutt-Schmidt (KS) law is another fundamental relation that involves the SFR and structural parameters via the relationship between the total surface density of gas ($\Sigma_{\rm gas} = \Sigma_{\rm HI} + \Sigma_{{\rm H}_2}$) and the SFR surface density \citep{Schmidt+1959,Kennicutt+1998, Kennicutt+2012,de_los_Reyes_Kennicutt2019,Kennicutt_de_los_Reyes2021}. One of the quantities directly related to the global KS law is the radius of the galaxy, which naively can be thought as the surface area  %computed 
$A \propto R_e^2$ (e.g. \citealp{Lin+2020}, but see \citealp{Peeples_Shankar+2011} who used the 90th percentile $z$-band isophotal radius, $R_{90,z}$). 
If the gas mass is constant for a given \ms,\footnote{By fixing \ms\ we eliminate any correlation between mass and the size-SFR relationship.} a careful look at the global KS law quickly reveals that more compact and higher-density galaxies are more star-forming (${\rm SFR}\propto 1/R_e^{2(n-1)}$, where $n=1.42$ is the exponent of the KS law). In other words, if the gas mass of SF galaxies is the same for a given \ms, a negative {\it monotonic relation} is expected between size and SFR from the global KS law.
%, consistent with the track hypothesis by \citet{vanDokkum+2015}. 
The above was first envisioned by \citet{Lin+2020}, who looked at the residuals between \re\ and SFRs on both the SDSS and CANDELS/3D-HST, but found a correlation that was weaker than expected for constant gas content. The authors concluded that the gas fraction in small-radius galaxies must be smaller than the gas fraction in large radius galaxies at fixed mass. 

Motivated by the results discussed above and the existence of a global KS law, our goal  is to revise previous results and perform our own analysis on the relationship between SFR and the structural parameters (which can be considered as proxies for morphologies) in the local Universe, $z\sim0.03$. Thus, this paper studies the SFR and structural parameters such as half-light radius, $R_{e}$, half-mass radius, $R_{e,\ast}$, central mass density within 1 kpc,  $\Sigma_{1}$, radial colour gradient (CG), $\nabla_{(g-i)}$, and S\'ersic index in the $r$ band, $n_{r}$, in order to get a coherent picture of the dependence of the SFR on galaxy structural properties.  In contrast with previous studies, we investigate all these properties in a systematic way as a function of the distance of the galaxy from the SFMS ridge line. Particularly, we note that for colour gradients, $\nabla_{(g-i)}$, such systematic study has not been reported previously in the literature. Colour gradients are key because they add another dimension to studying the properties of the stellar population as a function of galaxy's radius. In addition, unlike the previously mentioned papers, we study only central galaxies in order to avoid complex processes that exclusively affect satellite galaxies, which could hide important trends between the variables of interest. Finally, one of the main drivers of this paper is to understand the role that the global KS law plays in the interplay between SFR, mass, and structure/morphology. %After all, these quantities are related, directly or indirectly, to this law. 
 
 In this paper, we  show that SDSS galaxies follow a non-monotonic relationship between half-light/half-mass radius and specific SFR (sSFR) for a given stellar mass. We also observe a similar trend among other structural parameters, such as $\Sigma_1$, $n$ and color gradient. However, when studying the sSFR as a function of half-light radius, a negative monotonic relation emerges, consistent with the correct predicted relation from the global KS law, SFR$\propto R_e^{-0.28}$, when assuming the gas mass is constant for a given stellar mass. One of the major conclusion of this paper is that there is characteristic mass associated to the bend of the star-forming main sequence, $M_c = 2\times 10^{10}M_{\odot}$, below which structural parameters of central galaxies exhibit a notable change in behavior across the SFMS. We find that galaxies with masses lower than $M_c$, as we move up across the SFMS at a fixed stellar mass, galaxies are  compact, dense, with blue centres, and structurally similar to green valley galaxies (GV). In contrast, as we move down across the SFMS galaxies are larger and more discy with redder centres. We conclude that for low-mass galaxies, $\ms \leq M_c$, there are two mechanisms that could quench them. One is related to a change in morphology that happens fast, and the other is related to secular processes that happen more slowly. At higher masses, $\ms > M_c$, SFMS galaxies near the GV are structurally closer to those of the GV, as expected. In other words, galaxies changed their morphology before leaving the SFMS. Finally, we develop an empirical model based on the KS law that shows that the above trends are the result of the efficiency of the galaxies in transforming atomic into molecular hydrogen and forming stars.  

This paper is organized as follows.  Our methods are described in \S2, including the SDSS galaxy sample used and how we divide the SFMS into sub-samples based on sSFR.  \S3 describes the interplay we find between SFR, \ms, and galaxy structural parameters.  \S4 ranks the importance of various galaxy parameters in predicting the distance $\Delta_{\rm MS}$ from the SFMS ridgeline.  \S5 describes our empirical model for the connection between $R_e$ and SFR for a given $M*$.  \S6 discusses the empirical evidence for two paths for galaxy quenching. \S7 discusses results and implications for the galaxy-halo connection.  Finally, \S8 summarizes our conclusions.

In this paper we use the standard $\Lambda$CDM model with values of $\Omega_{\rm M} = 0.3$ and
$\Omega_{\Lambda} = 0.7$ with $h=0.7$, an Initial Mass Function, IMF, of \cite{Chabrier2003} and $R_e$ refers to the half-light radius along the semi-major axis. 
Finally, in this paper, the words {\it small} and {\it large} are reserved for the physical size of the galaxies while 
{\it low, high}, and {\it massive} are reserved for the stellar mass of the galaxy. 

%%%%%%%%%%%%%%%%%%%%%%%%%%%%%%%%%%%%%%%%%%%%%%%%%%%%%%%%%%%%%%
\section{Methods}

\subsection{The galaxy sample}
\label{secc:gal_data}

In this paper we use the catalogue of 2D photometric S\'ersic 
fits for the $r$ band, reported in \citet{Meert+2015} as well as catalogs for the $g$, and $i$ bands reported in \citet{Meert+2016} for galaxies in the
Sloan Digital Sky Survey (SDSS) Data Release 7. From these catalogs, we first select galaxies that have $m_r<17.77$, $b/a > 0.5$, and a redshift range of $0.005<z<0.2$. In addition, we select galaxies with good total magnitudes and sizes, that is, we discard all the galaxies with flag bits 20 and above, see \cite{Meert+2013} and \cite{Meert+2015}.
For each galaxy in the catalog, magnitudes and colours were K+E corrected at a redshift rest-frame $z=0$; for more details
see \citet{Rodriguez-Puebla+2020}. Since we are interested in resolving the central kiloparsec of each galaxy in our sample, we found that the PSF radius of the SDSS ($\sim0.75$ arc secs) allows a redshift limit of $z<0.0674$.

\begin{figure*}
    \includegraphics[height=2.5in,width=3in]{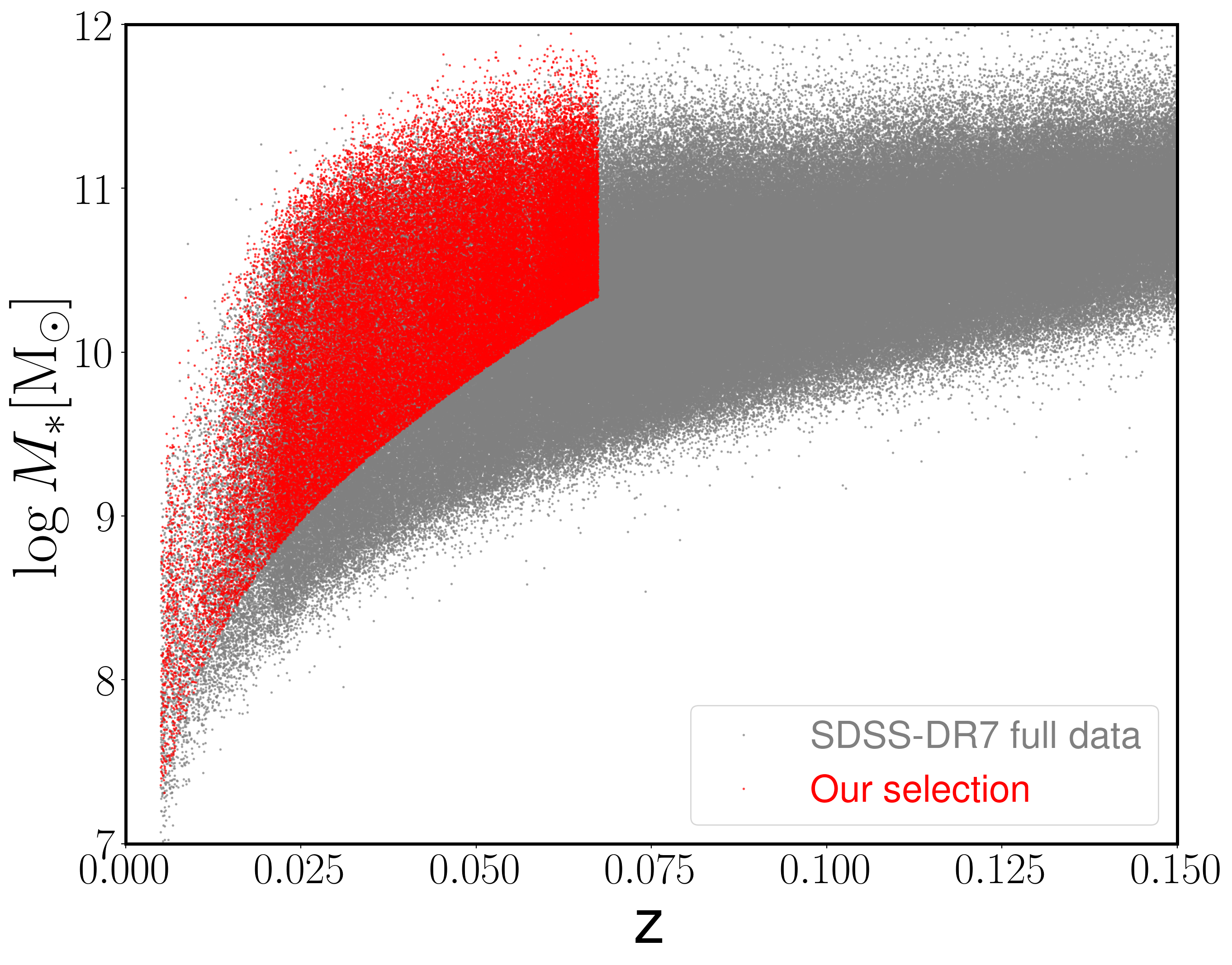}
    \includegraphics[height=2.5in,width=3in]{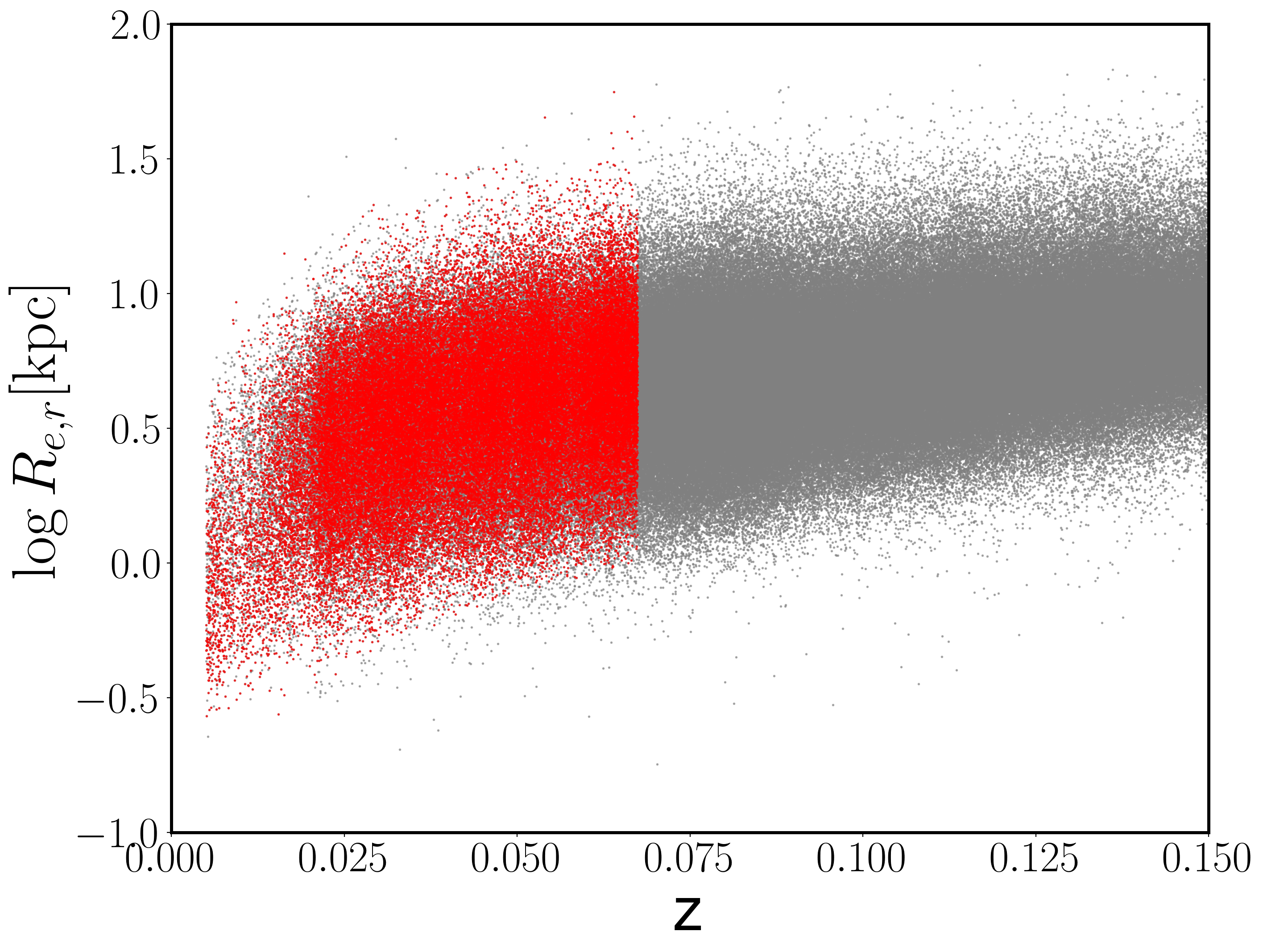}
    \caption{Stellar mass as a function of redshift and $r$-band effective radius as a function of redshift, shown in left and right panels respectively. Our fiducial sample is represented in red while the full SDSS-DR7 sample is represented in gray. We estimated a stellar mass limit that depends on colour and redshift and above which our sample is complete in \ms. We notice that above that limit the fraction of quiescent galaxies does not evolve significantly. The methodology is described in \citet[][see also, \citealp{vandenBosch+2008}]{Rodriguez-Puebla+2020}.}

\label{fig:re_ms_z}    
\end{figure*}

Figure  \ref{fig:re_ms_z} shows the stellar mass (left) and effective radius (right) as a function of redshift of the full SDSS-DR7 data as gray dots. We follow \citet[][see also \citealp{vandenBosch+2008}]{Rodriguez-Puebla+2020} to estimate a stellar mass limit that depends on colour and redshift, above which the sample is complete in stellar mass. We notice that by defining volume-limited sub-samples above that limit, the fraction of quiescent galaxies does not evolve significantly with redshift \citep{vandenBosch+2008}. In addition, we required that the half-light radius of every galaxy in the sample be larger than the SDSS PSF. The final selected galaxies we use for this work are shown as red dots.

We also use the SFRs tabulated from the GALEX-SDSS-WISE Legacy Catalog \citep[GSWLC;][]{Salim+2016,Salim+2018}. In particular, we use their GSWLC-D2 data release, which is the deepest catalog suitable for studying local star-forming and quiescent galaxies. The authors combined UV, optical, and 22$\mu$m to derive robust SFRs by using the spectral energy distribution fitting distribution technique. Thus, our final sample of galaxies with reported SFRs consists of 39,873 galaxies objects. 

Finally, this paper focuses on central (i.e., non-satellite) galaxies and the interplay between SFR, \ms, and morphology. To do so, we cross-correlated our catalog with 
\citet[][which represents an update to the \citealp{Yang+2007}]{Yang+2012} group catalog, to divide it into central and satellite galaxies.
In this paper, we define central galaxies as those with the highest stellar mass in each group as
done in \cite{Rodriguez-Puebla+2015}. Our final sample of central galaxies consists of 29,294 galaxies.  
 
%%%%%%%%%%%%%%%%%%%%%%%%%%%%%%%%%%%%%%%%%%%%%%%%%%%%%%%%%%%%%%

\subsection{Structural parameters}
\label{secc:struct_param}

 In this section, we describe how we calculate the mass profile, half-mass radius $R_\ast$, the average mass density at 1 kpc, $\Sigma_{1}$, and colour gradients (CGs) using data from the photometric catalog from \citet{Meert+2015, Meert+2016} based on their S\'ersic parameters for the $g$, $r$,  and $i$ bands. In order to calculate those quantities we must first derive the colour profiles. We make use of the S\'ersic parameters given in the photometric catalog and use their relationships to calculate surface brightness and luminosity profiles.
%The surface brightness profiles of galaxies are often well described by the \citet{Sersic1963} profile
The \citet{Sersic1963} surface brightness profile is
defined as
\begin{equation}
	\begin{split}
        I_{\lambda}(R)
        =I_{0,\lambda}\exp\left[-\beta _{n}\left(\frac{R}{R_{e,\lambda}}\right)^{1/n_\lambda}\right]
        =I_{e,\lambda}\exp\left[-\beta_{n} {\left(\frac{R}{R_{e,\lambda}}\right)^{1/n_\lambda}-1}\right],
    \end{split}
	\label{eq:sersic1}
\end{equation}
where $I_{0,\lambda}$ is the central surface brightness, $R_{e,\lambda}$ is the radius that encloses half of the total light (typically referred to as the effective radius) along the semi-major axis of the galaxy, $n_{\lambda}$
is the S\'ersic index, $ \beta_{n} =  1.9992n-0.3271, $ for $0.5<n<10$ \citep{Capaccioli1989} and $I_{e,\lambda}$ is the surface brightness at $R_{e,\lambda}$. The subscript $\lambda$ represents the $g$, $r$,  and $i$ bands. The %correlation
relation between 
$I_{e,\lambda}$ and $I_{0,\lambda}$ is given by
\begin{equation}
    I_{e,\lambda} = I_{0,\lambda}e^{-\beta_n}.
	\label{eq:ie}
\end{equation}

We obtain the luminosity profile by integrating Equation (\ref{eq:sersic1}) over any given radius
\begin{equation}
    L_{\lambda}(R) = 2\pi \int_{0}^{R} I_{\lambda}(R')R'dR',
	\label{eq:lumprof}
\end{equation}
and the total luminosity $L_{\lambda}$ by simply letting 
$R\rightarrow\infty$.\footnote{Notice that the some authors prefer to compute luminosity by truncating the fitted profiles up to 7.5 the half-light radius. The impact of truncating the luminosity impacts the stellar mass of massive galaxies, $\ms\gtrsim 10^{11.5} M_{\odot}$, to an uncertainty smaller than $\sim0.1$ dex \citep{Bernardi+2017}. Given that this uncertainty is small and most of our sample is below $\ms \sim 10^{11.5} M_{\odot}$ we ignore this effect.} Notice that the luminosity profile for each $g$, $r$,  and $i$ bands were K+E corrected as described at the beginning of this Section. 

In terms of the central surface brightness $I_{0}$ and
the effective radius \re\ the total luminosity is 
\begin{equation}
    L_{\lambda} = \frac{2\pi n_{\lambda}\Gamma (2n_{\lambda})}{(\beta_{n})^{2n_{\lambda}}} I_{0,\lambda}R_{e,\lambda}^{2},
	\label{eq:ltot}
\end{equation}
where $\Gamma (x)$ is the complete gamma function. At this point, from the \citet{Meert+2015,Meert+2016} catalogs we have constructed luminosity and magnitude profiles in the $g$, $r$,  and $i$ bands as well as color profiles for $(g-i)$ and $(g-r)$. We use these photometric profiles to derived five stellar mass profiles, $M_*(R)$, from five different mass-to-light ratios following \citet[][see their Appendix A, eq.34]{Rodriguez-Puebla+2020}, where we also took the geometrical mean of the determinations to be our fiducial $M_*(R)$ profile and total \ms.

In this paper, we are interested in studying trends in the SFR--\ms\ plane, especially the deviations from the SFMS, with the morpho-structural parameters of the galaxies as described in the Introduction. In particular, the half-mass radius, $R_{\ast}$, at a given \ms\ (compactness) and the stellar surface mass density within 1 kpc, 
$\Sigma_1$, are quantities of interest that were not available for our sample. Thus, for each stellar mass profile described above we derived $R_{\ast}$ and $\Sigma_1$ as described next. The half-mass radius $R_{\ast}$ was computed by solving the following equation   
\begin{equation}
    0.5\times M_{\ast} - M_{\ast}(R_{\ast}) = 0,
    \label{eq:re}
\end{equation}
and the stellar mass density within $R_1 =1$  kpc, $\Sigma_1$, as 
\begin{equation}
    \Sigma_{1}=\frac{M_{\ast}(R_1)}{\pi R_1^2}. %= \frac{M_{\ast}(R_{\ast})}{\pi}, 
    \label{eq:S1}
\end{equation}    
Similar to our stellar masses, we define our fiducial $\Sigma_{1}$ as the geometrical mean of the five different determinations while for $R_{\ast}$ we compute the mean of the five estimations.

%Joel added "colour gradient"
Finally, we define the colour gradient (CG) for the $(g-i)$ optical colour as the differences in colour between the radius $R_{0.1} = 0.1\times \re$ and $\re$, as done in \citet{Tortora+2010},
\begin{equation}
        \nabla_{(g-i)} = \frac{ (g-i)_{R_{0.1}} - (g-i)_{R_{e}} }{\log(R_{0.1}/\re)}.
\end{equation}
In a similar way we define $\nabla_{(g-r)}$. Notice that \citet{Meert+2015,Meert+2016} profiles were deconvolved by the point spread function of the SDSS. 
  
In this paper, we look specifically at the $(g-r)$ and $(g-i)$ optical colours due to the fact that these colours are primarily correlated with the temperature of the dominant stellar population, and therefore are mainly related to the age of the population rather than to the metallicity \citep{Gonzalez-Perez+2010}. Thus, red colour implies older stellar populations while blue colours suggest more recent star formation. We define a galaxy to have a negative colour gradient if the centre is redder as the outskirts turn blue. Positive gradients, therefore,  imply galaxies with red outskirts and a bluer centre.

After studying both colours $(g-i)$ and $(g-r)$ we have concluded that there is no significant difference between the trends. CGs tend to be steeper in $(g-i)$ but the behaviour is always the same. In the next sections, we show only our results in $(g-i)$, although every calculation was also  performed for $(g-r)$. In addition, we have repeated our results regarding the colour gradients but using those reported from the literature and found the same results (Morell in prep.).  

%%%%%%%%%%%%%%%%%%%%%%%%%%%%%%%%%%%%%%%%%%%%%%%%%%%%%%%%%%%%%%%%

\subsection{The Star-Forming Main Sequence}
\label{secc:SF_MS}

%%%%%%%%%%%%%%%%%%%%%%%%%%%%%%%%%%%%%%%%%%%%%%%%%%%%%%%%%%%%
\begin{figure}
    \includegraphics[height=2.8in,width=3.5in]{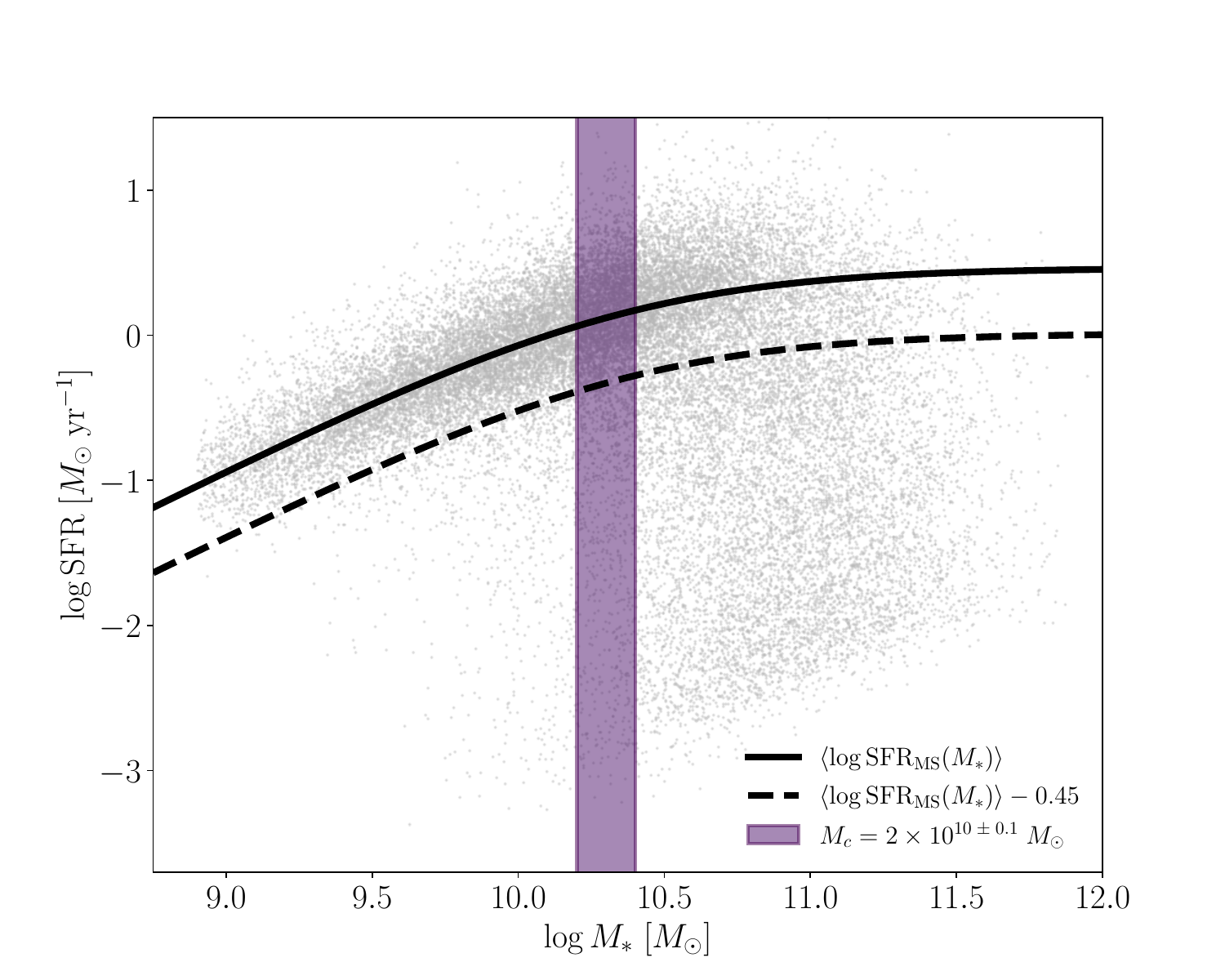}
    \caption{Star formation rate as a function of mass for our sample of central galaxies shown as gray dots. The black solid line represents our best-fit relation to $\langle \log {\rm SFR}_{\rm MS}(M_{\ast})\rangle$ using a double power law function. The dashed line separates the SFMS from the remaining galaxies, including the GV and quiescent galaxies. Our analysis reveals a characteristic mass of $M_c = 2\times 10^{10} M_{\odot}$, above which the mean of the SFMS {\it bends}. This is represented as a shaded area of $\pm0.1$ dex width around $M_c$.}

    \label{fig:sfr_ms_sample_and_fit}    
\end{figure}
%%%%%%%%%%%%%%%%%%%%%%%%%%%%%%%%%%%%%%%%%%%%%%%%%%%%%%%%%%%%

In this section, we describe how we separate the galaxy sample into several sub-samples according to the distance to the mean SFMS. 

As mentioned above, in this paper we use the SFRs from \cite{Salim+2018}. 
To obtain each subsample, we compute in an iterative way the mean \ms--SFR relation of SF galaxies, by initially using a slightly modified relation from \cite{Speagle+2014} adapted to be closer to the mean SFMS of our data. Then, all the galaxies -0.45 dex below this relationship were excluded and this was our initial guess for the SFMS. We then performed a power-law fit to the SFMS for our trimmed sample. We use the latter fit to redefine our new sample of SFMS galaxies by using only the galaxies -0.45 dex above this relation. We repeat the above process until the parameters in the power-law fits do not change more than a tolerance of $10^{-3}$. We note that the above is similar to the procedure described in \citet[][see also \citealp{Rodriguez-Puebla+2020a}]{Fang+2018}. Using our sample of SFMS galaxies, we find that the best-fitting model that describes the mean relationship is given by the generalised equation proposed by \citet[][see also, \citealp{Leslie+2020}]{Lee+2015}:
\begin{equation}
    \langle \log {\rm SFR}_{\rm MS}(M_{\ast})\rangle = \log {\rm SFR}_{0} - \log\left[1+\left(\frac{M_{\ast}}{M_{0}}\right)^{\gamma}\right]
    \label{eq:SFMS_best_fit}
\end{equation}
where  the best fit parameters are $(\text{SFR}_0/M_\odot\text{yr}^{-1}$, $\log(M_0/M_\odot)$, $\gamma$) = (0.464, 10.3815, -1.006). Figure \ref{fig:sfr_ms_sample_and_fit} presents the distribution of SFRs as a function of mass for our sample of central galaxies, along with the best-fit relation to Eq. (\ref{eq:SFMS_best_fit}). One crucial observation is the characteristic mass of \ms{}, beyond which the mean of the SMFS \emph{bends}. This bend in the SFMS has been a subject of discussion in previous studies \citep[e.g.,][]{Salim+2007, Whitaker+2014,Lee+2015, Schreiber+2015, Leslie+2020}, with similar mass ranges to what we report here. Below $M_c$ the SFMS follows a power law ${\rm SFR}\propto M_\ast^{0.85}$, whereas at higher masses the SFMS flattens out, with ${\rm SFR}(\ms)\sim{\rm const}$. When converting $M_0$ to a dark matter halo mass, it has been noted to be consistent with the halo mass scale $M_{\rm vir} \sim 10^{11.8}$, above which virial shock formation occurs, leading to a reduction in available cold gas in galaxies \citep{Daddi+2022}. We will revisit this point in Section \ref{secc:quenching}. Due to the significance of this characteristic mass within the SFMS, we will utilize it to examine changes along the SFMS. Henceforth, we define the characteristic mass centered at $M_c = 2\times 10^{10} M_{\odot}$ with a conservative uncertainty of $\pm 0.1$ dex around it due to random errors in stellar mass \citep{Conroy+2013}, and represent it in the relevant figures as a shaded area. 

Once we have characterised the %\sout{main} 
mean SFMS, Eq. (\ref{eq:SFMS_best_fit}), the next step in our program is to define nomimal sub-samples according to the distance of each galaxy from the mean SFMS relationship:
\begin{equation}
    \Delta_{{\rm MS},i} = \log{\rm sSFR}_{i}-\langle \log {\rm SFR}_{\rm MS}(M_{\ast})\rangle,
\end{equation}
where sSFR$_{i}$ is the observed sSFR for the $i$th galaxy. 
%and $\log{\rm sSFR}_{\rm fit, MS}$ is our best fit model given by equation \ref{eq:SFMS_best_fit}. 

%%%%%%%%%%%%%%%%%%%%%%%%%%%%%%%%%%%%%%%%%%%%%%%%%%%%%%%%%%
\begin{table}
    \centering
 \caption{Selection of galaxy sub-samples according to the distance of each galaxy to the SFMS}
    \begin{tabular}{l c}
        \hline
        \rule{0pt}{8pt}Sub-sample & $\Delta$MS Bin [dex]\\
        \hline
        \rule{0pt}{8pt}Highly star-forming (HSF) & \phantom{-0.00<0}$\Delta$MS > 0.25\phantom{-}\\
        Upper main sequence (UMS) & \phantom{-0.0}0 < $\Delta$MS < 0.25\phantom{-}\\
        Lower main sequence (LMS) & -0.25 < $\Delta$MS < 0\phantom{-0.0}\\
        Bottom of the main sequence (BMS) & -0.45 < $\Delta$MS < -0.25\\
        Green valley (GV) & \phantom{.00}-1 < $\Delta$MS < -0.45\\
        Quiescent (Q) & \phantom{-0.00<0}$\Delta$MS < -1\phantom{.00}\\
        \hline
    \end{tabular}
 \label{tab:sfrs_sub-samples}
\end{table}

%%%%%%%%%%%%%%%%%%%%%%%%%%%%%%%%%%%%%%%%%%%%%%%%%%%%%%%%%%

Table \ref{tab:sfrs_sub-samples} summarizes the definition of each subsample. Here we will assume that the SFMS has a standard deviation of the order of $\sigma\sim 0.25$ as has been found previously in the literature \citep[e.g.,][]{Salim+2007,Speagle+2014}. The subsamples are as follows: 1) Highly star-forming, HSF; these are galaxies 1$\sigma$ above the mean SFMS that may include starburst galaxies, we note however, that starburst galaxies do not represent a large fraction of this galaxies; 2) Upper main sequence, UMS: these are galaxies above the mean SFMS but within the 1$\sigma$ distribution; 3) Lower main sequence, LMS: these galaxies correspond to those that are below the mean SFMS and within the 1$\sigma$ distribution; 4) Bottom of the main sequence, BMS, corresponds to those galaxies that are below the 1$\sigma$ distribution but still at a distance above -0.45 dex below the mean SFMS; these galaxies are presumably approaching to the distribution of green valley (GV) galaxies, or quiescent galaxies by fading their SF. GV galaxies are those between 0.45 dex and 1 dex below the mean SFMS, while quiescent galaxies correspond to all galaxies that are 1 dex below. 

%%%%%%%%%%%%%%%%%%%%%%%%%%%%%%%%%%%%%%%%%%%%%%%%%%%%%%%%%%%%
\begin{figure*}
    \includegraphics[height=2.8in,width=3.5in]{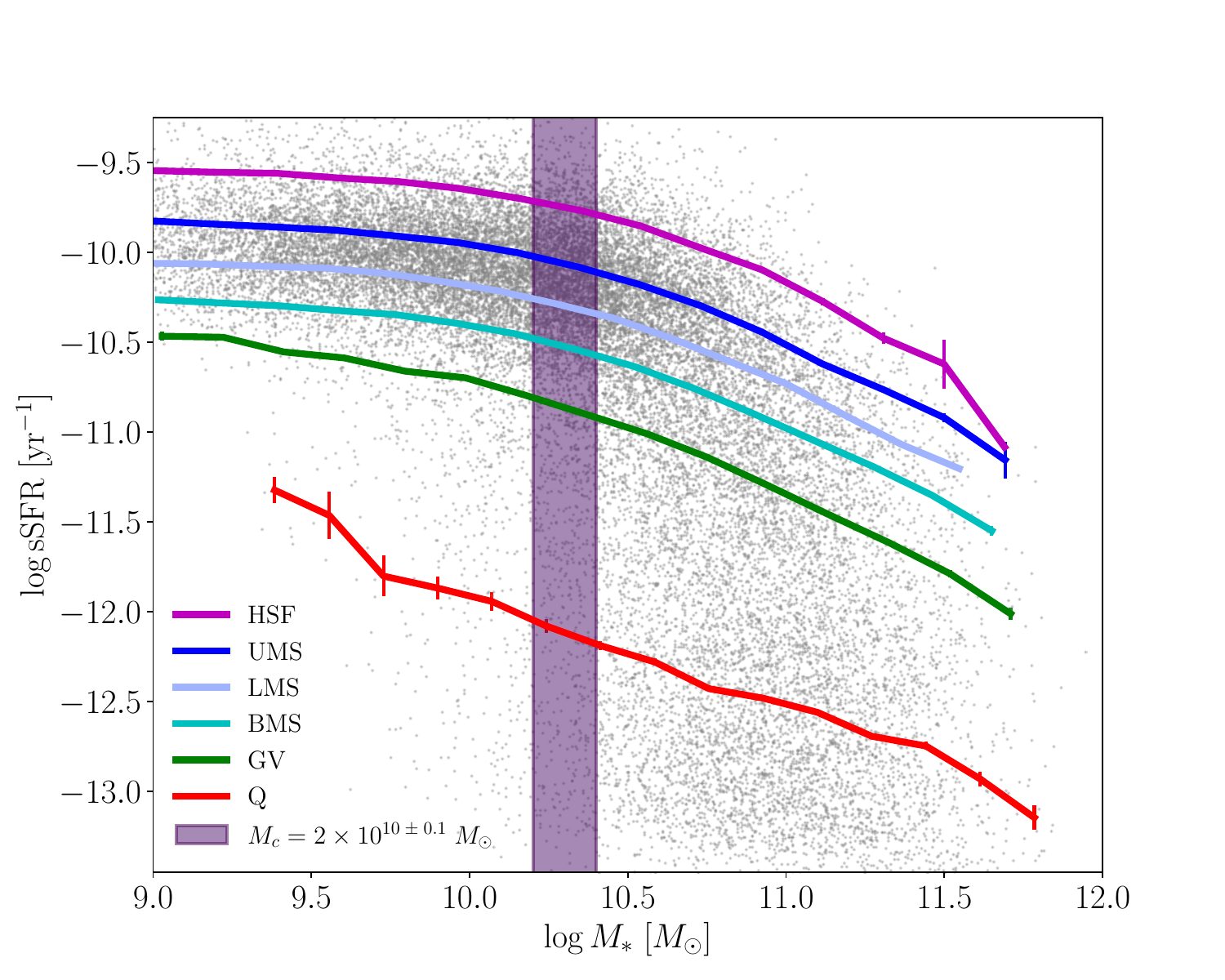}
    \includegraphics[height=2.8in,width=3.5in]{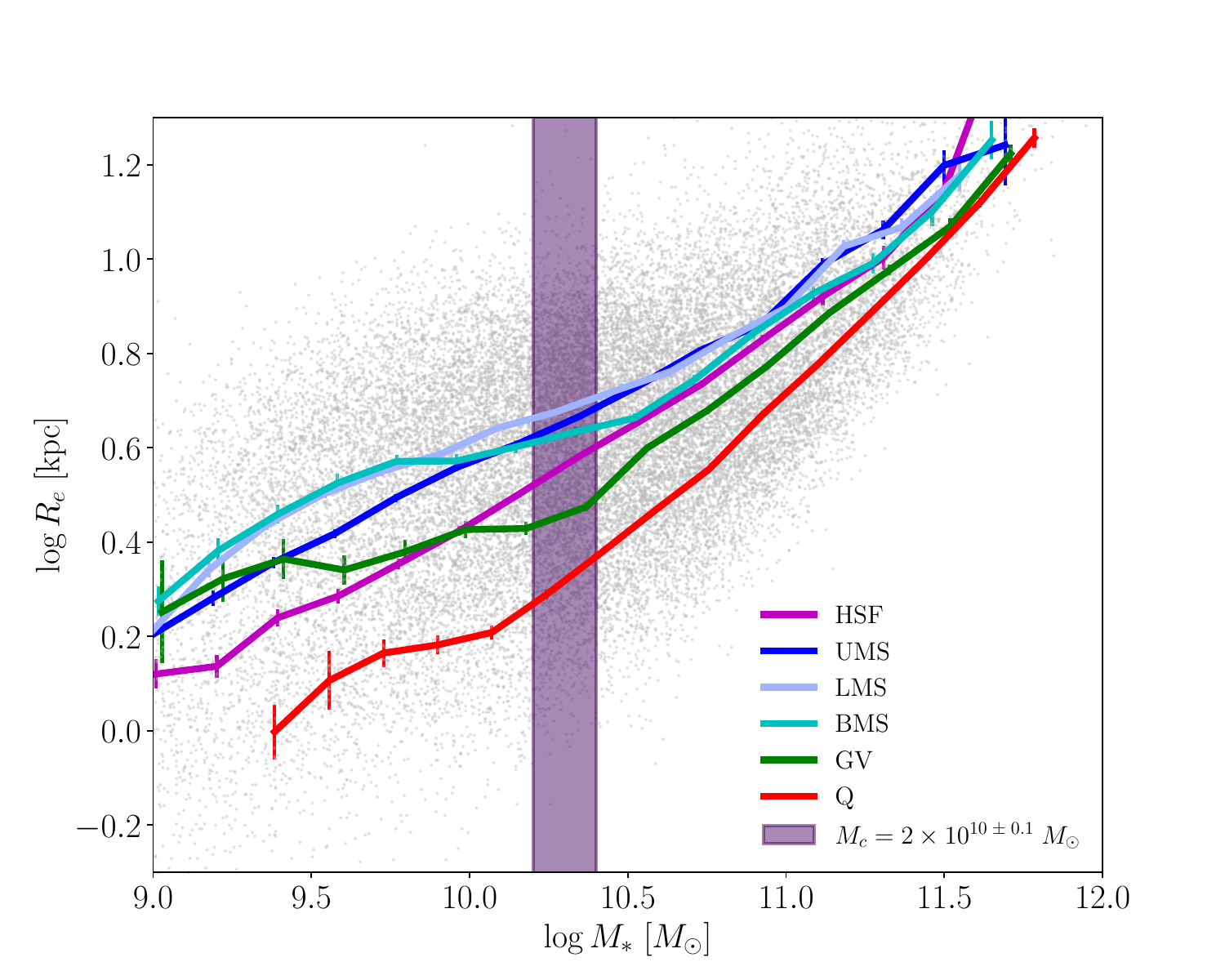}
    \caption{ From left to right we show sSFR  and $ \re$ as a function of mass. On each panel, every coloured line represents the average trend for each subsample of central SFMS galaxies: Highly star-forming (HSF), upper main sequence (UPS), lower main sequence (LMS), bottom main sequence (BMS), green valley (GV), and quiescent. Left panel: the sSFR as a function of \ms. In this panel, the mean for each sub-sample is shown for clarity and reference. Right Panel: $r$-band half-light radius vs stellar mass.
    %, notice that the label in the Figure specifies that we are using the $r$-band. 
    Note that HSF galaxies have smaller $\re$ than UMS, LMS, and BMS (i.e., all the other SF sub-samples). In fact, at masses below $\rm M_{\ast} < 10^{9.5} M_{\odot}$, HSF galaxies have a $\re$ almost identical to that of GV galaxies. In general, the average $\re$ of HSF stays parallel to the average of GV galaxies for all masses. At $\rm M_{\ast} > 10^{10.5} M_{\odot}$, the half-light radii of all sub-samples begin to converge. Note that these high-mass HSF galaxies now resemble BMS galaxies in size. However, the $\re$ of HSF still remains only greater than GV and quiescent galaxies even at high masses.}

    \label{fig:ssfr_rer_mass}    
\end{figure*}
%%%%%%%%%%%%%%%%%%%%%%%%%%%%%%%%%%%%%%%%%%%%%%%%%%%%%%%%%%%%
\begin{figure*}
    \centering
    \includegraphics[height=4.in,width=7.5in]{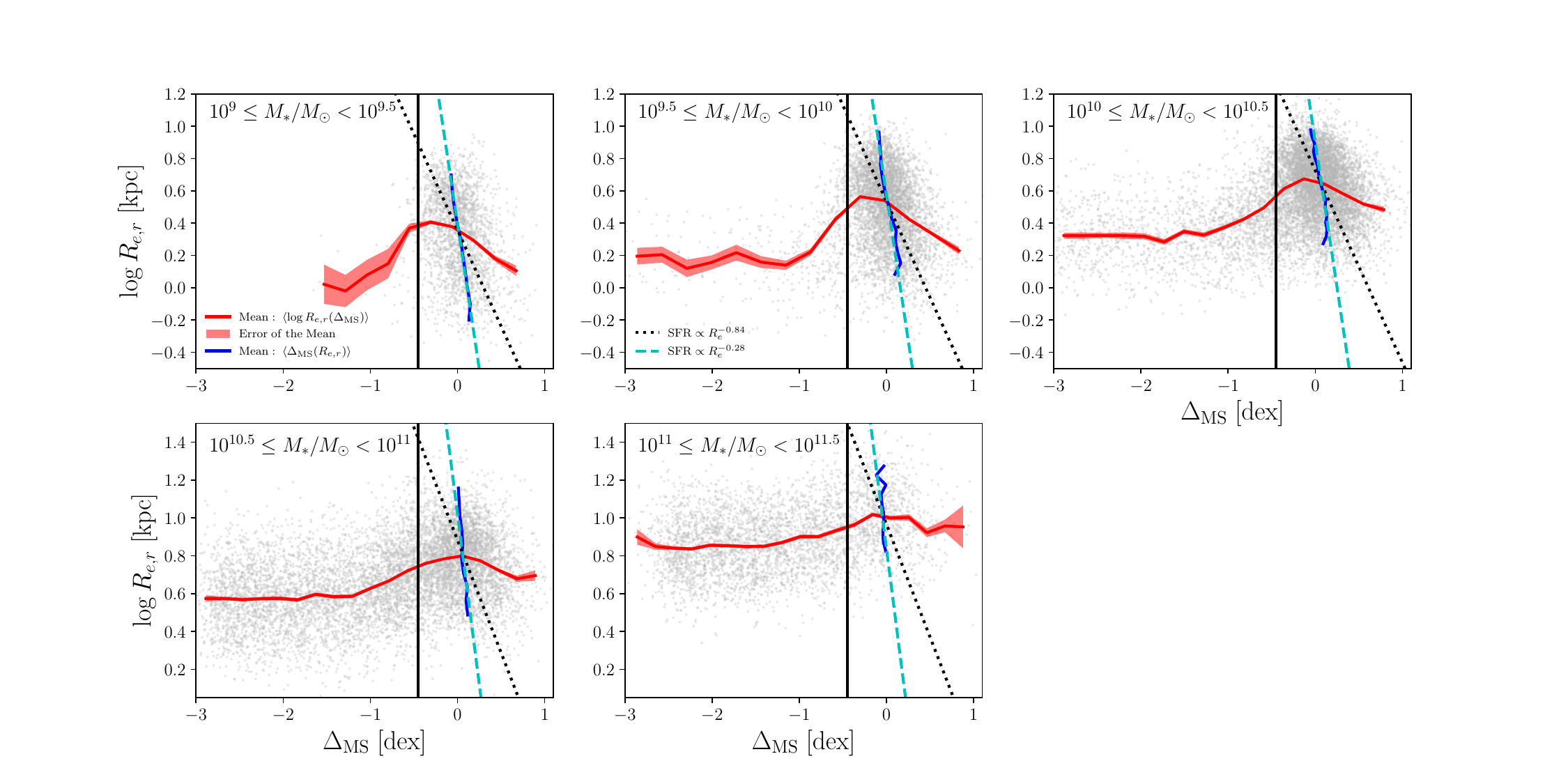}
    \caption{Galaxy half-light radius as a function of 
    %Joel their 
    the distance above or below the main sequence, $\Delta_{\rm MS}$, for five different stellar masses. Each panel represents the distribution of galaxies at a given mass bin (gray dots). The solid 
    %Joel
    red lines represent the mean $\langle \log R_{\rm e,r} \rangle$ as a function of $\Delta_{\rm MS}$ while the light red 
    %Joel shaded are 
    shading is the error of the mean. The vertical black line at $\Delta_{\rm MS}$=-0.45 dex separates GV and quiescent galaxies from the SFMS. The blue line in each panel shows the mean $\langle \Delta_{\rm MS}(R_{\rm e,r}) \rangle$ for SMFS galaxies which are the ones expected to 
    %Joel hold 
    follow the global KS law. The black dotted lines show the expected relationship between size and SFR from the global KS law; ${\rm SFR}\propto R_e^{-2(n-1)}$ if $n=1.42$ then ${\rm SFR}\propto R_e^{-0.84}$ or $R_e\propto {\rm SFR}^{-1.19}$. In each panel from low to high masses we find slopes of $\alpha \sim 0.4,0.4,0.3,0.2$ and $\sim0$ for $R_e\propto {\rm SFR}^{-\alpha}$ and galaxies with $\Delta_{\rm MS}>0$, and, $\beta \sim 0.3,0.3,0.4,0.2$ and $\sim0.1$ for ${\rm SFR}\propto R_e^{-\beta}$ and SFMS galaxies. These slopes might imply that the global KS law is inconsistent with the expected trends in this plane. As we will show in Section \ref{secc:model}, the correct trends, predicted by taking into account that the global KS law is defined within a star-forming region, imply a shallower relation of the form ${\rm SFR}\propto R_e^{-0.28}$, consistent with our observed slopes. This is represented by the cyan dashed lines.}

    \label{fig:re_MS_mass_bins}    
\end{figure*}
%%%%%%%%%%%%%%%%%%%%%%%%%%%%%%%%%%%%%%%%%%%%%%%%%%%%%%%%%%%%
\begin{figure}
    \includegraphics[height=2.8in,width=3.5in]{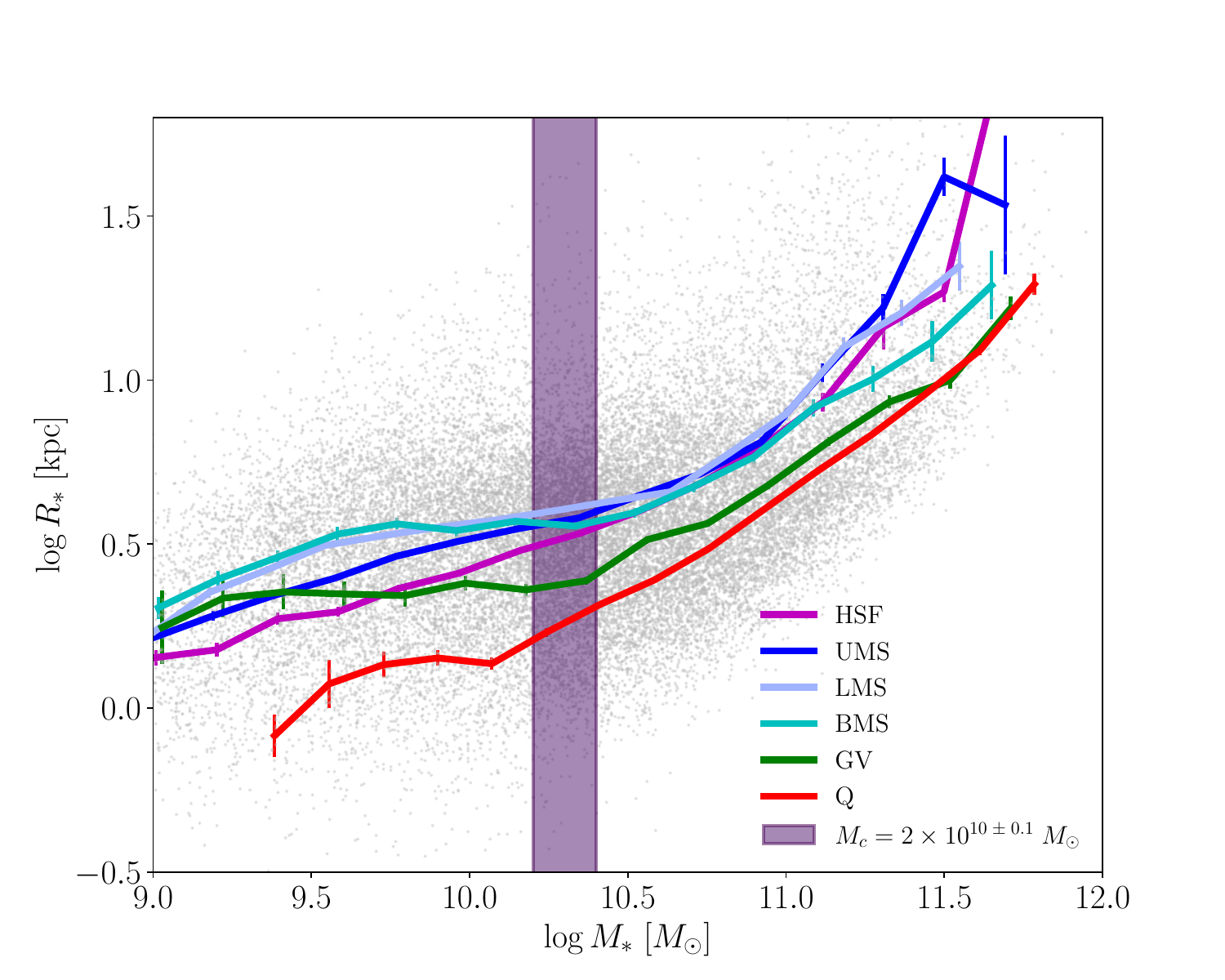}
    \caption{Half-mass radius $\re_{\ast}$ as a function of stellar mass. Every coloured line represents the average trend for each subsample of central SFMS galaxies, see Table \ref{tab:sfrs_sub-samples}. At low masses, BMS, and LMS galaxies tend to have the largest $ \re_{\ast}$. Low-mass HSF galaxies reproduce a very similar trend to that of GV galaxies. Quiescent and GV galaxies have the smallest half-mass radius at any given mass. Notice that HSF galaxies have radii smaller than all SF sub-samples below $\ms \sim 2 \times 10^{10} M_{\odot}$. Above $\ms \gtrsim \sim 2\times 10^{10} M_{\odot}$, GV galaxies are, on average, similar to quiescent galaxies and HSF are no longer as compact as in lower masses. BMS galaxies have the smallest $\re_{\ast}$ of all SF sub-samples followed closely by HSF galaxies.}

    \label{fig:ssfr_res_mass}    
\end{figure}
%%%%%%%%%%%%%%%%%%%%%%%%%%%%%%%%%%%%%%%%%%%%%%%%%%%%%%%%%%%%

\section{The interplay between SFR, \texorpdfstring{\ms}{M*}, \texorpdfstring{$R_{e}$}{Re}, and structural parameters}
\label{secc:Results}

The goal of this paper is to study whether there is, 
%Joel for a 
at fixed \ms, a correlation between the SFR and the structure of the galaxies. As discussed in the Introduction, previous authors have shown that, broadly speaking, there is a positive correlation between SFR and galaxy structure (size, density, etc.), when galaxies are divided into two broad populations: star-forming and quiescent. Here we diverge from the previous studies in two ways:  
\begin{enumerate}
    \item by studying the SFR and structure as a function of the distance from the main SFMS, Table \ref{tab:sfrs_sub-samples},
    \item and by studying this from the perspective of the Kennicutt-Schmidt law, which implies that for a fixed \ms\ and gas mass, small and compact, high central density galaxies are more star-forming, see Section \ref{secc:model}. 
\end{enumerate}
As such, we will not only describe the quantitative/qualitative aspects of the structural properties in the SFR--\ms\ plane, but also gain insight into the underlying physics of galaxy evolution by understanding how gas is transformed into stars from the global KS law. This will be discussed in more detail in Section \ref{secc:model}. 

On the other hand, it is not the SFR-radius relation that we study here, but rather $\Delta_{\rm MS}$ vs radius. Given that the radius (either the half-light or half-mass radius) is one of the quantities directly related to the KS law (see Section \ref{secc:model} below and also \citealp{Lin+2020}), we will spend more time below discussing the SFR-radius relationship than the other relations. 

Finally, the best way to understand how the morpho-structural properties of galaxies are related to their location in the sSFR--\ms\ plane is by studying the locations of the sub-samples defined in Table \ref{tab:sfrs_sub-samples} in the size-mass, CG-mass, $\Sigma_1$-mass, and S\'ersic index-mass relationships. More precisely, for each of our $\Delta_{\rm MS}$ sub-samples, we compute its mean relations in those planes. In addition, given that the global KS law involves the radius, we also study the continuous distribution of sizes as a function of the distance from the SFMS for a given $M_{\ast}$. Notice that the dependence with stellar mass is removed as our selection depends only on the distance to the SFMS in such a way that, when describing our results, the differences between the sub-samples, if there are any, should be interpreted for a given \ms.

\subsection{The dependence of \texorpdfstring{$R_e$}{Re} as a function of the SFR: Highly star-forming (HSF) low-mass galaxies are compact}
\label{secc:size_delta_ms}

Figures \ref{fig:ssfr_rer_mass} and \ref{fig:re_MS_mass_bins} present one the main results of this paper by showing how the mean sSFR-\ms\ relationships  for each of the $\Delta_{\rm MS}$ sub-samples defined above (see Table \ref{tab:sfrs_sub-samples}) are projected into the $R_{e}$-\ms\ relation. 
%We find the following for the half-light radius:
The right panel of Figure \ref{fig:ssfr_rer_mass} shows the mean $R_{e}$ of these sub-samples as a function of mass.  On average, the sizes of the SFMS galaxies are larger than those of the GV and quiescent galaxies. The sizes of the GV galaxies are also larger than those of the quiescent galaxies. These trends are consistent with results previously reported in the literature \cite[][and more references therein]{Omand+2014,vanderWel+2014,Lange+2015}. Naively, at this rough level of segregation one could conclude that, on average, more extended galaxies should be more star-forming than compact galaxies, for a given stellar mass. Does this imply a monotonic relationship, except for some random scatter, between size and sSFR? According to the results shown in the right panel of Figure \ref{fig:ssfr_rer_mass} and Figure \ref{fig:re_MS_mass_bins}, this is not the case and the answer is mass dependent. 
 %Our results are pointing that the above it is not the case and the answer depends on mass.

If we look at galaxies with stellar masses below $M_c = 2\times10^{10}M_{\odot}$, we find that, on average, the sizes of the HSF galaxies, $\Delta_{\rm MS}>0.25$ dex, are smaller than any other counterpart in the SFMS, i.e., galaxies with $-0.45<\Delta_{\rm MS} [{\rm dex}]<0.25$. For example in Figure \ref{fig:re_MS_mass_bins}, in the first mass bin it goes from $\re\sim 2.5$ kpc at $\Delta_{\rm MS}\sim-0.3$ dex to $\re \sim 1.3$ kpc at $\Delta_{\rm MS}\sim0.5$ dex. In other words, sizes have decreased nearly a factor of $\sim2$ across the SFMS. In the next two mass bins we found factors of $\sim2$ and $\sim1.5$ respectively. All these results are consistent with \citet{Wuyts+2011}. Interestingly, HSF galaxies are smaller than GV galaxies but somewhat larger than quiescent galaxies. This observation suggests that HSF galaxies with masses less than $M_c$ may serve as the progenitors of GV and quiescent galaxies \citep[see also][]{Wuyts+2011}; further discussion on this in Section \ref{secc:quenching}. Additionally, UMS galaxies are, on average, smaller than LMS and BMS galaxies for masses below $M_{\ast}\sim10^{10}M_{\odot}$ and exhibit similar half-light radii to GV galaxies. This further supports the notion that galaxies above the SMFS, particularly HSF galaxies, are structurally closer to galaxies below the SFMS, especially the GV galaxies. 

For galaxies more massive than $M_c$
%$\ms\sim2\times10^{10}M_{\odot}$, 
but smaller than $\ms\sim10^{11}M_{\odot}$, we find that HSF and BMS galaxies are similar in size. In addition, BMS and HSF are structurally closer to GV galaxies. In contrast UMS and LMS galaxies are more extended and larger than BMS and HSF. Above $\ms\sim10^{11}M_{\odot}$ the differences are not that obvious and the size-mass relations of all our sub-samples seem to converge to the same relationship. 

In Figure \ref{fig:re_MS_mass_bins} we now look at 
the distribution of $R_{e}$ as a function of the distance from the MS, $\Delta_{\rm MS}$, in different mass bins. We show the mean value of this relationship for each subsample at each mass bin with red solid lines; the red shaded area represents the error of the mean. As noticed in Figure \ref{fig:ssfr_rer_mass}, galaxies that are in the lower SFMS (LMS) are larger, on average, while the ones that have the highest SFRs (HSF) tend to have smaller radii. This is consistent with the results obtained by \citet[][see also \citealp{Lin+2020}  who found weaker trends]{Wuyts+2011}. This trend is most obvious in low-mass bins. As galaxies get more massive, the difference in size for each subsample begins to disappear. We also highlight that for galaxies smaller than $\ms \sim 10^{10} M_{\odot}$ the mean largest size is reached closer to the GV, vertical black line in Figure \ref{fig:re_MS_mass_bins}. 

If we compute the slopes of $R_e \propto {\rm SFR}^{-\alpha}$ for galaxies with $\Delta_{\rm MS}>0$, which are the ones that clearly display trends between sizes and SFR, we find $\alpha \sim 0.4$, 0.4, 0.3, 0.2, and $0$ in each mass bin, confirming the trends described above. Notably, these slopes are shallower than what would be expected by the global KS law\footnote{As described in the Introduction Section, we assumed that the gas mass is the same for a given \ms\ as in \citet{Lin+2020}.} (with a slope of $n=1.42$), specifically $R_e\propto {\rm SFR}^{-1.19}$. It is important to note that the SFRs vary from galaxy to galaxy, and shallower slopes can be the result of inverting the KS without considering this source of scatter. The blue solid line in each panel represents the value $\langle \Delta_{\rm MS} \rangle$ as a function of $R_e$. Consistent with our previous results, we also find a negative monotonic trend between SFR and size. We report slopes of $\beta \sim 0.3$, 0.3, 0.4, 0.2, and $0.1$ for ${\rm SFR} \propto R_e^{-\beta}$ in each mass bin. Again, these slopes are shallower than what would be expected by the global KS law, ${\rm SFR}\propto R_e^{-0.84}$, black dotted-line. Therefore, the shallower slopes found in the inverted relation cannot be solely attributed to the scatter in SFRs. We will revisit this point in Section \ref{secc:model}.

So far we have studied our results based on the half-light radius. At this point, an obvious question is whether our results are biased due to the colour gradient in galaxies.\footnote{IMF-driven gradients could also potentially bias our results \citep[e.g.,][]{Bernardi+2023}. Corrections due to effects of IMF gradients are out of the scope of this paper.} For that reason, the next step is to investigate if our main conclusions are consistent when analyzing the half-mass radius (Figure \ref{fig:ssfr_res_mass}).  

%Joel - minor changes in this paragraph
Similarly to the half-light radius, we again find that, on average, the half-mass radii of SFMS galaxies are larger than those of the GV and quiescent galaxies. The sizes of the GV galaxies are larger than those of the quiescent galaxies, particularly at low masses.
%Joel, as well. 
The major differences with the half-light radius results occur when dissecting the size-mass relation into different sub-samples by SFR. 

We begin by describing low-mass galaxies, $\ms\lesssim M_c$, from Figure \ref{fig:ssfr_res_mass}. %$\ms\lesssim2\times10^{10}M_{\odot}$. 
We observe once more that HSF galaxies are, on average, smaller than all other star-forming galaxies and exhibit radii closer to the average of GV galaxies. Notably, for masses below $\ms\sim 3\times10^{9}M_{\odot}$, HSF galaxies appear even smaller than GV galaxies. Similarly, UMS galaxies are smaller than LMS and BMS galaxies, displaying greater resemblance to GV galaxies. LMS and BMS have similar sizes. Below $M_{\ast}\sim 3\times10^{9} M_{\odot}$, on average, BMS galaxies are the largest galaxies of all the sub-samples.   

We now turn our analysis to galaxies more massive than $M_c$. %$\ms\gtrsim2\times10^{10}M_{\odot}$. 
For those galaxies, we observe that massive BMS and HSF galaxies are, on average, smaller than any other SF sub-samples. However, all SFMS galaxies appear to have very similar average sizes. Additionally, while GV galaxies exhibit larger half-light radii to those of quiescent galaxies, their half-mass radii at higher masses are also comparable to those of quiescent galaxies.

In general, we find that the main conclusions obtained for the half-light radius vs $M_\ast$ are also applicable when analyzing the half-mass radius instead, see Figure \ref{fig:ssfr_res_mass}. Nonetheless, the {\it trends are weaker}. Therefore, the answer to our question of whether our results are biased due to the colour gradients (CG) of the galaxies is no, although we did find small differences between the two radii that could potentially be attributed to colour gradients. In the following section, we will analyze CG as a function of mass and distance to SFMS galaxies.
%\var{I suggest to plot the compactness \ms/\re, too} 
%\aldo{Figure 5 is grad$_$color vs \ms.}

Finally, we have checked that our main conclusions were recovered when using $r-$band half-light radii from single Sersic fits of \citet{Simard+2011} as well as when using their sizes for a two Sersic fit component instead of Meert's (Morell et al. in prep.). 
Furthermore, we have questioned ourselves whether our SFRs are another source of spurious trends. Considering that the SFRs from \citet{Salim+2018} were based on WISE data, it is possible that certain galaxies, especially in their outer regions, in particular in massive galaxies, may exhibit shallower WISE colors, potentially leading to underestimated SFRs. By employing alternative empirical dust-corrected SFR indicators based on spatially resolved IFU spectroscopic data from the Mapping Nearby Galaxies at APO (MaNGA) survey \citep{Bundy+2015}, we have repeated Figure \ref{fig:re_MS_mass_bins} and found that this new SFR estimator replicated the same results (Faber et al. in prep). Therefore, the trends described in this section remain robust independently of the sizes and SFRs indicators used in this work. On the other hand, trends weaken at high masses and therefore the above does not change our general conclusions.  

\subsection{The colour gradients: low-mass HSF galaxies tend to have bluer centres}

\begin{figure}
    \includegraphics[height=2.8in,width=3.5in]{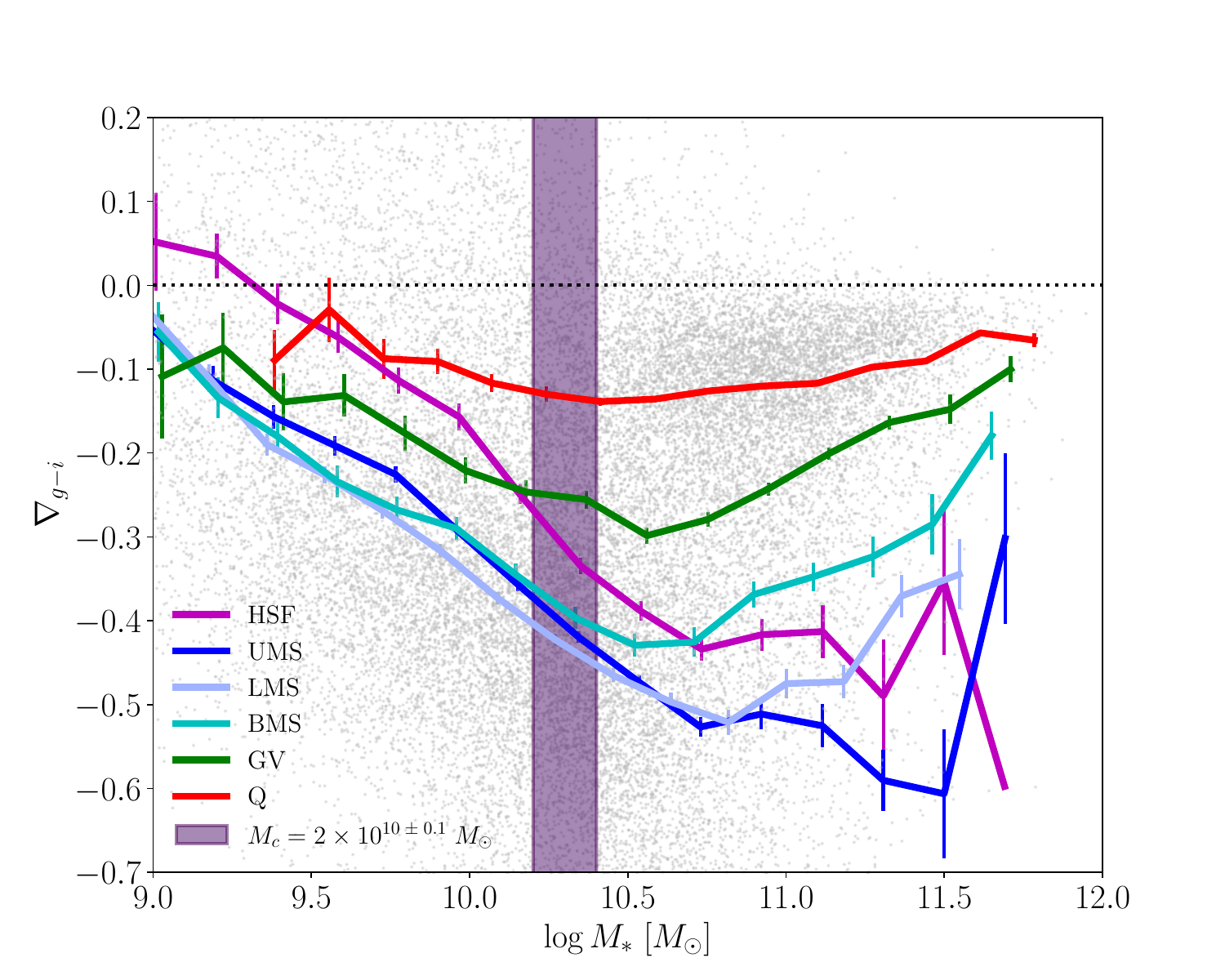}
    \caption{
    %\sout{This figure shows the} 
    Average colour gradient $\rm \nabla_{(g-i)}$ for each central galaxy subsample of SFMS galaxies, see Table \ref{tab:sfrs_sub-samples}. Negative gradients represent galaxies with redder centres than their outskirts. Positive gradients represent bluer centres. Quiescent galaxies have, on average, near-zero colour gradients at all stellar masses. From GV galaxies to UMS galaxies, the negative colour gradients get steeper as the SFR increases. However, note that HSF galaxies do not have the most negative gradients. Especially at low masses, HSF galaxies have near-zero colour gradients, and the trend remains above that of the GV galaxies below $ M_{\ast} \sim 10^{10} M_{\odot}$.  LMS galaxies show the steepest colour gradient at low masses while UMS galaxies display the steepest gradients at high masses.}

\label{fig:gradgi_avg_Ms_MS}    
\end{figure}

Figure \ref{fig:gradgi_avg_Ms_MS} shows the mean colour gradient $\nabla_{(g-i)}$-\ms\ relation divided into our six sSFR sub-samples. We notice that our colour gradients are consistent with the results found by previous authors \citep[e.g.][]{Tortora+2010,Gonzalez-Perez+2010,Pan+2015}, i.e., colour gradients depend on mass and SFR; star-forming galaxies have on average more negative colour gradients as they get more massive, while quiescent galaxies have approximately zero colour gradients. Recall that 
negative colour gradients correspond to redder centres and bluer outskirts, while positive gradients correspond to bluer centres and redder outskirts. As described in Section \ref{secc:struct_param}, bluer colours will be associated with younger ages, and thus, with recent episodes of star formation. We now study in detail the differences between our SFR divisions.  

 In the case of low mass SFMS, $\ms\lesssim M_c (= 2\times10^{10}M_{\odot})$, our results show that as the SF increases, on average, colour gradients are less negative. Indeed, the colour gradients of HSF are consistent with zero or even positive values for the lowest masses. HSF galaxies have a similar trend to the GV galaxies at low masses, below $\ms\ \sim 10^{10} M_{\odot}$, both approach to positive gradients. We highlight that similar to the result seen in the half-light and half-mass-\ms\ relations, HSF and GV galaxies remain similar in their properties. UMS, LMS and BMS have similar negative color gradients. 

 At high masses, $\ms \gtrsim M_c$, BMS galaxies have less negative gradients compared to other SFMS galaxies. In fact, for masses $\ms > M_c$, BMS colour gradients are only more negative than GV and quiescent galaxies. We notice that for almost all masses $\lesssim10^{11}M_{\odot}$, LMS and UMS resulted in similar colour gradients. 
 
 Our results point out that while being within the $1\sigma$ SFMS the colour gradient of the galaxies are mainly negative. However being a low mass and HSF galaxy, necessarily implies having a radial difference in the stellar population properties, in the direction of having a bluer centres. Also, high-mass BMS galaxies tend to have flatter colour gradients. Next, we investigate how the internal structure of the galaxies is correlated with the above results.

%%%%%%%%%%%%%%%%%%%%%%%%%%%%%%%%%%%%%%%%%%%%%%%%%%%%%%%%%%%%
%%%%%%%%%%%%%%%%%%%%%%%%%%%%%%%%%%%%%%%%%%%%%%%%%%%%%%%%%%%%

%%%%%%%%%%%%%%%%%%%%%%%%%%%%%%%%%%%%%%%%%%%%%%%%%%%%%%%%%%%%%%%%%%%%%
\begin{figure}
    \includegraphics[height=2.8in,width=3.5in]{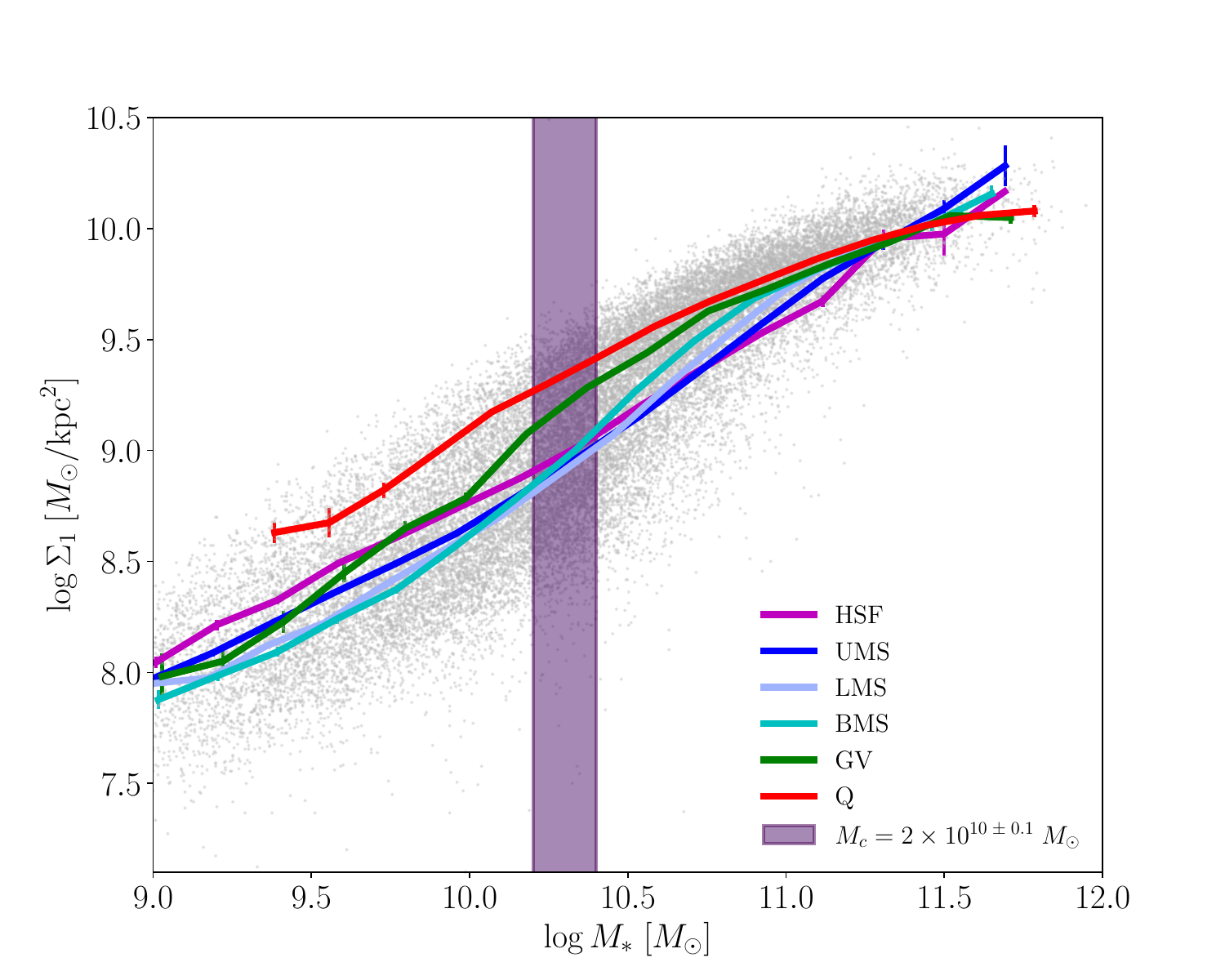}
    \caption{
    %This figure shows the 
    Average central mass density $\Sigma_{1}$ as a function of mass for each sub-sample of SFMS galaxies, see Table \ref{tab:sfrs_sub-samples}. As mass increases, the central density increases as well for all galaxies. Predictably, we see quiescent galaxies being the most centrally dense at any given mass. Something interesting to note is that HSF galaxies at very low masses have the same central mass density as GV galaxies. At $M_{\ast} \sim 10^{9.5} M_{\odot}$, GV galaxies become the second most centrally dense and HSF galaxies continue to decrease their central density with respect to the other sub-samples as they reach higher masses. At $M_{\ast} \sim 10^{11.5} M_{\odot}$ however, all galaxy sub-samples converge in central mass density.}

\label{fig:S1_avg_Ms_MS}    
\end{figure}
%%%%%%%%%%%%%%%%%%%%%%%%%%%%%%%%%%%%%%%%%%%%%%%%%%%%%%%%%%%%%%%%%%%%%
%%%%%%%%%%%%%%%%%%%%%%%%%%%%%%%%%%%%%%%%%%%%%%%%%%%%%%%%%%%%%%%%%%%%%
\begin{figure}
    \includegraphics[height=2.8in,width=3.5in]{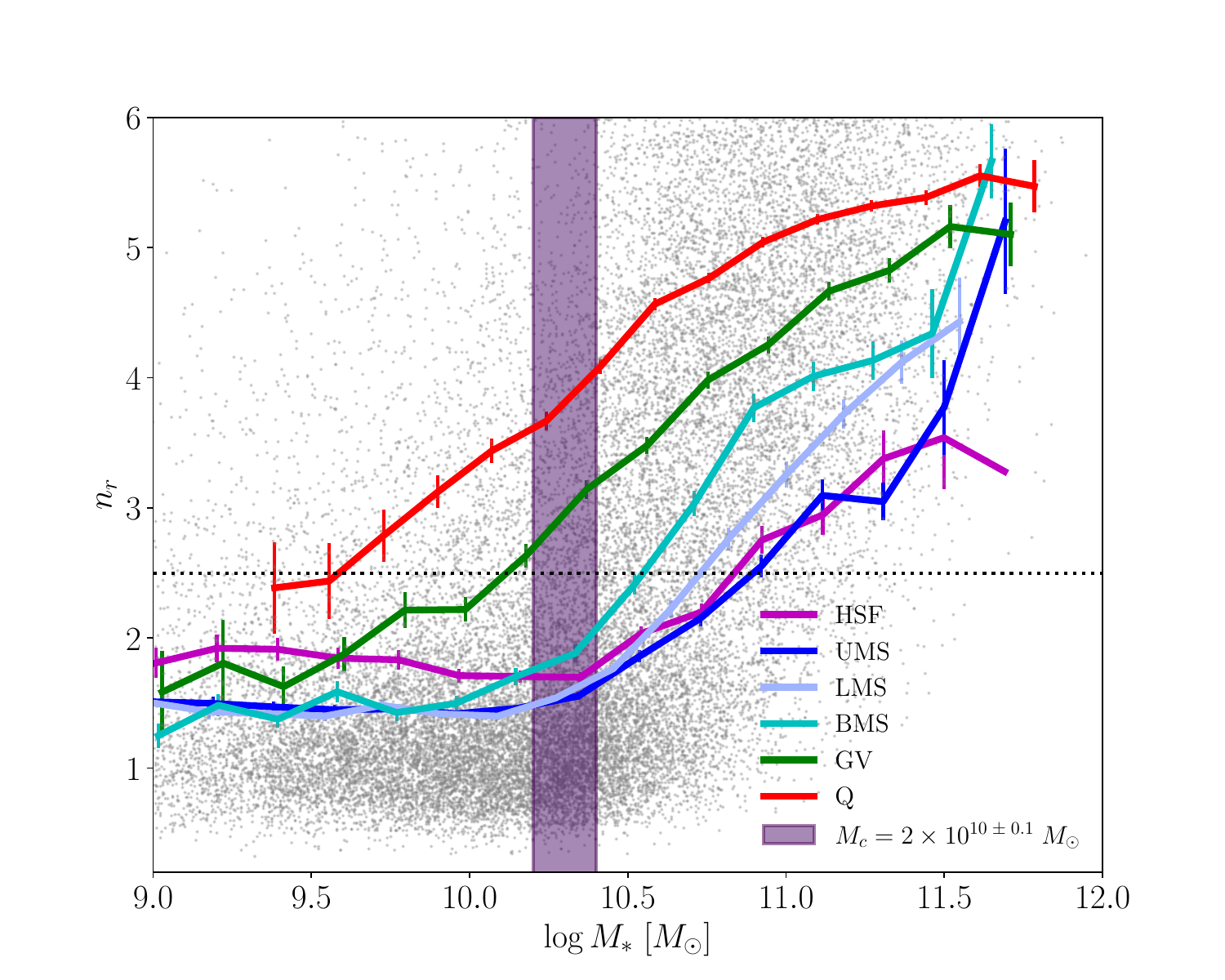}
    \caption{ %This figure shows the 
    Average S\'ersic index $n_{r}$ for each central galaxy subsample of SFMS galaxies, see Table \ref{tab:sfrs_sub-samples}. Notice the same pattern of galaxies with masses below $\ms \sim 2\times10^{10}M_{\odot}$ in HSF galaxies being, on average, similar to GV galaxies, with a S\'ersic index of $n_{r} \sim$ 2. Above this mass, BMS galaxies have the largest $n_{r}$ of all the SF sub-samples while UMS and HSF have the lowest.
    }
  \label{fig:sersic_avg_Ms_MS}    
\end{figure}
%%%%%%%%%%%%%%%%%%%%%%%%%%%%%%%%%%%%%%%%%%%%%%%%%%%%%%%%%%%%%%%%%%%%%

\subsection{Central mass density and S\'ersic index: low-mass HSF galaxies are denser}

To look more deeply into the results presented above, that is, the non-monotonic behaviour of size and colour gradients with SFRs, we now look at the $\Sigma_{1}$-\ms\ relationship. In the past, $\Sigma_{1}$ has been used as a surrogate for the bulge (and its type) of the galaxy and thus of the central supermassive black hole \citep{Fang+2013,Chen+2020,Luo+2020}. The next step of our program is to study the $\Sigma_{1}$ -\ms\ relationship. 

Figure \ref{fig:S1_avg_Ms_MS} shows the mean $\Sigma_{1}$-\ms\ relationships for each subsample.
We find that, on average, quiescent and GV galaxies are the most centrally dense. In more detail, for low-mass galaxies, $\ms \lesssim M_c$, we find a similar trend to all our previous results: galaxies above the SFMS exhibit a trend more similar to GV galaxies, on average, than those below the SFMS. In other words, SFMS galaxies get, on average, more centrally dense as their SFR increases at low masses. Indeed, BMS galaxies are the least centrally dense of all subsets at low masses. For more massive galaxies, $\ms \gtrsim M_c$, the opposite happens as mass increases. BMS and LMS galaxies increase their central density. Near $\ms \sim 10^{11} M_{\odot}$, the central density increases for all sub-samples as sSFR decreases. Finally, above  $\ms \sim 10^{11} M_{\odot}$ all galaxies have similar values of $\Sigma_1$.

Figure \ref{fig:sersic_avg_Ms_MS} shows the relationship between S\'ersic index in the $r$-band, $n_r$, and \ms. For galaxies with masses below $\ms \sim M_c$, we have persistently observed a pattern in which HSF galaxies are structurally similar to GV galaxies, with both exhibiting a S\'ersic index of $n_r \sim 2$, while all other SFMS subgroups have more similar and lower S\'ersic index, $n_r\sim1.5$. At higher masses, $\ms \gtrsim M_c$, BMS galaxies have the largest S\'ersic index of all the SFMS subgroups while the UMS and HSF galaxies have the lowest values.  As we move to higher masses ($\ms \gtrsim M_c$), we observe that the BMS galaxies have the largest S\'ersic index of all the SFMS subgroups, while the UMS and HSF galaxies exhibit the lowest values. This trend suggests that as galaxies become more massive, their light profiles become more centrally concentrated as we descend across the SFMS, contrary to the results at lower masses. 

 While the structural parameters studied in this section provide insights into their correlation along the SFMS, the above analysis does not indicate their ranking of importance as predictors for SFRs at a given \ms. We will address this question in the next section. 
 
 Finally, while in this section we study mean trends as a function of mass and distance from the MS, we have checked that all the above results remain very similar when using medians instead of means. We do not show medians for the sake of the space but they can be downloaded here.\footnote{\url{https://drive.google.com/drive/folders/1vu30ofFeL7lBatYBVWSDjf1lceyvnaWQ?usp=sharing}}

%%%%%%%%%%%%%%%%%%%%%%%%%%%%%%%%%%%%%%%%%%%%%%%%%%%%%%%%%%
\begin{table}
\begin{center}
 \caption{Parameters used to train the EBM method implemented in the InterpretML package \citep{Nori+2019}.}
\begin{tabular}{ccc}  \hline
  Parameter & Description & Value\\
  \hline
  \ttfamily{binning}                & Binning method & \emph{quantile}\\
  \ttfamily{max\_bins}              & Maximum bin number for $f_j$ & $256$\\ 
  % \ttfamily{max\_interaction\_bins} & Maximum bin number for $f_{j,k}$ &  $32\times32$\\
  \ttfamily{interactions}           & Number of pairwise interactions & $0$\\
  \ttfamily{outer\_bags}            & Number of outer bags & $100$\\
  \ttfamily{inner\_bags}            & Number of inner bags & $100$\\
  \ttfamily{validation\_size}       & Validation sample size for boosting & $15\%$\\
  \ttfamily{learning\_rate}         & Learning rate for boosting& $0.01$\\
  \hline
 \label{tab:ebm_method}
 \end{tabular}
\end{center}
\end{table}
%%%%%%%%%%%%%%%%%%%%%%%%%%%%%%%%%%%%%%%%%%%%%%%%%%%%%%%%

\section{What matters for predicting \texorpdfstring{$\Delta_\mathrm{MS}$}{delta MS} of SFMS galaxies?}
\label{sec:ga2m}
So far, we have analysed the behaviour of $\re$, $\Sigma_1$, $n_r$ and $\nabla_{(g-i)}$ as a function of the distance from the MS, $\Delta_{\rm MS}$. We aim now to understand which among these 
%aforementioned 
features is the most relevant in inferring $\Delta_\mathrm{MS}$. 
%Together with these properties, we investigate also the role of the probability for a galaxy of having a bar, $P_\mathrm{bar}$. The latter, drawn from the recent catalogue presented in DS+23, can give us some information about secular processes. 
To address this question, we exploit the so-called Generalised Additive Models with pairwise interactions, GA$^2$Ms, a machine learning technique that allows inferring an unknown variable and expressing it as a composition of functions.
The original Generalised Additive Models (GAMs) were proposed by \citet{HastieTibshirani1990}. A GAM is an additive model able to capture the relevance of different predictive features, using a composition of smooth functions, usually splines. The idea of a GAM is quite simple: it allows connecting some predictors to a dependent variable. A GA$^2$M improves the accuracy of a classical GAM, accounting for the cross-correlations between two features \citep{Lou+2013,Caruana+2015}.\footnote{Briefly, assume a given dataset $\mathcal{D}={(\mathbf{x}_i, y_i)}_1^N$, where $N$ is the size of the dataset, $\mathbf{x}_i=(x_{i,1},...,x_{i,p})$ is the set of $p$ features, while $y_i$ is the target quantity to be inferred. By indicating with $x_j$ the $j$-th variable within the space of features, a GA$^2$M can be expressed as
\begin{equation}
    g(E[y])= y_0 + \sum_j f_j(x_j) + \sum_{j\neq k} f_{jk}(x_j, x_k).
\label{eq:ga2m}
\end{equation}
In Equation (\ref{eq:ga2m}), $y$ is the dependent variable of the problem, $E[y]$ denotes the expected value, and $g$ is the so-called \emph{link function}. For each $f_j$ or $f_{jk}$, namely  the $1D$ and $2D$ \emph{shape functions}, respectively, $E[f_j] = E[f_{jk}] = 0$. The non-parametric nature of the shape functions ensures that their shapes are totally determined by the data, guaranteeing a degree of flexibility in the estimate of the model.}

To train our GA$^2$M, we rely on the machine learning (ML) method called Explainable Boosting Machine (EBM), a C++/Python  implementation of the GA$^2$M algorithm. EBM is part of the InterpretML library  \citep{Nori+2019}, an open-source Python-based package of different ML methods.
The big plus of EBM is that it has very high accuracy in the inference, comparable with those of random forest or boosted tree techniques, but, at the same time, is very informative like classical linear or logistic regressions. In EBM, the link function $g$ receives the summation of each shape function and provides an estimate of the target. The role of each feature in the inference is sorted according to the overall importance it has in making the final prediction.\footnote{To determine the importance of each feature (and of pairwise interactions, if considered), the method computes the mean of the absolute values of the corresponding shape function. For each $1D$ shape function, the $x$-axis spans over the range of the related feature, while the $y$-axis corresponds to the so-called \emph{score}, a quantity in units of the dependent variable $y$ to be added to make the final inference of $y$ considering all the other functional terms as well. Analogously, in the $2D$ shape functions the $x$- and $y$-axes correspond to two features, while the score is a third quantity, usually represented as a coloured heatmap. \label{fn:score_def}}

We consider here the SFMS subsample discussed in the previous Section \ref{secc:SF_MS} by limiting the analysis only to galaxies above the SFMS (which comprises HSF and UMS) and below (which comprises LMS and BMS). Each subsample is then subdivided into two stellar mass bins: $9\leq\log(M_*/\msun)<10.3$, and $\log(M_*/\msun)\geq10.3$. The main motivation for having two mass bins is simply because in the preceding sections we notice a change in the behaviour between structural properties with $\Delta_{\rm MS}$ if galaxies are above or below $\sim M_c$. In the following, we show the results of the analysis not accounting for the terms due to the pairwise interactions, so reducing from a GA$^2$M to a classical GAM (hence removing the second-order term in Equation \ref{eq:ga2m}), because we found their contribution negligible. In table \ref{tab:ebm_method}, we list the parameters used to train EBM. 

\begin{figure*}
    \centering
    \includegraphics[width=0.49\textwidth]{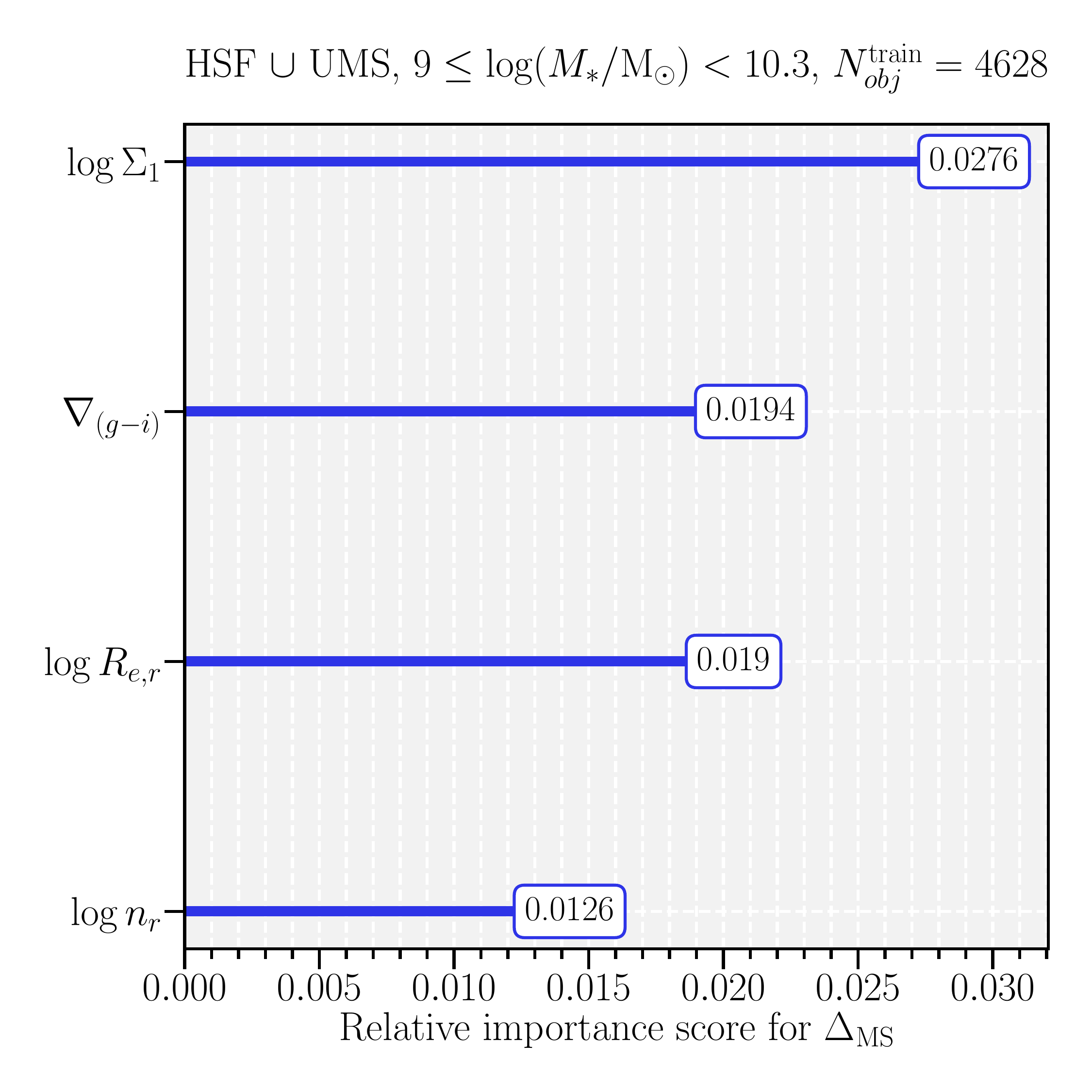}
    \includegraphics[width=0.49\textwidth]{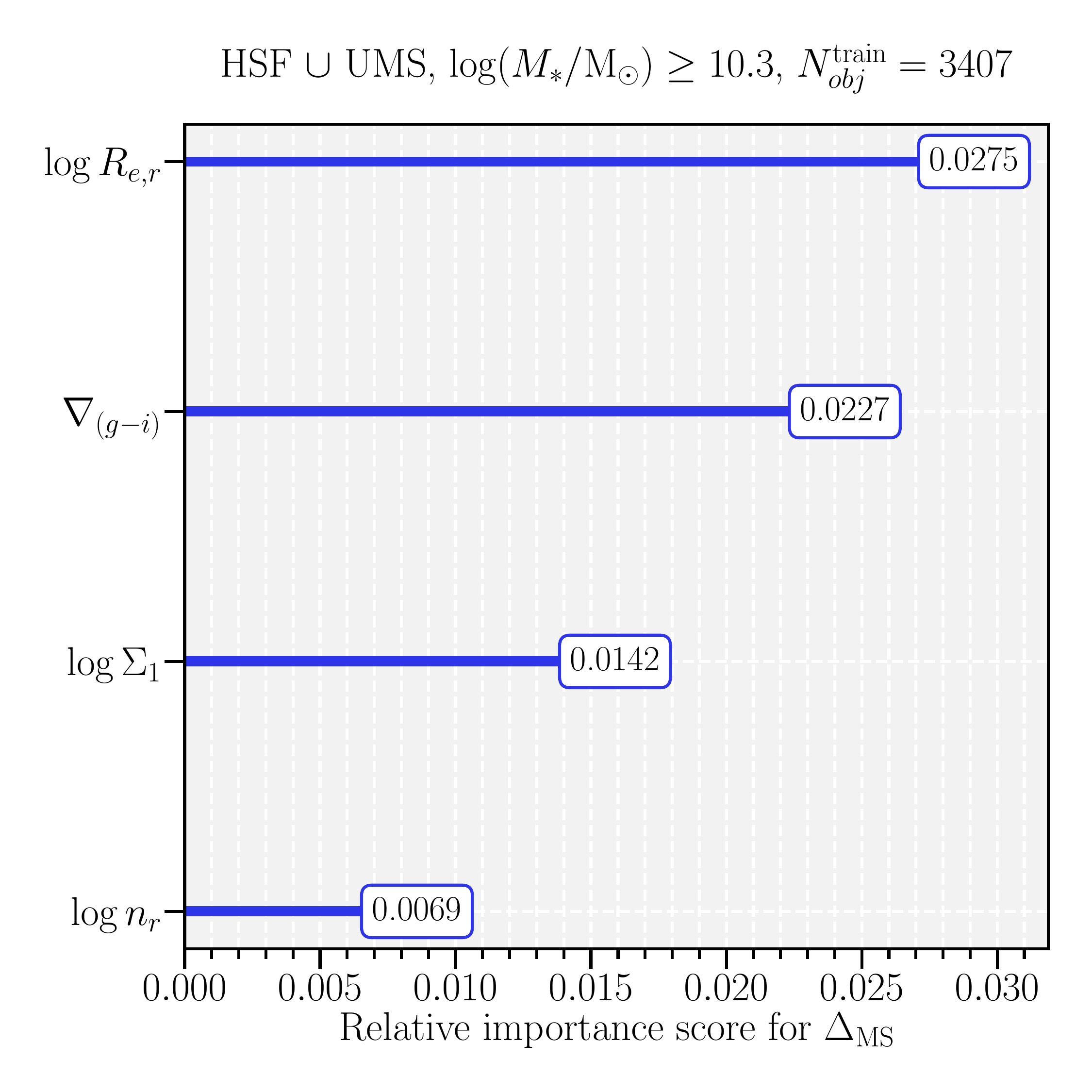}
    \\

    \includegraphics[width=0.49\textwidth]{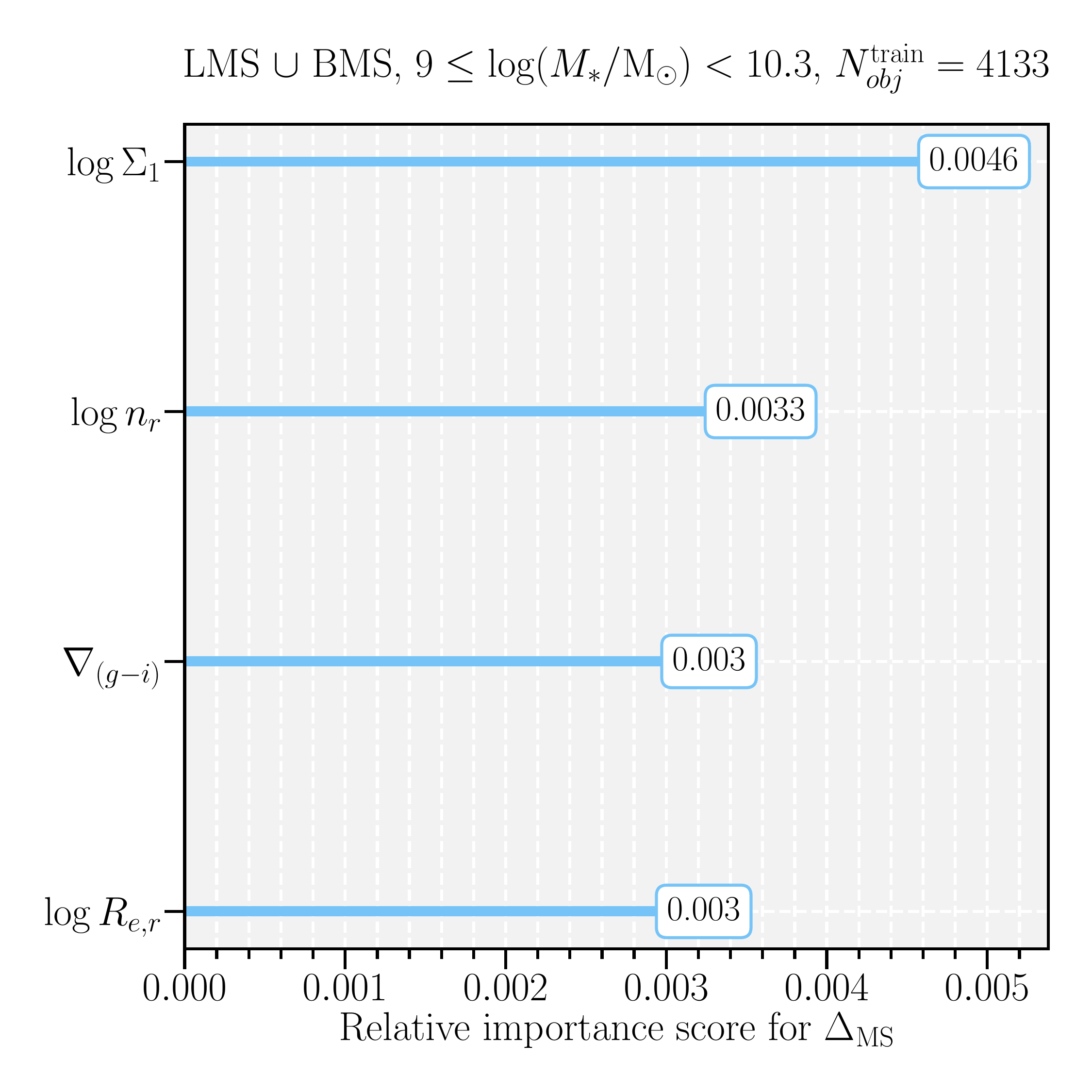}
    \includegraphics[width=0.49\textwidth]{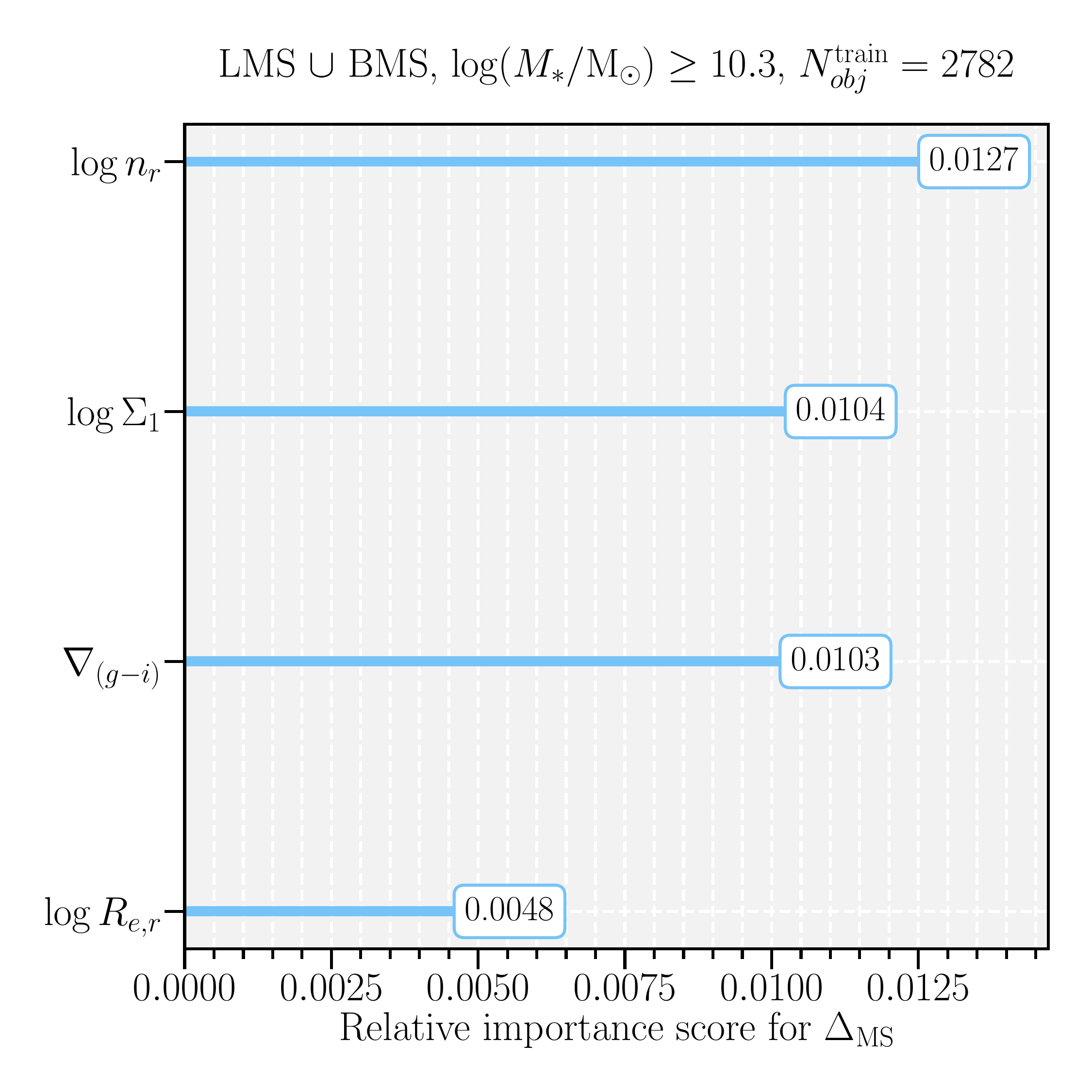}
    \\
    \caption{Relative importance scores for $\Delta_\mathrm{MS}$ of $\log\Sigma_1$, $\nabla_{(g-i)}$, $\log R_{e,r}$, and $\log n_r$. The top panels refer to SF galaxies above the SFMS (i.e. HSF and UMS), while the bottom panels refer to SF galaxies below the SFMS (i.e. LMS and BMS). For each galaxy subset, we show the results at $9\leq\log(M_*/\mathrm{M_\odot})<10.3$ (left panels) and at $\log(M_*/\mathrm{M_\odot})\geq10.3$ (right panels). The length of each bar is obtained by computing the mean absolute value of the corresponding 1D shape function.}
    \label{fig:gal_prop_importance}    
\end{figure*}

Figure \ref{fig:gal_prop_importance} shows galaxy properties according to the rank ordering of their importance to predict sSFRs for each subsample. The numbers indicated in the rectangles show the {\it score} of each property (see footnote \ref{fn:score_def}): the higher the score, the more relevant the property for predicting SFR. In general, we do not see strong differences between the structural properties used in this work. In detail, however, for low-mass galaxies above the SFMS $\Sigma_1$ (become more centrally denser) is more important followed by the CG. For galaxies below the SFMS, again $\Sigma_1$ is on the top but this time followed by the S\'ersic index. At higher masses, galaxies above the SFMS $R_e$ (being compact) is more important, while for galaxies that are below the SFMS the S\'ersic index is more important. The former is followed by the CG while the latter is followed by $\Sigma_1$.
%CGs are important (i.e., having a blue centre), while for UMS $\Sigma_1$ (become more centrally denser) is more important and for LMS and BMS $Re$ (being extended) is more important. It is interesting that for any of these properties, bars do not appear as relevant phenomena for the SFRs. In particular, for low-mass BMS which are galaxies that are less centrally concentrated, low $\Sigma_1$, and presumably transitioning into the GV. These galaxies are excellent candidates for secular processes. \aldo{Carlo, can we say something about redundant variables.}

%At higher masses, it is unclear what matters for HSF galaxies but our results suggest that $R_e$ is the first on the rank and the same for UMS while for LMS and BMS we found that $n_r$ and $\nabla_{(g-i)}$ are the most relevant quantities. 

As we have seen in the previous sections, the structural properties $\Sigma_1$, $n_r$, and $R_e$ display similar behaviours as a function of $\Delta_{\rm MS}$ for a given \ms. The above is indeed expected since they are correlated through the Sérsic profile. Therefore, one could question whether the results shown here are robust since $\Sigma_1$, $n_r$, and $R_e$ are expected to be redundant variables. Note, however, that redundant variables are expected to have similar scores. While Figure \ref{fig:gal_prop_importance} shows that the trends are weak, we notice that indeed there are clear differences between galaxy properties ranked first and second. In Appendix \ref{app:EBM_statistics} we present further information from EBM, such as the correlation matrix, the 1D-shape score functions, and the results of the analysis performed on a set of test samples. All these figures confirmed the results obtained in the preceding sections, but using an alternative technique: as we move higher from the mean of the SFMS, lower-mass galaxies are smaller, more centrally concentrated, and have bluer centres; as we move lower from the mean of the SFMS, lower-mass galaxies are more extended with redder centres, see Figures (\ref{fig:lms_bms_9_10.3}) and (\ref{fig:lms_bms_10.3_15}).

%%%%%%%%%%%%%%%%%%%%%%%%%%%%%%%%%%%%%%%%%%%%%%%%%%%%%%%%%%

\section{The interplay between the radius, SFR and the gas content for a given \texorpdfstring{\ms}{M*}}
\label{secc:model}

In this paper, we find empirical evidence indicating that the effective radius \citep[$R_e$, see also][]{Wuyts+2011} and the structural properties, namely $\Sigma_1$, $n_r$, and $\nabla_{(g-i)}$, exhibit on average a non-monotonic 
%Joel behavior concerning 
behaviour as a function of their distance to the SFMS $\Delta_{\rm MS}$. Notice that it is anticipated that $R_e$ shows similar trends for a given mass, along with $\Sigma_1$ and $n_r$, as these parameters are statistically correlated. 

As discussed in the Introduction section, assuming that the gas mass is constant for a given $M_\ast$ (indeed, $M_g$ correlates tightly with $M_\ast$; see e.g., \citealp{Calette+2018} and below), the global KS law implies a negative monotonic relation between size and SFR. That is, more compact galaxies and higher density galaxies are more star-forming, ${\rm SFR}\propto R_e^{-0.84}$ or $R_e\propto {\rm SFR}^{-1.19}$. This is not what we observe on average in Figure \ref{fig:re_MS_mass_bins} and instead 
%Joel of 
shallower slopes, $\sim-0.3$, are found for %this 
the former relation. Is this an indication that the relationship between size and SFR for a given mass is inconsistent with the global KS law? In this Section we develop a phenomenological model based on the global KS law to empirically understand the above and the observed trends along the SFMS described in the preceding Sections. 
%In particular, in this section, we use galaxy size given \ms\ as the primary property to derive a phenomenological model that can empirically explain what drives the observed trends along the SFMS described in the preceding sections. 

Our discussion starts by introducing the global KS law for the total gas mass\footnote{Notice that the total gas mass from the original Kennicutt paper is given by $M_{\rm gas} = M_{\rm HI} + M_{\rm H_2}$, i.e., no correction for helium and metals were made.} \citep{Kennicutt+1998,Kennicutt+2012,de_los_Reyes_Kennicutt2019} %\var{no estoy seguro. 1) Schmidt hablaba de densidad volumétrica, no superficial. 2) Kennicutt98 usó creo el radio efectivo y tanto para gas como para SFR}. 
This law establishes a connection between a global characteristic gas surface density and a global characteristic surface SFR, 
\begin{equation}
    \Sigma_{\rm SFR} = A_{\rm KS} \Sigma_{\rm gas}^{n},
\end{equation}
where the normalization factor has a value of $A_{\rm KS} = 2.5\times10^{-4} \pm0.7 \; M_{\odot} {\rm yr}^{-1} {\rm kpc}^{-2}$ for a \citet{Salpeter1955} IMF and the exponent $n=1.42$ \citep[][for a recent determination but similar to the orginal one, after accounting differences in IMF, see \citealp{de_los_Reyes_Kennicutt2019}]{Kennicutt+1998}. The global surface density quantities in the KS law are derived by computing the mass in gas and the SFR contained within a star-forming region of radius $R_{\rm SF}$ encompassing approximately $\sim95\%$ of the H$_\alpha$ flux \citep{de_los_Reyes_Kennicutt2019},
$\Sigma_{\rm gas} = M_{\rm gas}(<R_{\rm SF}) / \pi R_{\rm SF}^2 $ and $\Sigma_{\rm SFR} = {\rm SFR}(<R_{\rm SF}) / \pi R_{\rm SF}^2 $. By definition ${\rm SFR}(<R_{\rm SF}) = 0.95 \times {\rm SFR}$.
Notice that the star formation radius can be, on average, as large as a factor of $\sim1.7$ compared to the half-light radius $R_e$ for SFMS galaxies, but it is a trend that can vary with mass \citep[see Fig \ref{fig:R_halpatoRe_vs_Ms} Appendix \ref{app:fits_to_KSlaws} and also][]{Salim+2023}. The expected relationship between SFR and the star-forming radius is described by the equation: 
 \begin{equation}
     {\rm SFR} = \frac{A_{\rm KS}}{\pi^{n-1}} \frac {M_{\rm gas}^{n}( < R_{\rm SF})} {10^{6n}R_{\rm SF}^{2(n-1)}} = \frac{A_{\rm KS}}{\pi^{0.42}} \frac{ M_{\rm gas}^{1.42}( < R_{\rm SF})} {10^{8.52} R_{\rm SF}^{0.84}}.
     \label{eq:ks_law_all_terms}
 \end{equation}
 Here, we utilized the value of the KS exponent $n=1.42$ and the term $10^{6n}$ is related to the fact that sizes are used in units of kpcs. Notice that all the quantities involved in Equation (\ref{eq:ks_law_all_terms}) depend on stellar mass, particularly emphasizing the relationship with gas mass. 
 
\begin{figure}
    \includegraphics[height=2.96in,width=3.6in]{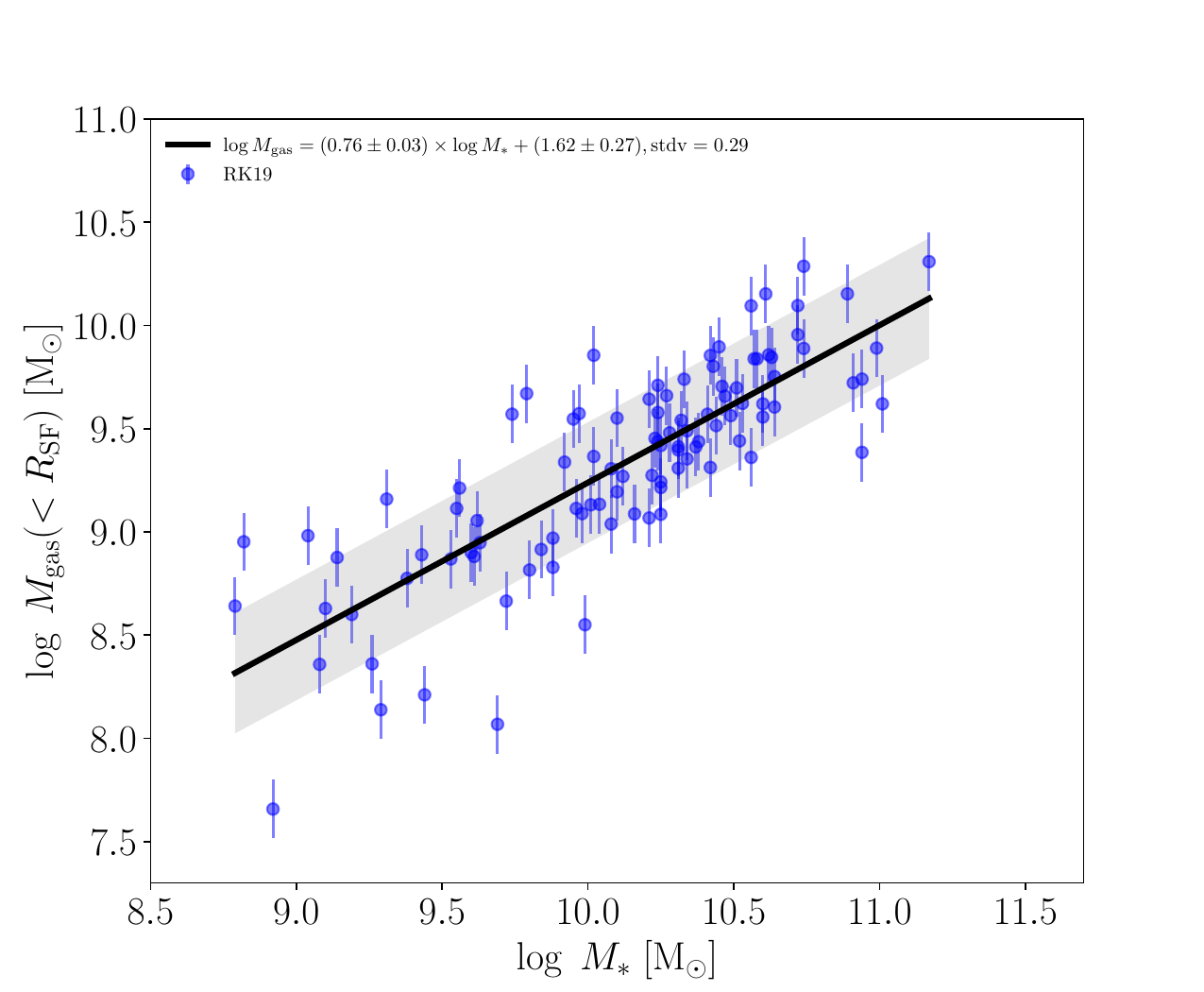}
    \caption{The gas mass within the star-forming radius $R_{\rm SF}$ is shown as a function of stellar mass from \citet{de_los_Reyes_Kennicutt2019} for SFMS. There is a strong correlation between gas mass and stellar mass, with small dispersion (including random errors) around it ($\sigma_{\rm gas} = 0.29$ dex). This indicates that, for a given stellar mass, the gas mass (witihin $R_{\rm SF}$) remains approximately fixed.}
\label{fig:Mgas_within_RSF_Ms}    
\end{figure}

 We utilize the data provided by \citet{de_los_Reyes_Kennicutt2019} from nearby spirals to analyze the gas mass within $R_{\rm SF}$ relative to stellar mass. To ensure consistency with our adopted IMF, we adjust the stellar mass by subtracting 0.24 dex and the SFRs by subtracting 0.05 dex from the values reported by \citet{de_los_Reyes_Kennicutt2019}. They originally employed a \citet{Salpeter1955} IMF using the estimator proposed by \citet{Eskew+2012} for stellar mass determination and a \citet{Kroupa2001} IMF with the estimator from \citet{Murphy+2011} for SFR estimation. Additionally, we select SFMS galaxies as described in Section \ref{secc:SF_MS}. Figure \ref{fig:Mgas_within_RSF_Ms} shows the strong correlation between $M_{\rm gas}( < R_{\rm SF})$ and $M_\ast$ for SFMS given by 
 \begin{equation}
     \log \left( \frac{M_{\rm gas}}{M_{\odot}} \right) = (0.76\pm 0.03) \times \log \left( \frac{M_{\ast}}{M_{\odot}} \right) + (1.62 \pm 0.27),
 \end{equation}
 and with a small dispersion around it of $\sigma_{\rm gas} = 0.29$ dex.
 Consequently, when fixing the stellar mass, the gas mass remains approximately constant \citep[see also][]{Calette+2018}. Additionally, fixing the stellar mass helps eliminate any extra correlation between mass and the SFR and size. In conclusion, we can reasonably assum that Equation (\ref{eq:ks_law_all_terms}) reduces to ${\rm SFR} \propto R_{\rm SF}^{-0.84}$ for a given stellar mass. Since our objective is to correlate the SFR with $R_e$, the next step is to establish a relationship between $R_{\rm SF}$ and $R_e$. 
 
 As discussed in \citet{de_los_Reyes_Kennicutt2019}, $R_{\rm SF}$ is a factor of $\approx 1.83$ smaller than the \citet{Holmberg1950} radius in nearby spirals. According to the fit provided by \citet{Salim+2023}, the Holmberg radius is given by $ R_{\rm 25} \propto R_e^{0.333}\times M_\ast^{0.188}$. The above implies that $R_{\rm SF} \propto R_{\rm 25}$ and therefore at fixed $M_\ast$ the KS law, Eq. (\ref{eq:ks_law_all_terms}), gives
\begin{equation}
    {\rm SFR}\propto R_e^{-0.28}. 
    \label{ec:SFR_Re_law}
\end{equation}
This relation is indeed closer to the ones found in Figure \ref{fig:re_MS_mass_bins} (cyan dashed lines) and consistent with shallower trends with the half-light radius. Therefore, we conclude that the global KS law is indeed consistent with the relationship between size and SFR described in 
%Joel the preceding 
Section \ref{secc:Results}. 

%%%%%%%%%%%%%%%%%%%%%%%%%%%%%%%%%%%%%%%%%%%%%%%%%%%%%%%%%%%%%%
%%%%%%%%%%%%%%%%%%%%%%%%%%%%%%%%%%%%%%%%%%%%%%%%%%%%%%%%%%%%%%

Next, we expand further our discussion and we employ the global KS law to calculate gas masses from SFRs and sizes, that is, we will use the inverted global KS law to obtain $M_{\rm gas}$, to further explore the implications of the global KS law. Due to scatter in the KS law, inverting this relationship involves considering the scatter around the relation. Unfortunately, the literature lacks a comprehensive characterization of the inverted global KS law. To address this, we utilize data from \cite{de_los_Reyes_Kennicutt2019}, who revisited the global KS law for spiral galaxies, and the compilation and homogenization from Calette et al. (in prep).\footnote{In our study, we utilize a sub-sample compiled by \cite{Calette+2018}, which heavily relies on the xCOLDGASS sample by \citet{Saintonge+2017}. This sub-sample is defined in a manner similar to that described in the main body of this paper,  selecting SFMS galaxies and requiring detections in both $M_{\rm HI}$ and $M_{\rm H_2}$. It is worth noting that in our forthcoming work, Calette et al. (in prep.), we employ the conversion factor between CO luminosity and H$_2$ mass from \citet{Accurso+2017}.} Based on the information from these two papers, we perform an unweighted linear regression, expressed as 
\begin{equation}
    \Sigma_{\rm gas} = \alpha_{\rm KS} \Sigma_{\rm SFR}^{\eta}.
\end{equation}
The results show $\eta =0.44\pm 0.02$ and $\log ( \alpha_{\rm KS} / M_{\odot} \; {\rm pc}^{-2}) = 2.09\pm 0.04$ with a dispersion, including random errors, of 0.19 dex. It is important to note that $n \neq 1 / \eta$. For further details, refer to Appendix \ref{app:fits_to_KSlaws}. 

\begin{figure}
    \includegraphics[height=2.93in,width=3.6in]{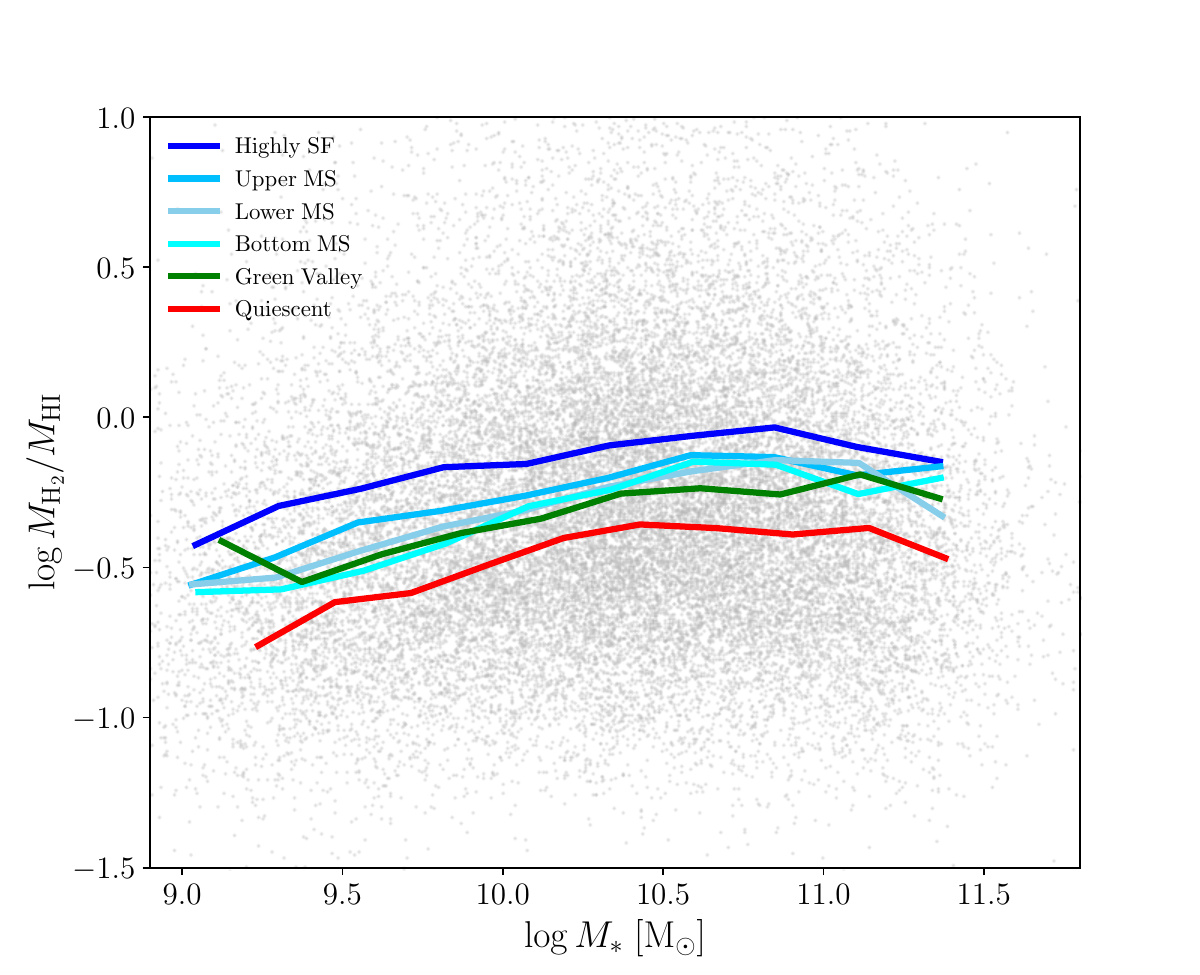}
    \caption{Predicted H$_{2}$/HI ratios as a function of \ms, see the text and Appendix \ref{app:fits_to_KSlaws} for details. The gray dots show the distribution of SDSS central galaxies while the solid lines show the avearge H$_{2}$/HI ratios for each central galaxy subsample of SFMS galaxies, see Table \ref{tab:sfrs_sub-samples}. HSF galaxies exhibit larger H$_{2}$/HI ratios followed by UMS. No clear trends are observed between LMS BMS and GV galaxies while quiescent galaxies have the lowest ratios.}
\label{fig:H2_to_HI_ratio_predicted}    
\end{figure}

\begin{figure*}
    \centering
    \includegraphics[height=4.in,width=7.5in]{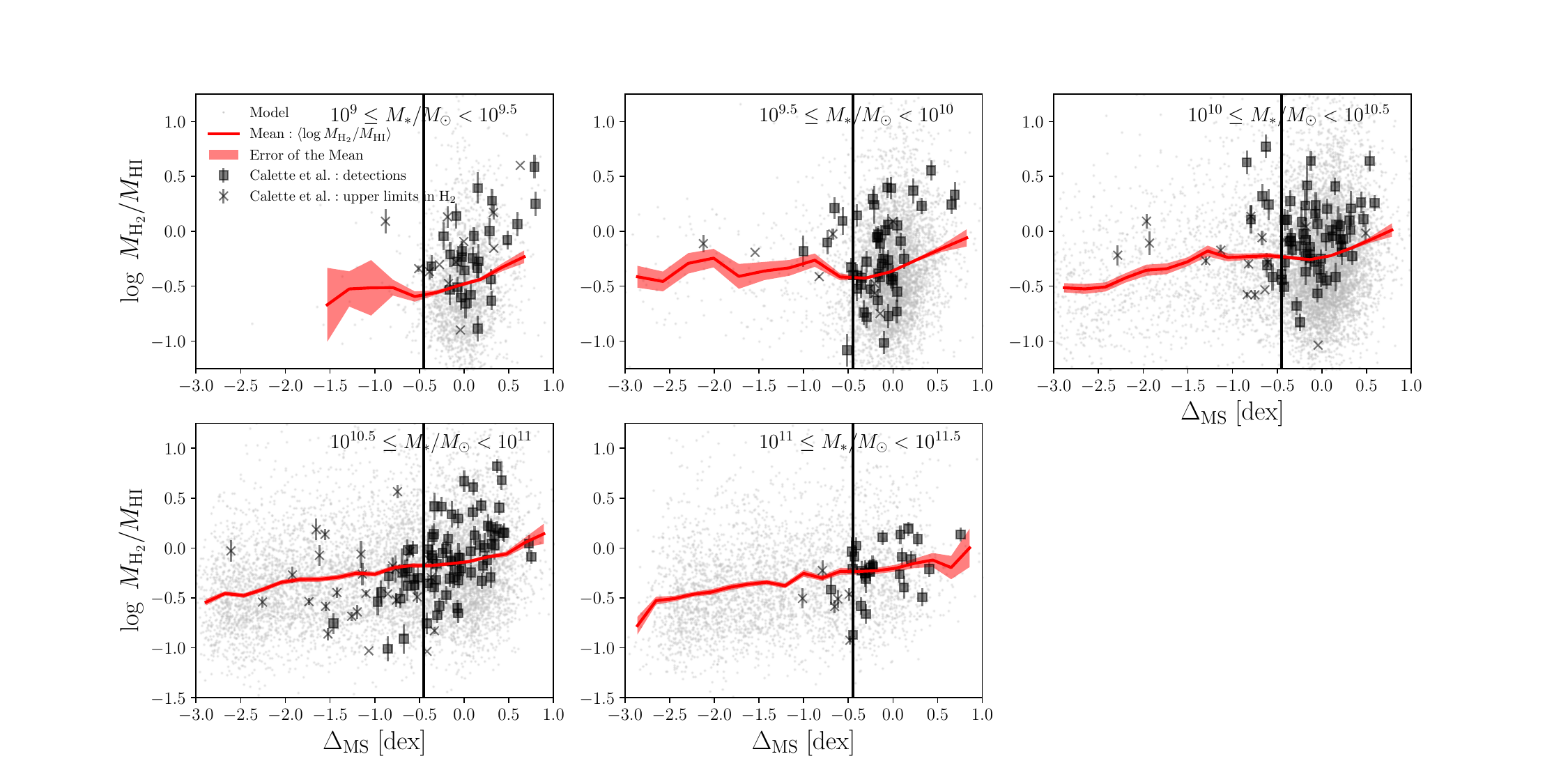}
    \caption{Predicted $\rm H_{2}$-to-$\rm HI$ mass ratio from the inverted global KS law and the $\Sigma_{\rm H_2}-\Sigma_{\rm SFR}-\Sigma_\ast$ relationship as a function of their distance to the SFMS, $\Delta_{\rm MS}$, for five different stellar masses. In each panel, gray dots are individual galaxies while the red solid lines show the mean $\log_{\rm H_2}/ M_{\rm HI}$ and the light red shaded are the error this mean. The solid line is the boundary that divides GV from SFMS galaxies. The filled square symbols show observations from the Calette et al. compilation with detections in both H$_2$ and HI gas masses while the skeletal symbols are those with upper limits in H$_2$ masses only. Note the monotonic trend between $\rm H_{2}$-to-$\rm HI$ and $\Delta_{\rm MS}$ for SFMS. This is more evident for galaxies with masses below $M_\ast<10^{10.5}\msun$. }

    \label{fig:MH2MHI_MS_mass_bins}    
\end{figure*}

As defined by the global KS law, the gas mass is composed of atomic, HI, and molecular, H$_2$, hydrogen. Molecular clouds, which account for most of the molecular hydrogen in galaxies, are the sites where the star formation occurs. To understand the relationship between SFR, gas mass and size 
%Joel
it is key to study the fraction of molecular hydrogen in galaxies. Our next step is to derive an empirical correlation between $M_{\rm H2}$, SFR and $M_{\ast}$ to predict the mass of molecular hydrogen in galaxies  \citep[see e.g.,][]{Feldmann2020}. Moreover, a study of this relationship has led \citet{Dou+2021} to find evidence  that the star formation laws observed in both starburst and SFMS galaxies may have similarities to those in quiescent galaxies \citep[see also][but see \citealp{Daddi+2010}]{Tacconi+2018, Tacconi+2020}. While we primarily focus on SFMS galaxies in this section, we use the above argument and show some results for quiescent galaxies in Figures \ref{fig:H2_to_HI_ratio_predicted} and \ref{fig:MH2MHI_MS_mass_bins}, although we do not discuss them in detail. We compute H$_2$ masses by using the following empirical relationship (details can be found in Appendix \ref{app:fits_to_KSlaws}),
\begin{equation}
    \Sigma_{\rm H_2}({\rm SFR},M_{\ast}) = \alpha_{\rm H_2} \times \Sigma_{\rm SFR}^{0.57\pm0.03} \times \Sigma_{\ast}^{0.46\pm0.05}.
    \label{eq:M_h2}
\end{equation}
Here $\log ( \alpha_{\rm H_2} / M_{\odot} \; {\rm pc}^{-2}) = -1.75\pm 0.47$ with a dispersion, including random errors, of 0.19 dex and as above we define the following surface densities: $\Sigma_{\rm H_2} = M_{\rm H_2} / \pi R_{\rm SF}^2 $ and $\Sigma_{\ast} = M_{\ast} / \pi R_{\rm SF}^2 $. 

It is interesting to note that adding the two exponents in Equation (\ref{eq:M_h2}) the result is $\sim1$. The above has two important consequences: 
\begin{itemize}
    \item H$_2$ mass does not depend or depends very weakly on radius. In general, if $ \Sigma_{\rm H_2} \propto \Sigma_{\rm SFR}^{\alpha} \times \Sigma_{\ast}^{\beta}$ and $\alpha + \beta = 1$ then this is equivalent to $M_{\rm H_2} \propto {\rm SFR}^{\alpha} \times M_\ast^{\beta}$ such that there is no dependence with $R_{\rm SF}$ (for the exact values from our fits, $M_{\rm H_2} \propto {\rm SFR}^{0.57\pm0.03}  M_{\ast}^{0.46\pm0.05} \times R_{\rm SF}^2/R_{\rm SF}^{2.06\pm 0.12} $). In other words, the H$_2$ depletion times, $t_{\rm dep, H_2} \equiv \Sigma_{\rm H_2} / \Sigma_{\rm SFR}$, are independent of radii for SFMS galaxies \citep{Tacconi+2018},
    
    \item Dividing by \ms\ in both sides of the equation, then $M_{\rm H_2}/\ms \propto M_{\ast}^{0.03} {\rm sSFR}^{0.57}$, that is, the ratio $M_{\rm H_2}/\ms$ depends strongly on the sSFR of the galaxy, consistent with the results obtained by \citet{Dou+2021}.
\end{itemize}

So far, we have (i) a way to estimate the gas mass, $M_{\rm gas}$, of galaxies by means of the inverted global KS law using the SFR, $R_e$ and \ms\ of galaxies, and (ii) a way to estimate the molecular gas mass, $M_{\rm H_2}$, by means of Eq. (\ref{eq:M_h2}) using the SFR and \ms\ of galaxies. From both quantities, we can calculate the molecular gas fraction, $f_{\rm H_2}=M_{\rm H_2}/M_{\rm gas}$ or the H$_{2}$-to-HI mass ratio, $M_{\rm H_2}/M_{\rm HI}= f_{\rm H_2}/(1-f_{\rm H_2})$. Since depletion times are independent of radii, we anticipate that $f_{\rm H2}$, and therefore the H$_2$-to-HI ratio, depends mostly on SFR and the size of the star-forming region $R_{\rm SF}$ (which at the same time depends on \ms\ and $R_e$, see above and \citealp{Salim+2023}) as shown by the first equality in the equation below 
\begin{eqnarray}
    \label{eq:fH2_size_sfr}
    f_{\rm H_2}(\ms,{\rm SFR},R_e) &=& \frac{t_{\rm dep, H_2}(\ms,{\rm SFR})}{\alpha_{\rm KS}} \Sigma^{1-\eta}_{\rm SFR}\\ \nonumber
    & = &\frac{\alpha_{\rm H_2}}{\alpha_{\rm KS}} 
    \frac{\Sigma^{0.59}_{\rm SFR}}{{\rm sSFR}^{0.46}}, %\left(\frac{0.95\; \rm SFR}{\pi R_{\rm SF}^2} \right)^{1-\eta} 
\end{eqnarray}
where we used the definition of the H$_2$ depletion time, Eq. (\ref{eq:M_h2}), and the best values obtained for the inverted global KS law.

%%%%%%%%%%%%%%%%%%%%%%%%%%%%%%%%%%%%%%%%%%%%%%%%%%%%%%%%%%%%%%
%%%%%%%%%%%%%%%%%%%%%%%%%%%%%%%%%%%%%%%%%%%%%%%%%%%%%%%%%%%%%%

%%%%%%%%%%%%%%%%%%%%%%%%%%%%%%%%%%%%%%%%%%%%%%%%%%%%%%%%%%%%
%%%%%%%%%%%%%%%%%%%%%%%%%%%%%%%%%%%%%%%%%%%%%%%%%%%%%%%%%%%%

%%%%%%%%%%%%%%%%%%%%%%%%%%%%%%%%%%%%%%%%%%%%%%%%%%%%%%%%%%%%%%
\begin{figure*}
    \centering
    \includegraphics[height=4.in,width=7.5in]{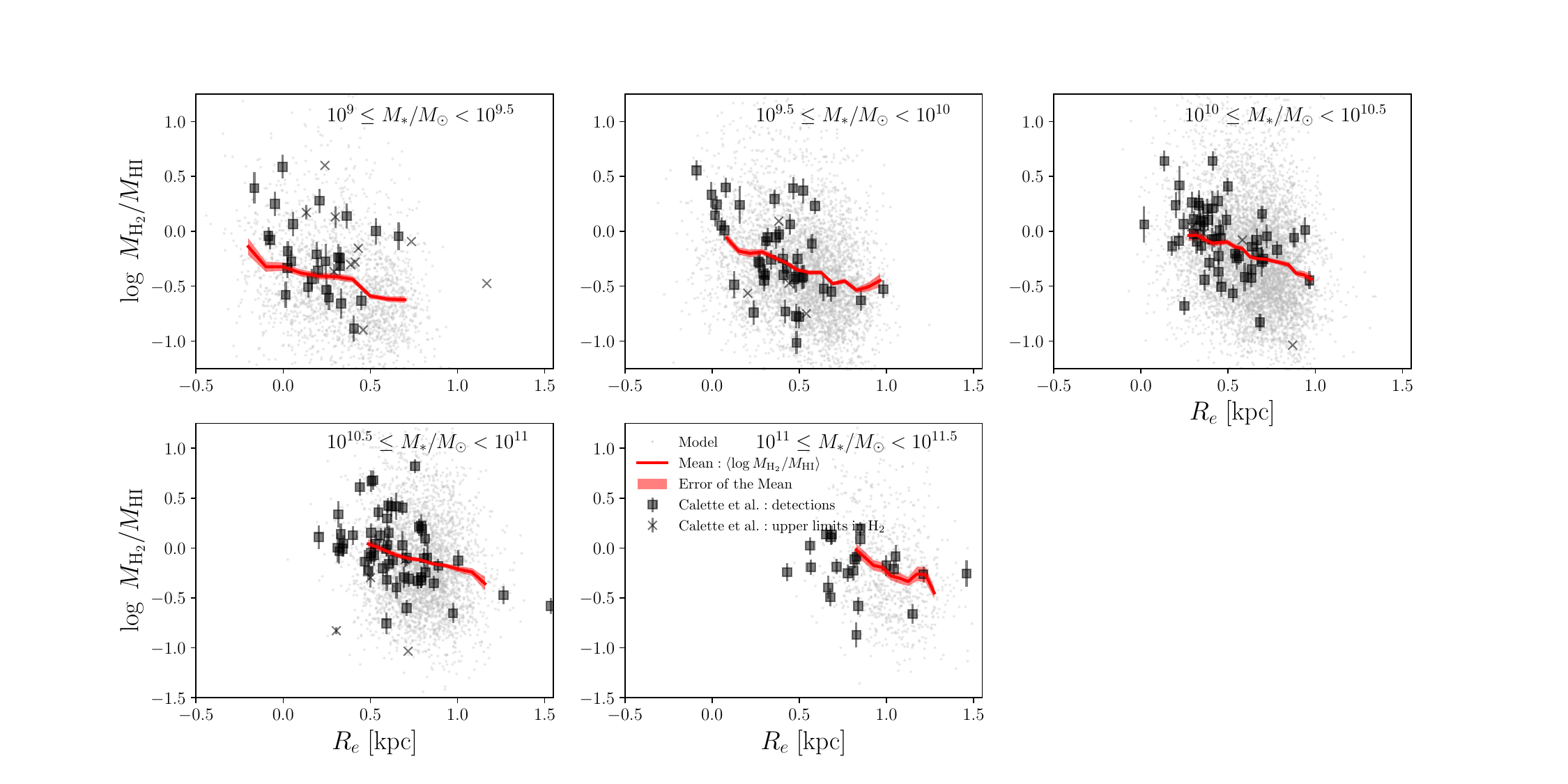}
    \caption{Predicted $\rm H_{2}$-to-$\rm HI$ mass ratio and the $\Sigma_{\rm H_2}-\Sigma_{\rm SFR}-\Sigma_\ast$ relationship as a function of $R_e$ for five different stellar masses and for SFMS only. Similarly to Fig. \ref{fig:MH2MHI_MS_mass_bins}, the solid lines are the mean of the log, the shaded area the error of the mean, and the symbols are from the Calette et al. compilation. On average, there is a negative monotonic trend between the $\rm H_{2}$-to-$\rm HI$ mass ratio and $R_e$. This is a consequence of the fact that in compact galaxies the conversion of HI into H$_2$ is more efficient.}
\label{fig:H2_to_HI_ratio_predicted_as_a_func_of_Re}    
\end{figure*}
%%%%%%%%%%%%%%%%%%%%%%%%%%%%%%%%%%%%%%%%%%%%%%%%%%%%%%%%%%%%%%

The predicted $M_{\rm gas}$ and $M_{\rm H_2}$ masses from our models are compared to the measured ones in Figure \ref{fig:model-vs-observations-gas}. The predictions nicely 
%Joel recovers 
recover the gas and molecular masses, $M_{\rm gas}$ and $M_{\rm H_2}$, respectively, for the majority of the galaxies with a random dispersion of $\sim0.2$ dex. 

Once we have constrained the best fit parameter for $\alpha_{\rm KS}$ and $\eta$, our attention turns to the SDSS catalog. Utilizing the inverted KS relation, $\Sigma_{\rm gas} \propto \Sigma_{\rm SFR}^{0.44\pm 0.02}$ and the molecular relation $\Sigma_{\rm H_2} \propto \Sigma_{\rm SFR}^{0.57\pm0.03} \times \Sigma_{\ast}^{0.46\pm0.05}$ we can deduce $M_{\rm gas}$, $M_{\rm H_2}$ and $M_{\rm HI}$ if the SFR, \ms\ and $R_{\rm SF}$ are known. 
%\citep[similarly to,][but with the difference that they used the 90th percentile $z$-band isophotal radius, $R_{90,z}$ and we are using the slope and normalization of the inverted KS law]{Peeples_Shankar+2011}. 
Notice that $M_{\rm gas}$ and $M_{\rm HI}$ represent the total gas and HI mass enclosed within the radius $R_{\rm SF}$, which are either lower or equal to their global masses within the entire galaxy. In the case of H$_2$ it represents the total mass. For our purposes, this distinction is sufficient, as our focus is on understanding the impact of the global KS law on the relationship between SFR, gas mass and size within the star forming regions.

Figure \ref{fig:H2_to_HI_ratio_predicted} shows the H$_2$-to-HI mass ratio predicted by our model as a function of stellar mass for our SDSS sample. The gray dots are individual galaxies and the solid lines are the mean $\langle \log M_{\rm H_2}/M_{\rm HI}\rangle$ ratios as a function of mass for our subsamples defined in Table \ref{tab:sfrs_sub-samples}. At all masses, HSF galaxies, on average, exhibit larger $M_{{\rm H}{2}}/M_{\rm HI}$ ratios compared to other SF clases. HSF are followed by UMS galaxies while LMS, BMS, and GV galaxies show similar ratios. Quiescent galaxies have the lowest ratios. Recall that $M_{\rm HI}$ has been derived within $R_{\rm SF}$. In general the trends are weak, differences between HSF and BMS are not larger than a factor of $\sim2$, but there are hints that this ratio displays a monotonic behaviour that is expected from the global KS law, that is, HSF galaxies have elevated values of star formation because they have a larger fractions of H$_2$ masses. A better way to see this is in Fig. \ref{fig:MH2MHI_MS_mass_bins}. 

%%Aquí me quede en correcciones. Me parece que la fig. 10 aporta poco y confunde mucho; basta con la 9 para mostrar la cierta segregación que hay en H2/HI a paridad de masa estelar por el grado de sSFR.  La que sí es informativa y concluyente es la fig. 11. 
Figure \ref{fig:MH2MHI_MS_mass_bins} shows the H$_2$-to-HI mass ratio as a function of $\Delta_{\rm MS}$ for five different mass bins as shown in the panels. The gray points are the predicted H$_2$-to-
HI mass ratios for individual galaxies in our sample while the red lines show the mean values and the 
%Joel
shaded areas are the errors of the mean. The black filled symbols show the observed values from the Calette et al. (in prep.) compilation. If we focus on galaxies in the SFMS, i.e., $\Delta_{\rm MS}>-0.45$, and below $3\times10^{10}\msun$, we notice that on average the H$_2$-to-HI mass ratio trends are weaker at $-0.45<\Delta_{\rm MS}<0$ and increases towards higher SFRs, in particular for HSF galaxies, consistent with Figure \ref{fig:H2_to_HI_ratio_predicted}. It increases a factor of $\sim2$ from $\Delta_{\rm MS}\sim0$ to $\Delta_{\rm MS}\sim0.6$ in the first mass bin and $\sim1.5$ and $\sim1.4$ times in the second and third mass bin, respectively. Above $3\times10^{10}\msun$, the H$_2$-to-HI mass ratio, on average, remains roughly constant for the fourth and fifth mass bins.

Figure \ref{fig:H2_to_HI_ratio_predicted_as_a_func_of_Re} shows the H$_{2}$-to-HI mass ratio, predicted by our model as function of half-light radius, $R_e$, for five different stellar masses. Notice that this Figure exclusively displays SFMS galaxies. This is because GV and quiescent galaxies at higher masses exhibit similar or even larger sizes than SFMS galaxies. This exclusion is important to prevent potential misleading conclusions that could arise from flattening trends between the H$_{2}$-to-HI mass ratio and sizes due to their comparable sizes. Figure \ref{fig:H2_to_HI_ratio_predicted_as_a_func_of_Re} shows that for galaxies with masses lower than $ \ms\leq 3\times 10^{10}\msun$, 
there is a clear and monotonic trend in each panel between $R_e$ and the ratio H$_{2}$-to-HI mass ratio. This trend goes in the direction that at a given stellar mass, those galaxies with higher values of H$_{2}$-to-HI mass ratio, and thus larger molecular gas ratios, are more compact. For example, the ratio of smaller galaxies, $R_e\sim 1.6$ kpc, is a factor of $\sim 2.7$ lower than larger galaxies $R_e\sim 5$ kpc, at masses of $\sim 2\times 10^{9}\msun$. At the second and third bins the differences between the smallest and the largest galaxies are respectively a factor of $\sim2$ and $\sim1.5$. As we approach to higher masses the H$_2$-to-HI mass ratio, on average, increases by a factor of $\sim1.6$ from $R_e\sim3$ to $\sim 10$ kpc at the fourth mass bin and by the same amount at fifth bin between $R_e\sim6$ and $\sim 17$ kpc. Our findings agree in a general way with those of \citet{Lin+2020}, who also pointed out  higher molecular gas fractions in smaller-radii galaxies, though our results differ quantitatively owing to the different scaling laws used. 

The trends observed in Figures from \ref{fig:H2_to_HI_ratio_predicted} to \ref{fig:H2_to_HI_ratio_predicted_as_a_func_of_Re} reveal that the H$_{2}$-to-HI mass ratio, and thus the fraction of molecular gas,  depends more on $R_e$ than on the SFR. This can be understood as follows: if in Equation (\ref{eq:fH2_size_sfr}) we substitute $R_{\rm SF}\propto R_e^{0.333} M_*^{0.188}$, then $f_{\rm H_2}\propto M_{\ast}^{0.24} {\rm SFR}^{0.13} / R_e^{0.4}$. At fixed stellar mass, galaxies with smaller radii are compact and denser. Notice that
\begin{equation}
    f_{\rm H_2}\propto {\rm sSFR}^{0.13} \mathcal{C}^{0.4}, 
    \label{eq:fh2_sfr_compact}
\end{equation}
where $\mathcal{C} = M_\ast/R_e$ is the compactness (and we omitted a small dependence of $M_\ast^{0.02}$ in $f_{\rm H_2}$). In other words, the primary relation is between the fraction of molecular gas and the compactness of the galaxy while the sSFR introduces some scatter around this relation. Nonethless, we stress that the above is just a consequence of the relationship between the star forming region with the half-light radius of the galaxy and the physical correlation is with the size of the star forming region since 
\begin{equation}
    f_{\rm H_2}\propto {\rm SFR}^{0.56} R_{\rm SF}^{-1.12}.
\end{equation}
Finally, if we assume that in SFMS galaxies the star formation is regulated by the local density of the interstellar medium, ISM, and the pressure of the ISM is determined by the density of the gas and stars, then $M_{\rm H_2}/M_{\rm HI} \propto P^{\alpha} $ with $\alpha = 0.92$ \citep{Blitz_Rosolowsky_2006}. Therefore, in high-pressured galaxies (which correspond to those more compact), the conversion of HI into H$_2$ is more efficient, which is what Equation (\ref{eq:fH2_size_sfr}) predicts. 

The important results of this section are summarized in Figures \ref{fig:MH2MHI_MS_mass_bins} and \ref{fig:H2_to_HI_ratio_predicted_as_a_func_of_Re}. According to the inverted global KS law and the $\Sigma_{\rm H_2}-\Sigma_{\rm SFR}-\Sigma_\ast$ correlation (Equation \ref{eq:M_h2}), which predicts the gas and molecular masses, galaxies with higher H$_{2}$-to-HI mass ratios correspond mostly to compact galaxies for a given \ms\ (Fig. \ref{fig:H2_to_HI_ratio_predicted_as_a_func_of_Re}). On the other hand, the H$_{2}$-to-HI mass ratio as a function of the SFMS deviation (Fig. \ref{fig:MH2MHI_MS_mass_bins}) has a monotonic behavior for galaxies less massive than $\sim 3\times 10^{10}$ \msun\ and a weaker correlation for more massive galaxies. Therefore, (i) the galaxy radius $R_e$ at a given \ms\ correlates well with the H$_{2}$-to-HI mass ratio, and (ii) its complex interplay with the SFR (or the deviation from the SFMS) is mostly a consequence of the behavior of the H$_{2}$-to-HI mass ratio with the SFR. These results imply that any interplay between SFR and radius for a given \ms\ is not primary, but a consequence of the more direct interplay between radius (or compactness) and the H$_{2}$-to-HI mass ratio, see Eq. (\ref{eq:fh2_sfr_compact}); the galaxy compactness seems to be a relevant property for the process of transforming HI into H$_{2}$, which is the first step to later form stars.

%%%%%%%%%%%%%%%%%%%%%%%%%%%%%%%%%%%%%%%%%%%%%%%%%%%%%%%%%%%%%%
%\section{Discussion}
%\label{secc:discussion}
\section{Paths for galaxy quenching}
\label{secc:quenching}

%%%%%%%%%%%%%%%%%%%%%%%%%%%%%%%%%%%%%%%%%%%%%%%%%%%%%%%%%%%%%%%%%%%%%
%%%%%%%%%%%%%%%%%%%%%%%%%%%%%%%%%%%%%%%%%%%%%%%%%%%%%%%%%%%%%%%%%%%%%
%\begin{figure*}
%    \includegraphics[height=4in,width=5in]{Figures/sSFR_ms_re.pdf}
%    \caption{The specific star formation rate (sSFR) as a function of compactness, $\mathcal{C} = \ms/R_e$, for our sample of central SDSS galaxies. Red dots represent galaxies with $\Delta_{\rm MS}\leq -1$ and $-1<\Delta_{\rm MS}< -0.45$, indicating respectively quiescent and GV galaxies, while blue dots denote galaxies on the SFMS with $\Delta>-0.45$. The solid lines show the mean relations $\langle {\rm sSFR} (M_\ast/R_e) \rangle$ in narrow bins of 0.2 dex of constant stellar mass for SFMS. Dotted lines illustrate the predicted relation from the KS law for a fixed stellar mass and given gas mass. For galaxies below $M_{\ast}\sim 2\times 10^{10} \msun$, the predicted slope from the KS law aligns well with observations, while at larger masses, the observed slopes are shallower compared to the KS law. In addition, most of the quiescent galaxies are confine within the range $10 < \log \mathcal{C} < 10.5$.}
%  \label{fig:ssfr_ms_re_relation}    
%\end{figure*}
%%%%%%%%%%%%%%%%%%%%%%%%%%%%%%%%%%%%%%%%%%%%%%%%%%%%%%%%%%%%%%%%%%%%%

%%%%%%%%%%%%%%%%%%%%%%%%%%%%%%%%%%%%%%%%%%%%%%%%%%%%%%%%%%%%%%
\begin{figure*}
    \centering
    \includegraphics[height=4.in,width=8in]{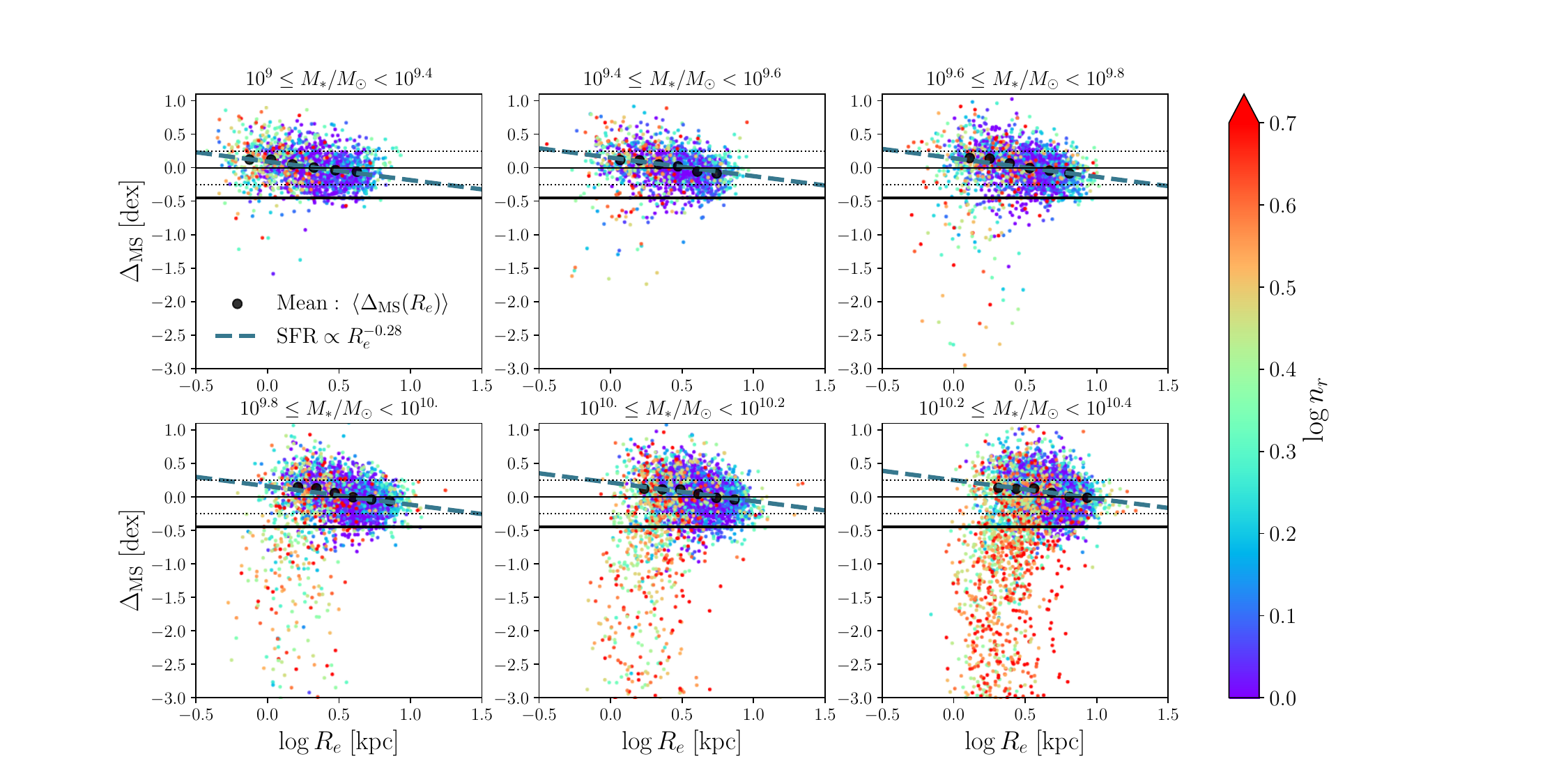}
    \caption{Distance to the main sequence, $\Delta_{\rm MS}$, as a function of the half-light radius color-coded by the sersic index for galaxies below $\sim M_c$. The thick horizontal black line at $\Delta_{\rm MS}=-0.45$ dex separates GV and quiescent stages from the SFMS, while the thin and dotted lines are respectively the $\Delta_{\rm MS} = 0$ and $\Delta_{\rm MS} = \pm 0.25$. The dashed line shows the expected relationship between size and SFR from the global KS law, SFR$\propto R_e^{-0.28}$, see Section \ref{secc:model}. The zero point of this relationship was calibrated using the mean trends of $\Delta_{\rm MS}(R_e)$, black circles. Galaxies positioned above the mean SFMS, particularly those categorized as HSF, tend to exhibit smaller sizes and higher Sersic indexes, $n_r>2.5$, similar to those of GV and quiescent galaxies. These galaxies are presumed to quench rapidly by changing undergoing a morpho-structural change as a SFMS. These galaxies are expected to maintain their structural parameters, serving as progenitors of quiescent galaxies with higher Sersic indexes.}
\label{fig:sersic_ms_SFR_Re}    
\end{figure*}
%%%%%%%%%%%%%%%%%%%%%%%%%%%%%%%%%%%%%%%%%%%%%%%%%%%%%%%%%%%%%%

%%%%%%%%%%%%%%%%%%%%%%%%%%%%%%%%%%%%%%%%%%%%%%%%%%%%%%%%%%%%%%
\begin{figure*}
    \centering
    \includegraphics[height=4.in,width=8in]{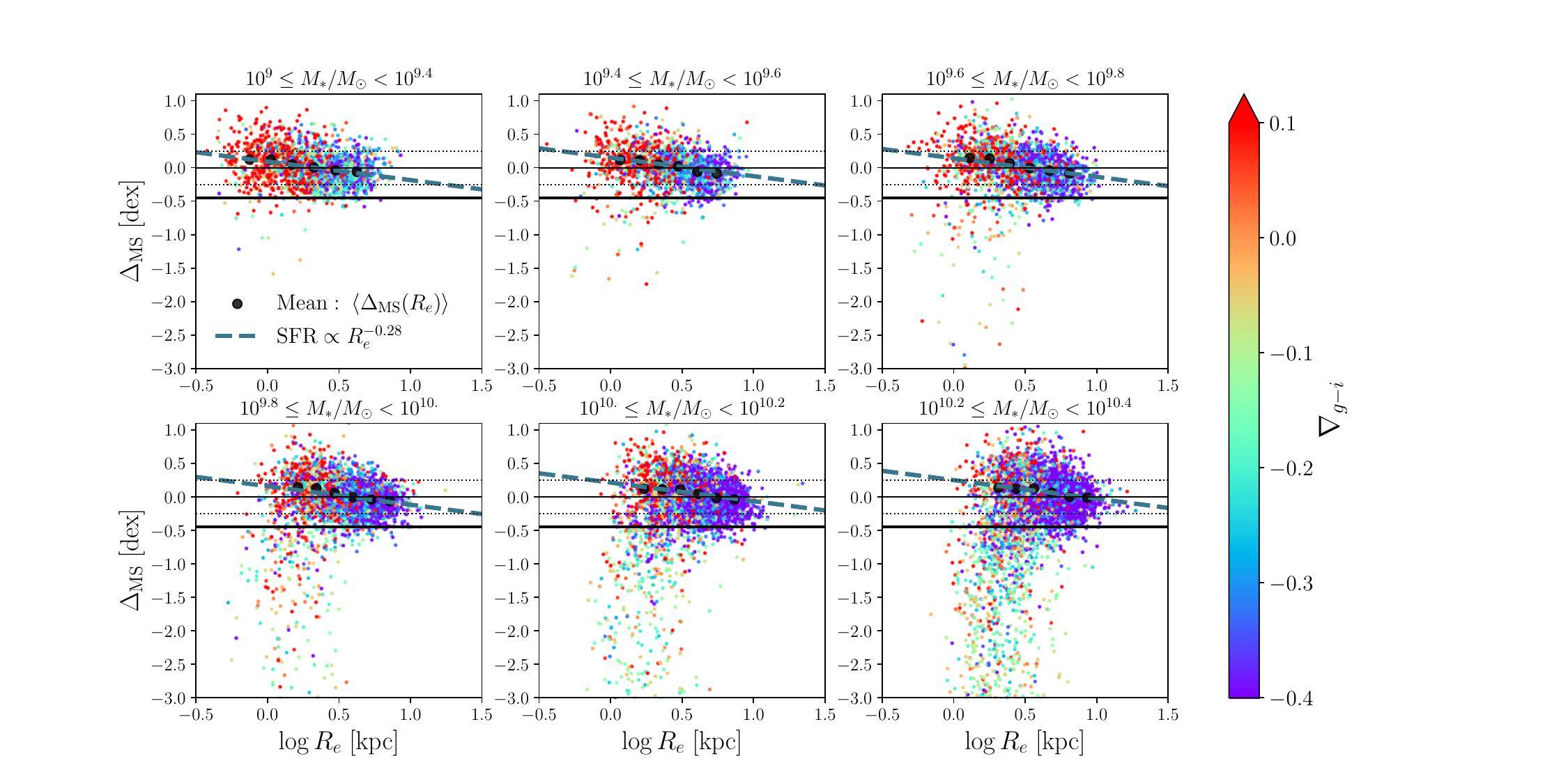}
    \caption{Similar to Figure \ref{fig:sersic_ms_SFR_Re}, we plot the distance to the main sequence, $\Delta_{\rm MS}$ as a function of the half-light radius with galaxies color-coded according to the color gradient for galaxies below $\sim M_c$. Galaxies above the SFMS tend to exhibit smaller sizes and bluer/star-forming centres, in particular those classified as HSF galaxies. Their sizes, Sersic indexes (see Figure \ref{fig:sersic_ms_SFR_Re}) and CGs resemble those of the GV and quiescent ones, reinforcing the idea that galaxies above the SFMS are progenitors of the GV and quiescent galaxies. As discussed in the text, these galaxies are expected to quench rapidly by undergoing morpho-structural changes as they transition from the SFMS to the GV or to a more quiescent phase, see Figure \ref{fig:sersic_ms_SFR_Re}}
\label{fig:grad_color_ms_SFR_Re}    
\end{figure*}
%%%%%%%%%%%%%%%%%%%%%%%%%%%%%%%%%%%%%%%%%%%%%%%%%%%%%%%%%%%%%%

%%%%%%%%%%%%%%%%%%%%%%%%%%%%%%%%%%%%%%%%%%%%%%%%%%%%%%%%%%%%%%
\begin{figure*}
    \centering
    \includegraphics[height=4.in,width=5.5in]{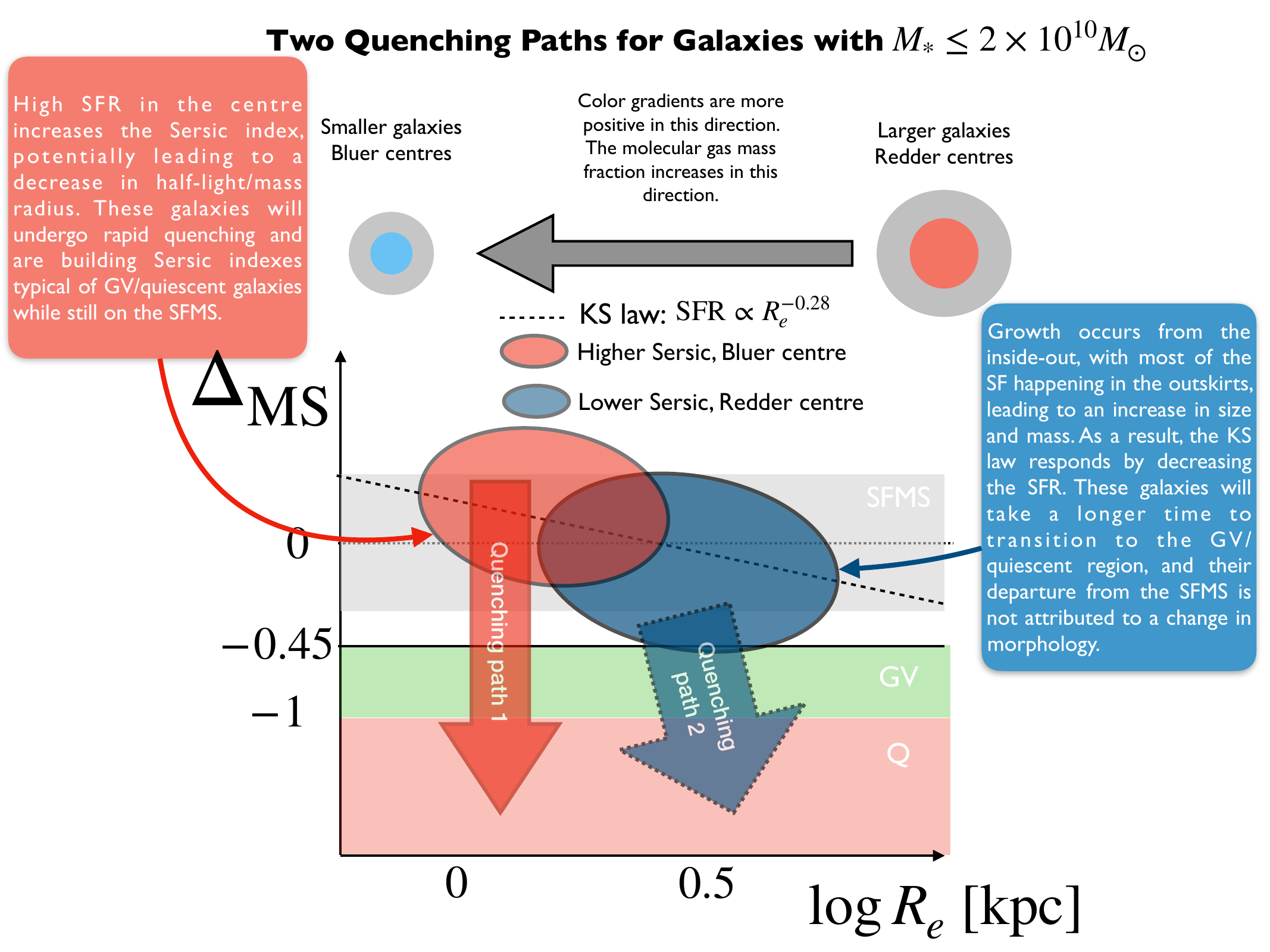}
    \caption{Schematic figure explaining the most important trends observed in Figures \ref{fig:sersic_ms_SFR_Re} and \ref{fig:grad_color_ms_SFR_Re}. Two paths for galaxy quenching are proposed for galaxies with $M_\ast\leq 2\times 10^{10} M_\odot$. {\bf Quenching Path 1:} Galaxies above the mean of the SFMS are smaller and compact, with bluer centers and higher Sersic indices and H$_2$-to-HI mass ratios. These galaxies exhibit structural similarities to GV and quiescent galaxies. This suggests that galaxies above the mean of the SFMS, particularly HSF galaxies, are the progenitors of GV and quiescent galaxies. If this is the case, the transition from the SFMS phase should occur quickly in these galaxies, in the direction indicated by the red vertical arrow. Additionally, they undergo a morpho-structural change before leaving the SFMS. {\bf Quenching Path 2:} Galaxies below the mean SFMS are larger, with smaller Sersic indices, redder colors, and lower H$_2$-to-HI mass ratios. Most of the SF occurs in the outskirts, leading to an increase in their half-light radii and mass. For these galaxies, it will take longer to transition from the SFMS to the GV quiescent region. These galaxies are expected to leave the SFMS, as indicated by the blue arrow with the dotted border. The tilt of the arrow is due to the slow nature of this second path, causing galaxies to gain more mass before transitioning into the GV/quenching region at higher masses. Their departure is not attributed to a change in morphology. }
\label{fig:cartoon_quenching_paths}    
\end{figure*}
%%%%%%%%%%%%%%%%%%%%%%%%%%%%%%%%%%%%%%%%%%%%%%%%%%%%%%%%%%%%%%

%%%%%%%%%%%%%%%%%%%%%%%%%%%%%%%%%%%%%%%%%%%%%%%%%%%%%%%%%%%%%%
\begin{figure*}
    \centering
    \includegraphics[height=4.in,width=8in]{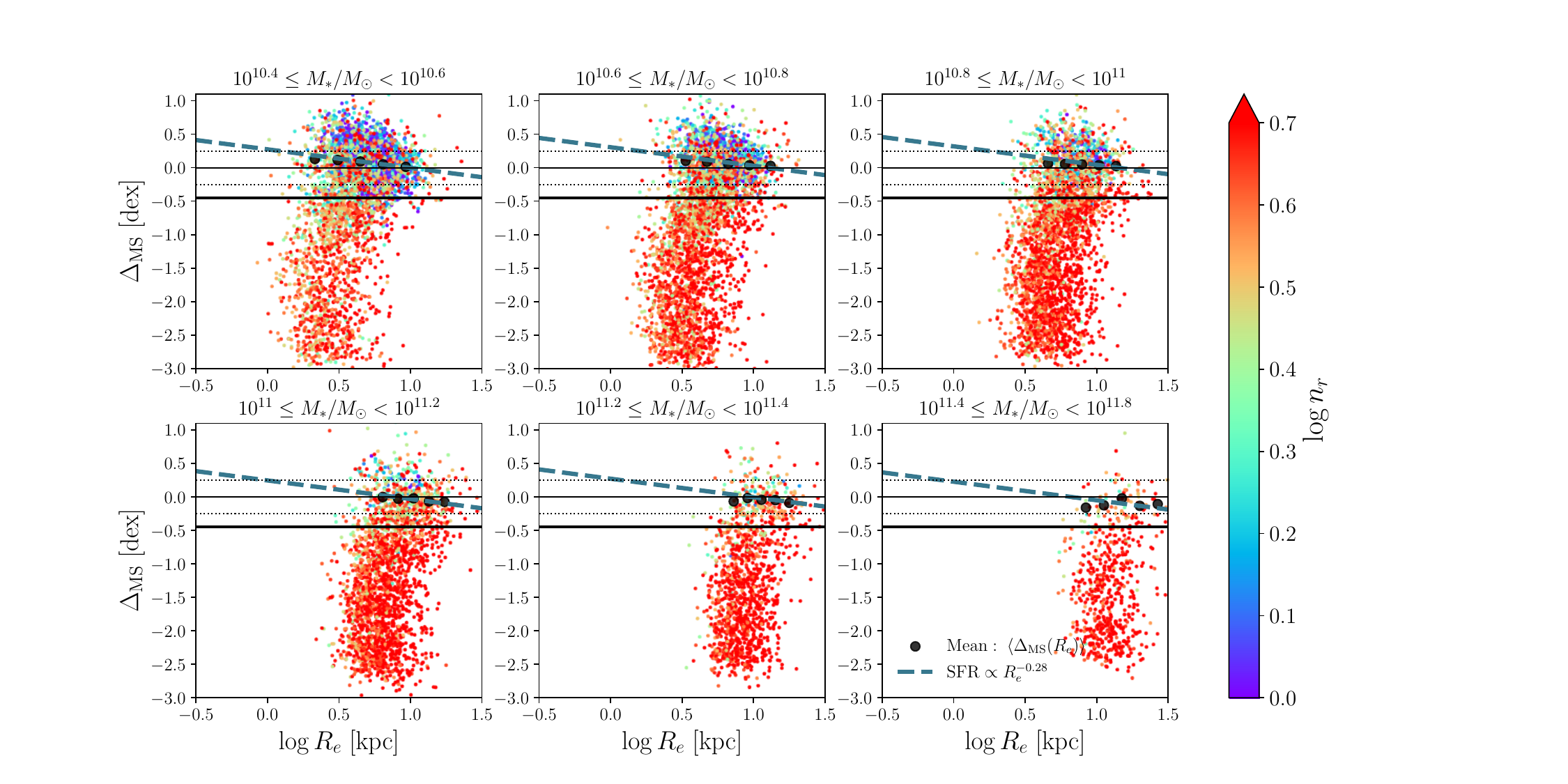}
    \caption{Similar to Figure \ref{fig:sersic_ms_SFR_Re}, the distance to the main sequence, $\Delta_{\rm MS}$, as a function of the half-light radius is shown by color-coding by galaxies sersic index for galaxies above $\sim M_c$. High-mass galaxies exhibit shallower trends between the sSFR and $R_e$ compared to the slope predicted by the KS law. Moreover, a significant portion of galaxies above $\ms \sim 3 \times 10^{10} M_{\odot}$ display higher Sersic indexes consistent with those of GV and quiescent galaxies. This observation underscores that structural transformations occur even within the SFMS.}
\label{fig:sersic_ms_SFR_Re_high_mass}    
\end{figure*}
%%%%%%%%%%%%%%%%%%%%%%%%%%%%%%%%%%%%%%%%%%%%%%%%%%%%%%%%%%%%%%

%%%%%%%%%%%%%%%%%%%%%%%%%%%%%%%%%%%%%%%%%%%%%%%%%%%%%%%%%%%%%%
\begin{figure*}
    \centering
    \includegraphics[height=4.in,width=8in]{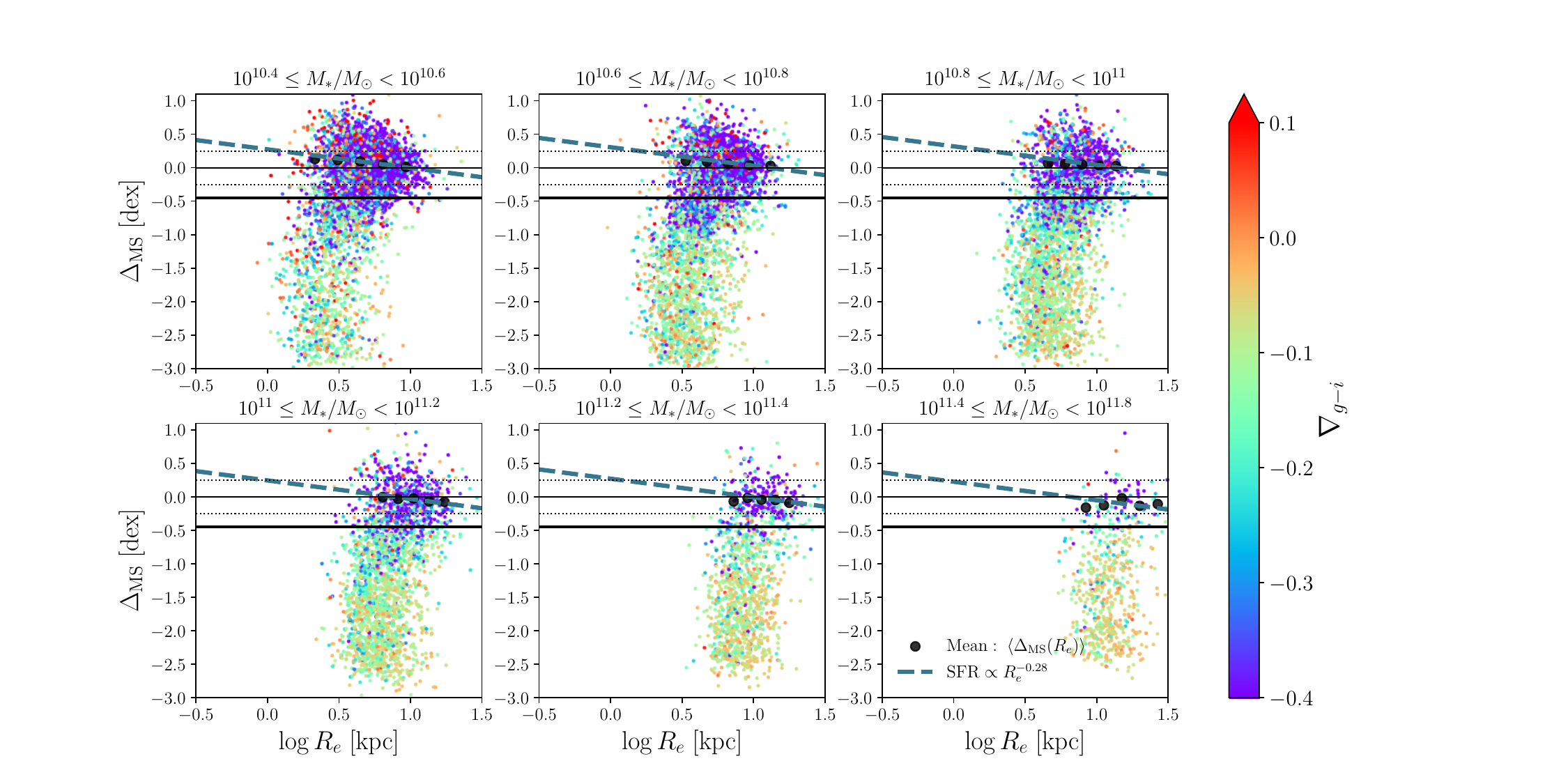}
    \caption{Similar to Figure \ref{fig:grad_color_ms_SFR_Re}, we plot the distance to the main sequence, $\Delta_{\rm MS}$ as a function of the half-light radius with galaxies color-coded according to the color gradient for galaxies above $\sim M_c$. High-mass galaxies not only have higher Sersic indexes but also very negative color gradients indicating that most of the star formation is occurring in the outskirts of these galaxies. Observe how color gradients are strongly bimodal.}
\label{fig:grad_color_ms_SFR_Re_high_mass}    
\end{figure*}
%%%%%%%%%%%%%%%%%%%%%%%%%%%%%%%%%%%%%%%%%%%%%%%%%%%%%%%%%%%%%%

%%%%%%%%%%%%%%%%%%%%%%%%%%%%%%%%%%%%%%%%%%%%%%%%%%%%%%%%%%%%%%]]
\begin{figure*}
    \includegraphics[height=4.in,width=8in]{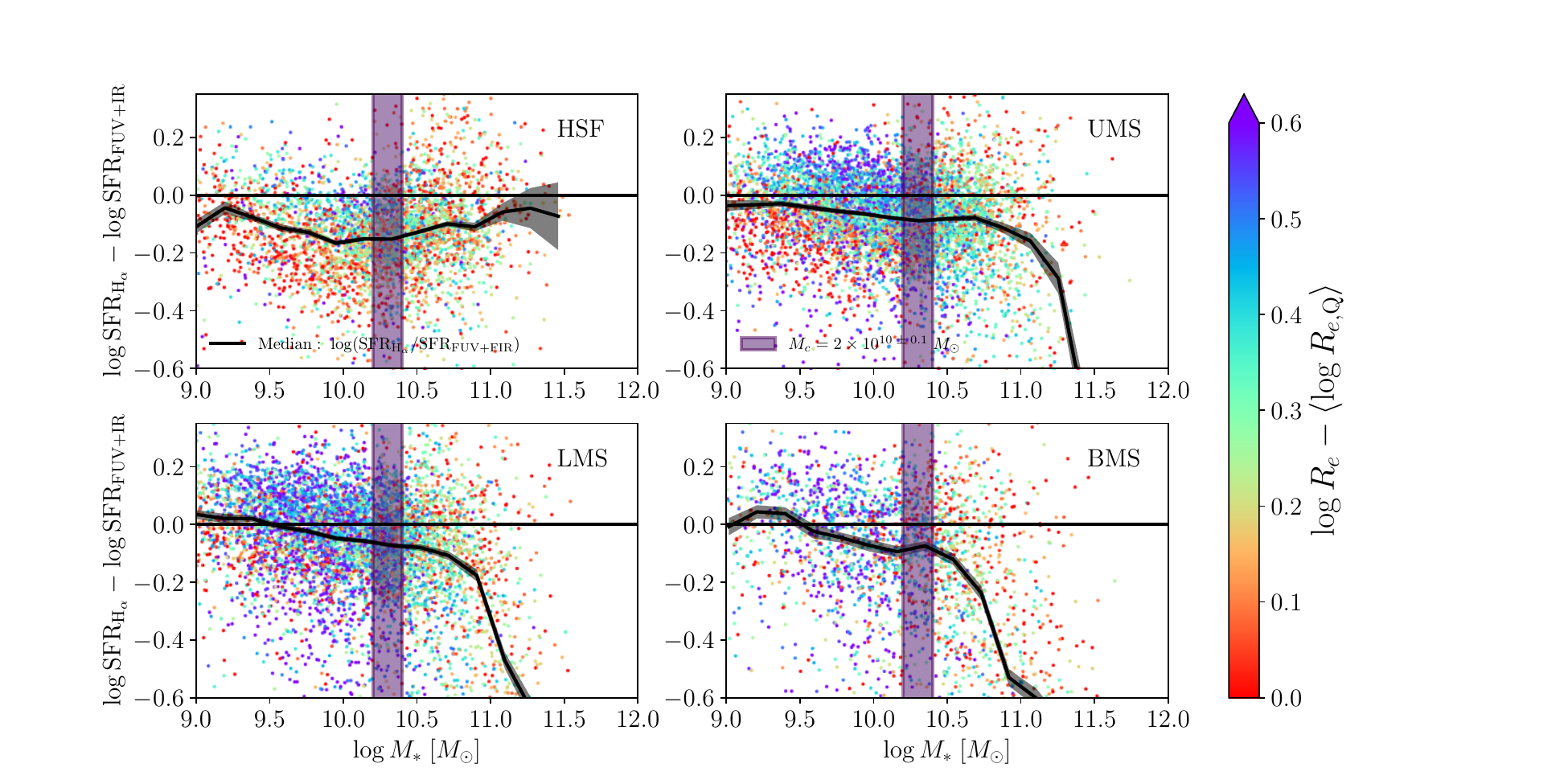}
    \caption{Difference between $\log{\rm SFR}$ determined from  H$\alpha$ and FUV+FIR for SFMS, with SFRs from \citealp{Brinchmann+2004} and \citealp{Salim+2018}, respectively. H$\alpha$ SFRs probe scales between $\sim 6-10$ Myrs, while FUV+FIR SFRs probe $\sim 100$ Myrs, representing different points in the star formation history of galaxies. Each galaxy is colour coded by their distance to the mean $\langle \log R_{e,{\rm Q}} (M_\ast)\rangle$ of quiescent galaxies, see Equation (\ref{eq:size_mass_Q}). Negative differences indicate a decline in recent star formation history, while positive differences indicate an increase. When the two are equal, the star formation history is constant. The solid red line shows the median relation while the shaded area represents a poissonian error. For masses below $M_{\rm c}$, high star-forming, HSF, galaxies tend to have decreasing star formation histories, SFHs, at all masses. Similarly, upper main sequence galaxies, UMS, have decreasing SFHs but to a lesser degree. Galaxies classified as lower main sequence, LMS, and at the bottom of the sequence, BMS, have increasing star formation histories at higher masses. AT higher masses than $M_{\rm c}$, all the four categories have decreasing SFHs, but notice that the most massive HSF galaxies
    %Joel seem to have be 
   have closer to constant SFHs.}
\label{fig:ssfr_halpha_fir_fuv}    
\end{figure*}
%%%%%%%%%%%%%%%%%%%%%%%%%%%%%%%%%%%%%%%%%%%%%%%%%%%%%%%%%%%%%%

%The main goal of this paper was to investigate whether there is a relationship between the distance to the SFMS and the size and morpho-structural properties of galaxies. One of the key findings of our study is that there is a non-monotonic relationship between half-light/mass radius and SFR, Figures \ref{fig:ssfr_rer_mass} and \ref{fig:ssfr_res_mass}. This behavior is also observed in other structural properties, such as the colour gradient (Figure \ref{fig:gradgi_avg_Ms_MS}), the central mass density (Figure \ref{fig:S1_avg_Ms_MS}) and S\'ersic index (Figure \ref{fig:sersic_avg_Ms_MS}, see also Figures \ref{fig:lms_bms_9_10.3} and \ref{fig:lms_bms_10.3_15}).\footnote{This is expected since these other quantities are closely correlated with \re.} Additionally, we found that the H$_{2}$-to-HI mass ratio increases for galaxies above the SFMS (Figure \ref{fig:MH2MHI_MS_mass_bins}). Next, we interpret these results as evidence for the existence of two distinct paths for galaxy quenching.

%%%%%%%%%%%%%%%%%%%%%%%%%%%%%%%%%%%%%%%%%%%%%%%%%%%%%%%%%%%%%%
%\subsection{Paths for galaxy quenching}
%\label{secc:quenching}

The results presented in this paper consistently demonstrate a characteristic mass of $M_c = 2\times 10^{10}M_{\odot}$ at which the trends in the structural parameters of the galaxies (i.e., morphology) exhibit a notable change in behavior across the SFMS. This characteristic mass is close to the observed {\it bend} of the SFMS as constrained in Eq. (\ref{eq:SFMS_best_fit}), see Figure \ref{fig:sfr_ms_sample_and_fit}. For masses below $M_c$ the SFMS follows a power law ${\rm SFR}\propto M_\ast^{0.85}$, whereas at higher masses the SFMS flattens out, with ${\rm SFR}(\ms)\sim{\rm const}$. It is noteworthy that $M_c$ aligns with characteristic masses identified in prior studies, albeit somewhat smaller \citep[see also][for a similar realisation]{Daddi+2022}. For instance, \citet[][see also \citealp{Bell+2012}]{Kauffmann+2003a} identified a characteristic mass of $3\times 10^{10} M_{\odot}$ through the analysis of galaxy star formation activity and structural properties. Others have proposed characteristic masses based on analyses of the galaxy stellar mass function or the fraction above which quiescent galaxies represent more than $50\%$ of the population, ranging from approximately $4-8\times 10^{10} M_{\odot}$ \citep[see, e.g.,][]{Bell+2003, Bundy+2006, DroryAlvarez2008, Pozzetti+2010, Peng+2010, Baldry+2012,Ilbert+2013,Muzzin+2013}. Additionally, the maximum of the stellar-to-halo mass ratio at a given stellar mass is reached around $\sim 3.2\times 10^{10} M_{\odot}$, \citep[according to equation 66 from][]{Rodriguez-Puebla+2017} resulting in haloes of $M_{\rm vir}\sim 10^{12} M_{\odot}$. Note, however, that all these characteristic masses refer to transitions between one population dominating over the other. In our case, we emphasize that $M_c$ refers to a characteristic mass revealing substructure among the relationships between galaxy properties {\it within} the SFMS. With this distinction in mind, we proceed with our discussion.  

Our characteristic mass $M_c$ corresponds to a halo mass of $M_{\rm vir}\sim 10^{11.8} \msun$ when considering SFMS galaxies only \citep[see figure 12 from][see also \citealp{Rodriguez-Puebla+2015}]{Kakos+2024}. This halo mass, is right on the position where a stable virial shock is expected to form through which the cosmological inflowing gas must cross, resulting in the heating of the infalling gas \citep{Dekel+2006}. As a result, star formation is expected to be strongly inefficient above this halo mass. The work conducted by \citet{Daddi+2022} arrives at a similar conclusion through the analysis of the bend in the SFMS between $0\lesssim z\lesssim 4$. However, it is important to note that this conclusion does not encapsulate the entire story.  Structural properties of galaxies behave differently above this characteristic mass, suggesting the involvement of other factors. Below $M_c$, HSF galaxies display structural properties similar to the more compact and denser GV galaxies. Above $M_c$, galaxies at the of the BMS now exhibit similarity to the structural properties of GV galaxies. Therefore, the capacity of galaxies to alter their structure is crucial in understanding the various mechanisms that cause galaxies to deviate from SFMS. 

Next, we divide our discussion into low and high mass galaxies according to $M_c = 2\times 10^{10}M_{\odot}$.

\subsection{Low-Mass Galaxies: Two paths for galaxy quenching}

Figures \ref{fig:sersic_ms_SFR_Re} and \ref{fig:grad_color_ms_SFR_Re} present the distance to mean SFMS, $\Delta_{\rm MS}$, as a function of the half-light radius, $R_e$ across 6 mass bins for galaxies with masses below $M_c$. The data is color-coded by the Sersic index, $n_r$, and color gradient, $\nabla_{(g-i)}$, respectively. In both figures, a horizontal black line is drawn at $\Delta_{\rm MS}=-0.45$ dex to separate the GV and quiescent stages from the SFMS. The horizontal dotted lines are $\Delta_{\rm MS} = \pm 0.25 $ dex as a representation of the $1\sigma$ distribution of the SFMS. Additionally, the blue dashed line represents the expected relationship between half-light radius and SFR from the global KS law: SFR $\propto R_e^{-0.28}$, see Equation \ref{ec:SFR_Re_law}. We calibrated the zero point of this relationship 
by fitting to the mean values of $\Delta_{\rm MS}$ vs $(R_e)$ (black circles). An important first conclusion drawn from Figures \ref{fig:sersic_ms_SFR_Re} and \ref{fig:grad_color_ms_SFR_Re} is that the \emph{size of galaxies contributes to the scatter around the SFMS}, as observed according to the KS law. 

Figures \ref{fig:sersic_ms_SFR_Re} and \ref{fig:grad_color_ms_SFR_Re} synthesize our findings for lower-mass galaxies, $\ms \lesssim M_c$. As we {\it move up} from the mean SFMS, galaxies are smaller and compact, with bluer centres and their light profile becomes more centrally concentrated, and have higher $f_{\rm H2}$ and therefore high H$_2$-to-HI mass ratios, see Figure \ref{fig:H2_to_HI_ratio_predicted_as_a_func_of_Re}. Moreover, galaxies above the SFMS exhibit structural similarities to those in the GV and quiescent galaxies. Particularly, HSF galaxies, see Figures \ref{fig:ssfr_rer_mass}-\ref{fig:sersic_avg_Ms_MS} and Figure \ref{fig:sersic_ms_SFR_Re} above. This suggests that HSF galaxies are progenitors of GV galaxies \citep[see also][for a similar conclusion]{Wuyts+2011}. In contrast, as we {\it descend} from the mean SFMS, galaxies exhibit a structural shift toward being more disk-like, $n_r\sim1$ with redder centres and accompanied by lower H$_2$-to-HI mass ratios, Figure \ref{fig:H2_to_HI_ratio_predicted_as_a_func_of_Re}. This outcome is counterintuitive, as one might expect galaxies at the bottom of the main sequence, BMS, to be natural candidates for progenitors of GV galaxies, or even quiescent galaxies, as they fade in SF. Our results thus strongly suggest that low-mass galaxies can be quenched through {\it two distinct processes}, one involving galaxies that will transition above the SFMS, and a second one involving galaxies that will transition at the BMS.

Figure \ref{fig:cartoon_quenching_paths} presents a schematic picture on how SMFS galaxies evolved into GV and quiescent galaxies, assuming that once SFMS galaxies cease their star formation, they do not change significantly their mass and size and Sersic index (i.e., mass distribution):

\begin{itemize}
    \item Low-mass galaxies above the SFMS as progenitors of GV galaxies: This process involves the production of a centrally concentrated mass and a bluer/star-forming centre implying that they undergo a morpho-structural change before transitioning out of the SFMS phase. The increase of SFR is due to the observed high H$_2$-to-HI mass ratios, Figure \ref{fig:H2_to_HI_ratio_predicted_as_a_func_of_Re}, most likely in the centre of the galaxy and smaller H$_2$ depletion times. This transition from the phase of actively forming stars to a more quiescent state is expected to occur relatively quickly in such galaxies. Indeed, as we will show in Figure \ref{fig:ssfr_halpha_fir_fuv} galaxies below the SFMS have decreasing SFHs. Mergers, tidal encounters \citep[e.g.,][]{Yesuf+2021}, or gravitational instabilities \citep[e.g.,][]{Cenci+2023} are physical processes that could lead to the enhancement of SFR in the central parts of galaxies and change their morphology, occurring over a timescale of $\sim 70$ Myrs \citep[]{Cenci+2023}. Additionally, HSF galaxies appear to host a significantly higher number of satellites compared to the average, approximately twice as many within their dark matter haloes \citep[]{Kakos+2024}.

    \item Low-mass galaxies below the SMFS as progenitors of GV and/or quiescent galaxies: The second process involves the maintenance of a low density in the centre, which results in an old/red centre that characterizes galaxies classified as BMS. This implies an inside-out evolutionary pattern, where the majority of star formation occurs in the outskirts, contributing to the ongoing increase in size. Consequently, the global KS law is expected to respond by decreasing the galaxy's SFR. Additionally, these galaxies are characterized by their low H$_2$-to-HI mass ratios and larger H$_2$ depletion times. As shown below in Figure \ref{fig:ssfr_halpha_fir_fuv}, galaxies at the BMS have nearly constant SFHs over the last $\sim$100 Myrs. This is consistent with fact that the transition from the SFMS phase to a quiescent stage should occur over a longer timescale in these galaxies. In this case, the departure from the SFMS is not related to a morphology change but to a process that does not involve the increase of the central density, see Figure \ref{fig:sersic_ms_SFR_Re} and \ref{fig:grad_color_ms_SFR_Re}, and also \citet{Woo+2019}. It is unclear how these galaxies will undergo a morpho-structural change that matches to those of GV and quiescent galaxies.   
\end{itemize}

Another implication of the above concerns the parallel track hypothesis \citep{vanderWel+2009,vanDokkum+2015,Chen+2020} discussed in the Introduction. If galaxy size is indeed the next key parameter for galaxy evolution after stellar mass, then our results suggest that the track hypothesis is consistent with our findings. Further studies are needed to fully understand this relationship.

\subsection{Higher-Mass Galaxies}

Similar to Figures \ref{fig:sersic_ms_SFR_Re_high_mass} and \ref{fig:grad_color_ms_SFR_Re_high_mass} in the preceding section, we present $\Delta_{\rm MS}$ as a function of $R_e$ across six mass bins for galaxies with masses above $M_c$. It is noteworthy that high-mass galaxies exhibit shallower trends between the SFR and $R_e$ compared to the slope predicted by the KS law, suggesting a deviation from the expected behavior. Therefore, in contrast to lower masses, the \emph{sizes of galaxies do not significantly contribute to the scatter around the SFMS}. Additionally, a significant proportion of galaxies above $\ms \sim 3 \times 10^{10} M_{\odot}$ display higher Sersic indexes, if those galaxies in the past had lower Sersic indexes and were within the SFMS, the above indicates structural transformations even within the SFMS \citep[also consistent with][]{Kauffmann+2003a,Bell+2012}. Notably, the structural parameters of SFMS galaxies are consistent with those of GV and quiescent galaxies. However, a significant difference is observed: \emph{color gradients are strongly bimodal}, with SFMS galaxies exhibiting very negative gradients compared to GV and quiescent galaxies, as shown in Figure \ref{fig:grad_color_ms_SFR_Re_high_mass}. This suggests that while these galaxies are centrally concentrated and may already host a classical bulge, most of their star formation occurs at the outskirts. Surprisingly, this seems sufficient to maintain these galaxies within the SFMS, even when their host dark matter haloes are more massive than the halo mass quenching threshold, $M_{\rm vir} \sim 10^{11.8} M_{\odot}$. Recently, \citet{Hafen+2022} demonstrated the existence of cooling flows with angular momentum, 'rotating cooling flows', that form when gas in haloes is shock-heated to the halo's virial temperature. A key finding in that study is that rotating cooling flows are the primary mode of gas accretion, leading to thin-disk galaxies at $z\sim0$. While rotating cooling flows can help explain why galaxies have a notable amount of star formation in the outskirts (negative color gradient), they do not fully elucidate the drastic change in morphology. 

In more detail, as we {\it descend} from the mean SFMS but still remain within the SFMS, galaxies are more compact, dense, with the least red centres and their light profile are more centrally concentrated. Indeed, BMS are structurally closer, nonetheless still different, to GV galaxies. HSF galaxies, on the other hand have light profiles that are slightly less centrally concentrated than the BMS and GV galaxies. It is unclear whether HSF galaxies at higher masses represent a stage prior to quenching in the same way that low mass galaxies behave (recall that high mass HSF have more negative colour gradient, redder centre, than their low mass counterpart, see Figure \ref{fig:gradgi_avg_Ms_MS}, i.e, they lack of a younger/bluer centre). We speculate that if HSF galaxies are the progenitors of GV and quiescent galaxies they should evolve different and more slowly than their low mass counterpart.

\subsection{Further empirical evidence supporting paths for galaxy quenching}

In this subsection we investigate whether additional data supports the pictures described above. In particular, we compare two different kinds of SFRs that can help to understand the most recent star formation history of galaxies as a function of their $R_e$ and $M_\ast$ as detailed below. 

Figure \ref{fig:ssfr_halpha_fir_fuv} shows the differences between \citet{Brinchmann+2004} and \citet{Salim+2018} SFRs as a function of \ms\ for our 4 different SFMS sub-samples,\footnote{We note that a factor of $-0.05$ dex has been applied to \citet{Brinchmann+2004} SFRs to be consistent with a \citet{Chabrier2003} IMF.} Table \ref{tab:sfrs_sub-samples}. In the figure, each galaxy has been colour coded by their distance to $\langle \log R_{e,{\rm Q}} (M_\ast)\rangle$ of quiescent galaxies given by
\begin{equation}
    \langle \log R_{e,{\rm Q}}\rangle = \frac{R_{e,0}}{2^{\beta-\alpha}} \left(\frac{M_\ast}{M_0}\right)^{\alpha} \left(1+\frac{M_\ast}{M_0}\right)^{\beta-\alpha}
    \label{eq:size_mass_Q}
\end{equation}
where $\log \left(R_{e,0}/{\rm kpc}\right) = 0.18 \pm 0.04$, $\log \left(M_0/M_{\odot}\right) = 9.95\pm 0.08$, $\alpha = -0.01 \pm 0.01$ and $\beta = 0.68\pm 0.02$.

\citet{Brinchmann+2004}\footnote{The catalog used in this paper is the updated version that can be found here:  \url{https://wwwmpa.mpa-garching.mpg.de/SDSS/DR7/sfrs.html}} modeled the emission lines, including H$_\alpha$, of SDSS galaxies by means of the \citet{Charlot_Longhetti_2001} model, which combine the Bruzual and Charlot (1997) models with emission line modeling from the code {\textsc{Cloudy}} \citep{Ferland1996}. Notice that we are using their global SFRs instead of the results from the fiber. The authors applied the empirical correction described in \citet{Salim+2007} by computing the light outside the fibre for each galaxy, and then fit stochastic models, following the color-color grid method described by \citet{Salim+2007} to this photometry. The authors showed that using the standard \citet{Kennicutt1998rev} conversion factor from H$_\alpha$ to SFR is a good average correction for most galaxies. Thus, hereafter we will assume that \citet{Brinchmann+2004} SFRs are closely related to H$_\alpha$ SFRs and thus probing time-scales $\sim 6-10$ Myrs \citep[see][]{Calzetti2013,Flores+2021}. In the case of \citet{Salim+2018} the authors used information from FUV and FIR, see Section \ref{secc:SF_MS}, probing time-scales of $\sim 100$ Myrs \citep{Calzetti2013,Salim+2016,Salim+2018,Flores+2021}. Therefore, Figure \ref{fig:ssfr_halpha_fir_fuv} shows the difference between two epochs in the SFH of the galaxies. If the difference is negative, the SFR is expected to have faded over the last $\sim 100$ Myr, while if it is positive, an increase in the SFR is expected for at least towards the last 5--10 Myr. The red solid lines and shaded areas in each panel show the median and the poissonian error. 

Figure \ref{fig:ssfr_halpha_fir_fuv} supports our previous discussion. Below $\ms\sim M_c (= 2\times 10^{10} M_{\odot}$, vertical shaded area), the median value of the recent SFHs of HSF galaxies have faded by a factor of $\sim 1.3$ at $\ms \sim 10^{9} M_{\odot}$ and by $\sim 1.5$ at $\ms \sim 10^{10} M_{\odot}$ over the last 100 Myrs. Furthermore, galaxies with half-light radius comparable to quiescent galaxies have experienced a significant decrease in their recent SFH, fading a median value of $\sim2$ at $\ms \sim 10^{10} M_{\odot}$. Overall, we observe a trend where low-mass HSF galaxies with half-light radii similar to quiescent galaxies are experiencing a decline in their recent SFH while larger galaxies are closer the zero line. The above results supports the speculation that the transition from active star formation to a quiescent phase occurs rapidly. The median values of UMS galaxies are slightly below the zero line, but once again, we observe a systematic trend: galaxies with half-light radii comparable to quiescent galaxies are experiencing an overall decline in their recent SFHs while larger galaxies are around the zero line. In contrast the median values of LMS and BMS galaxies are slightly above the zero line (i.e, constant SFH) and slightly below it around $\ms \sim 10^{10} M_{\odot}$. Overall, their half-light radii are larger compared to quiescent galaxies and their SFHs are approximately constant or even possitive. This trend implies that the quenching process from the LMS and BTM to the GV or quiescent may proceed more gradually. Notice that there are some low-mass galaxies in all the panels whose SFHs have decreased substantially,
%Joel
so that, in terms of H$_\alpha$ SFRs, these galaxies should be considered GV or even quiescent. Interestingly, at high masses $\ms\gtrsim M_c$, galaxies, on average, have experienced a decline in their SFHs. Specifically, massive UMS, LMS, and BMS galaxies have significantly reduced their SFHs to the extent that some are classified as quiescent galaxies based on the SFR H$_\alpha$ estimator. It is worth noting that, despite the decreasing SFHs of the HSF galaxies,
%Joel are decreasing, 
they are actually closer to being constant. This observation might indicate that the transition from actively forming stars to a more quiescent state is faster for SFMS UMS, LMS, and BMS galaxies but slower for HSF galaxies.

Finally, we acknowledge that interpreting Figure \ref{fig:ssfr_halpha_fir_fuv} involves some speculation, particularly due to the empirical corrections applied by \citet{Brinchmann+2004} based on colors. However, it is encouraging that the figure predicts trends consistent with the discussion at the beginning of this section and previous interpretations, as we will discuss further below. More work needs to be done by using spatially resolved imaging spectroscopy surveys \citep[e.g.,][]{Sanchez+2012,Bundy+2015} to make a direct comparison and avoid any aperture correction issues. 

%%%%%%%%%%%%%%%%%%%%%%%%%%%%%%%%%%%%%%%%%%%%%%%%%%%%%%%%%%%%%%
\section{Discussion}
%\label{secc:discussion}

The main goal of this paper was to investigate whether there is a relationship between the distance to the SFMS and the size and morpho-structural properties of galaxies. One of the key findings of our study is that there is a non-monotonic relationship between half-light/mass radius and SFR, Figures \ref{fig:ssfr_rer_mass} and \ref{fig:ssfr_res_mass}. This behavior is also observed in other structural properties, such as the colour gradient (Figure \ref{fig:gradgi_avg_Ms_MS}), the central mass density (Figure \ref{fig:S1_avg_Ms_MS}) and S\'ersic index (Figure \ref{fig:sersic_avg_Ms_MS}, see also Figures \ref{fig:lms_bms_9_10.3} and \ref{fig:lms_bms_10.3_15}).\footnote{This is expected since these other quantities are closely correlated with \re.} However, when analyzing trends of the SFR as a function of $R_e$, we observe a negative monotonic trend, consistent to what one would expect if the SFR were controlled by the KS law (${ \rm SFR} \propto R_e^{-0.28}$), as depicted in Figures \ref{fig:re_MS_mass_bins}, \ref{fig:sersic_ms_SFR_Re}, and \ref{fig:grad_color_ms_SFR_Re}. For galaxies below $M_c = 2\times 10^{10} M_{\odot}$ the SFMS follows a power-law relation, where the SFR scales as ${\rm SFR} \propto M_\ast^{0.85}$. Here $M_c$ is the mass at which the SFMS bends and corresponds to a dark matter halo mass where the formation of virial shocks occurs. Our findings imply that the sizes of galaxies contribute to the scatter around the SFMS. Additionally, we found that the H$_{2}$-to-HI mass ratio increases for galaxies above the SFMS (Figure \ref{fig:MH2MHI_MS_mass_bins}). Our results indicate two quenching paths for low-mass galaxies, $M\ast \leq M_c$: one related to a rapid change of morphology, and the other related to a more gradual, secular process (see Fig. \ref{fig:cartoon_quenching_paths}). At higher masses, $M_\ast > M_c$, the SFMS flattens out, with ${\rm SFR}(\ms) \sim {\rm const.}$, and galaxies appear to undergo structural transformations already within the SFMS. 

Next, we compare these results with previous works and discuss on the galaxy-halo connection.

%%%%%%%%%%%%%%%%%%%%%%%%%%%%%%%%%%%%%%%%%%%%%%%%%%%%%%%%%%%%%%
\subsection{Comparison with previous works}
\label{secc:comparison}

In this section, we provide a comparative analysis of our findings with recent existing literature on the topic. Recent studies that have investigated the relationship between the structure of the galaxies and sSFR include \cite{Wuyts+2011,Brennan+2017,Lin+2020,Woo+2019,Walters+2021}, and \citet{Yesuf+2021}.

\cite{Wuyts+2011} and \citet{Brennan+2017} specifically focused on the relationship between size and sSFR. The results of \citet{Wuyts+2011} are similar to ours; these authors found a non-monotonic relation between $R_e$ (also the Sersic index) and SFR. They also noted that moderate quiescent and highly star-forming galaxies coexist over an order of magnitude in $M_{\ast}$. Although the authors never explicitly specified the validity range of the aforementioned trend, by examining their Figure 3, it becomes apparent that the trend holds for $M_{\ast}\lesssim10^{10} M_{\ast}$. This aligns closely with our findings, where low-mass highly star-forming HSF galaxies exhibit structural similarities to low-mass green valley GV galaxies. In contrast, \citet{Brennan+2017} found a monotonic relation between $R_e$ and $\Delta_{\rm MS}$. The authors argued that after carefully eliminating galaxies with poor photometric fits, the trends observed by \citet{Wuyts+2011} nearly disappeared. Here we emphasise  that 1) we are using galaxies with good total magnitudes and sizes (Section \ref{secc:gal_data}), and 2) we had made multiple checks by using other sets of data for galaxy sizes, as discussed at end of Section \ref{secc:size_delta_ms}, and found the same trends. Therefore, we 
%Joel discard 
disagree that by removing galaxies with poor photometric fits the correlation between size and $\Delta_{\rm MS}$ disappears. 

The recent paper by \citet{Lin+2020} also analysed the relationship between size and sSFR for star-forming galaxies on the SFMS only. While their results  are generally consistent with ours, they conclude that there was no significant correlation between \re\ and sSFR, particularly for galaxies with masses below $\sim 10^{10}M_{\odot}$. For more massive galaxies, they found a weak correlation and suggested that this could be due to small (more compact) galaxies evolving slowly out of the main sequence. 
%This aligns with our finding that massive galaxies with masses around $\sim 10^{11}M_{\odot}$ evolve and quench more slowly. 
However, \cite{Lin+2020} did not emphasize the non-monotonic relationship between \re\ and sSFR despite there being some hints in their data, see their Figure 5. 
%The fact that the correlation is non-monotonic suggests that the key to understanding the relationship between SFR and \re\ is not the gas fraction of the galaxy, but rather the efficiency of converting HI into H$_2$.

%%%VAR: aquí me quedé %%%%

In our interpretation, the non-monotonic relationship indicates two distinct pathways to quenching. \citet{Woo+2019} investigated the relationship between morphology and quiescence in galaxies and we find that our results are consistent with theirs (note, however, that contrary to \citet{Woo+2019}, our paper deals primarily with SFMS galaxies). They proposed two different quenching paths, one consistent with the classical "inside-out" growth, and the other a "compaction-like" process that feeds the AGN and quenches the galaxy. Like us, they investigated the $\Sigma_{1}$ -\ms\ relation and found that galaxies on the upper part of this relation tend to have younger centres, and enhanced sSFR, but they also found that 
%Joel
they are metal-deficient. \citet{Woo+2019} suggested that these galaxies most likely go through a compaction-like process triggered by disk and bar instabilities \citep[see e.g.,][]{Lin+2017,Chown+2019}. These processes can rapidly deplete the available gas, leading to a uniform quenching of the galaxy. They also found that galaxies on the lower part of the $\Sigma_{1}$ - \ms\ relation are consistent with a secular disk growth. These galaxies 
%Joel with 
have low $\Sigma_{1}$ for their \ms\ have old, metal-rich, and sSFR suppressed in their centres. One of their main conclusions is that age gradient, sSFR, and metallicity depend on the galaxy's position on the $\Sigma_{1}$ - \ms\ relation. These results are consistent with our findings regarding the relationships between SFR and $\Sigma_{1}$, colour gradient, and \re. Among all these relationships, we find that galaxies with masses below $\ms\sim 10^{10} M_{\odot}$ and high SFR also have high $\Sigma_{1}$ and blue centres indicating the prescence of younger populations, and their trends are similar to those of green valley galaxies, hinting that they migrate relative quickly to the GV. For higher masses, we find a more consistent picture with the gradual "inside-out" growth and quenching pathway.

The \citet{Woo+2019} study sheds light on the existence of two quenching paths, including a "compaction-like" event that is still not fully understood in the local universe
%Joel added refs 
\cite[see also][discussed further below]{Walters+2021,Walters+2022}.
While we observe galaxies that may be consistent with this path, such as the low-mass HSF galaxies (but less clear for the high mass ones), the underlying mechanisms responsible for these events remain unclear. However, the \citet{Yesuf+2021} work on SFMS galaxies provides a possible insight.
In \citet{Yesuf+2021}, the focus was on investigating the structural parameters that predict whether a star-forming galaxy lies above or below the SFMS. 
Specifically, they found that asymmetry is the most important predictor for the position in the SFR-\ms\ plane; there is direct relationship between the asymmetry of a galaxy and its SFR. Asymmetry is related to spiral arms and lopsidedness in isolated galaxies, while in mergers or interactions it is related to structural perturbations. 
Therefore, in our context, it is plausible that HSF galaxies are expected to be more asymmetric compared to the BMS. 

\citet{Yesuf+2021} suggested two possible mechanisms to explain the correlation between high asymmetry and higher SFR on the SFMS: mergers and tidal encounters that enhance star formation, or enhanced diffuse gas accretion. We notice that the latter could be an efficient mechanism to promote the conversion of HI into H$_2$. An intriguing finding from \citet{Kakos+2024} reveals a higher number of satellite galaxies orbiting around low-mass HSF galaxies. This finding may support the tidal encounter theory proposed by \citet{Yesuf+2021}. It is still not clear, however, which mechanism dominates the observed correlation. 
%Joel added
\citet{Bottrell+2024} finds that in the TNG50 simulation, mini mergers help to explain the correlation between asymmetry and elevated SFR. Another interesting finding reported by \citet{Yesuf+2021} is that a large number of SFMS 
%Joel added
galaxies with higher asymmetry are bulge-dominated, concentrated and compact, similar to quiescent galaxies. This result is similar to our findings in low mass HSF galaxies, which tend to have higher S\'ersic indices, high $\Sigma_{1}$, and small \re, except that we find that these galaxies are more similar to GV ones. 

It is interesting to note that the study of simulated starburst galaxies by \cite{Cenci+2023} from the FIREbox cosmological volume shows that galaxies initiate bursts of star formation in two ways. Below $\ms \sim10^{10} M_{\odot}$, starbursts are often triggered by global gravitational instabilities, while in massive galaxies, tidal torques resulting from galaxy mergers are responsible for the initiation of star formation bursts. In both cases the burst of star formation was accompanied by a compaction event and the increase in the fraction of molecular gas, consistent with our results. The above seem consistent with the \citet{Yesuf+2021} picture and align with the interpretation of our findings.

Simulations offer an opportunity to better understand the processes that drive the evolution of the central density of star-forming galaxies. \citet{Walters+2021} looked at the structural evolution of isolated galaxies in the Illustris TNG100 simulation, specifically at the growth of the central density within 2 kpcs, $\Sigma_{2}$, in relation to total stellar mass \ms\ in galaxies at low $z$. The authors found that the Illustris TNG100 is able to reproduce the $\Sigma_{2} - \ms$ relation, the radial profile of the stellar ages, sSFRs, and metallicities observed in SDSS MaNGA galaxies when comparing to \citet{Woo+2019}.  
Similar to the results of \citet{Woo+2019}, they found that dense-core galaxies evolve parallel to the $\Sigma_{2} - \ms$ relation, whereas galaxies with diffuse cores evolve in shallower trajectories along the $\Sigma_{2} - \ms$ relation. Therefore, understanding the core formation in the Illustris TNG simulation was key for \citet{Walters+2021}. These authors find that sSFR gradients and BH accretion rate are not good predictors of future core growth in the Illustris TNG simulation; they are only indicators of past core growth. 
%Joel added
\citet{Walters+2022} showed that in TNG100 these bluer-centre galaxies quench rapidly.

Disc instabilities have been proposed as being drivers of structural evolution at redshifts $z>1$
%Joel added Lapiner
\citep{Dekel_Burket2014,Zolotov+2015,Inoue+2016, Avila-Reese+2018,Lapiner+2023}, as they cause fragmentation and the inflow of cold gas towards the centre of the galaxy with subsequent bursts of star formation that consume the gas and quench central star formation later. However, \citet{Walters+2021} showed that in the TNG simulation, disc instabilities do not lead to core buildup. On the contrary, rapid past core growth results in more stable disks due to the stabilizing effect of spheroids \citep[see e.g.,][]{Martig+2009}. Interestingly, they find that BH feedback sets a maximum rate of core growth in galaxies with $\ms\ > 10^{9.5} M_{\odot}$. This means that in galaxies below that mass, BH feedback does not suppress star formation and rapid core growth.
For massive galaxies, they found the total specific angular momentum of gas accretion to be the main predictor of future core growth. 

If those TNG galaxies are indeed analogous to ours, these results can help us understand better at least one of the two different quenching paths at low-masses. The low-mass, HSF galaxies that we observe have high $\Sigma_{1}$ and bluer centers. One explanation could be that structural evolution triggers the central star formation as the BH feedback would not be sufficiently efficient at these low masses to prevent star formation and therefore a rapid core buildup. However, these galaxies will quickly deplete their gas, and most likely, during that phase, the BH mass will grow enough to quickly transition them into the GV. Another possibility is that strangulation due to gas consumption could take them into the GV or even to quiescent region. In future studies it will be key to study the occupation of AGNs in HSF galaxies. For masses above $\ms\sim10^{9.5}M_{\odot}$, BH feedback prevents future central buildup and allows for a secular \citep{Walters+2021}, inside-out quenching; the fossil record analysis of elliptical galaxies from the MaNGA/SDSS-IV survey shows that most of them followed indeed this path \citep{Avila-Reese+2023}. 

Finally, other previous studies have proposed distinct pathways to explain the quenching of galaxies. Perhaps the closest one among them 
%Joel inserted
to ours is \citet{Schawinski+2014}, who investigated the evolutionary paths of late-type and early-type galaxies through the green valley (GV), utilizing morphological information. The authors discovered that these two types of galaxies undergo different mechanisms while transitioning through the GV. In early-type galaxies, quenching is triggered by the rapid depletion of the galaxy's gas reservoir, causing them to briefly stay above the SFMS before experiencing a rapid decline in their SFRs. Conversely, late-type galaxies transition more gradually toward the GV through secular processes. These findings align with our interpretation. However, it is important to note that the authors arrived at these conclusions by studying the GV of \emph{all} galaxies (i.e., they included satellites) and by morphology, rather than central galaxies by their properties according to their distance with respect to the SFMS. In that regard, equating early-type to quiescent galaxies might not be appropriate.  

\subsection{Implications for the galaxy-halo connection}

Galaxies formed and evolved within massive dark matter halos, and it is expected that galaxy properties are linked to the properties of their host halos. Two halo properties that are thought to regulate SFRs in galaxies are halo accretion and concentration. The former controls the incorporation of gas into the interstellar medium of the galaxy \citep{Avila-Reese+2000,DekelMandelker2014,Rodriguez-Puebla+2016a}, while the latter is related to the timing of gas infall into the halo, with more concentrated halos experiencing earlier infall \citep{Avila-Reese+1998,Wechsler+2002,Gao+2004,Dutton+2010a}. Additionally, high concentration is expected to lead to more centrally concentrated galaxies and thus smaller sizes \citep[see e.g.,][]{Jiang+2019,Liang+2024}. Moreover, \citet{Lin+2020} suggested that concentrations could potentially control the interplay between SFRs and radii. 

Recently, the importance of halo concentration in regulating galaxy size was challenged by \citet{Behroozi+2022}, who demonstrated, based on environmental information, that halo accretion, which will set the rate of infalling gas, is the primary determinant of galaxy size. Although our paper does not aim to disentangle the relative importance of these properties in controlling both SFRs and galaxy sizes, we note that the non-monotonic relationship between SFR and the effective radius $R_e$ may be embedded in the correlation between these two properties. We caution that if halo accretion and/or concentration dominate SFRs in a monotonic manner \citep[but see][]{Kakos+2024}, the same cannot be assumed for $R_e$ or the central density of the galaxy. Rather, it is likely that multiple halo properties (such as the angular momentum of the dark matter halo, $\lambda$, which in turn could be related to the angular momentum of the gas) contribute to the regulation of galaxy size and structure \citep[e.g.,][]{Dalcanton+1997,MMW98,Avila-Reese+2000,Firmani+2009,Walters+2021,Liang+2024}.

%%%%%%%%%%%%%%%%%%%%%%%%%%%%%%%%%%%%%%%%%%%%%%%%%%%%%%%%%%%%%%

%\subsection{Caveats}
%\aldo{
%\begin{itemize}
%    \item Sandy made a point that Salim's SFRs should be corrected by color gradients. 
%    \item Bad Sizes from Meert (we already checked that they are extremly similar to Simard, except for galaxies with the highest Sersic. Needs to be check, but the above difference is explained if Meert assumes a higher Sersic.)
%    \item Even if the above is a strong difference that could bias some results, it is not a problem since we are dealing mostly with SFMS galaxies with low Sersic Index. 
%\end{itemize}
%}

%%%%%%%%%%%%%%%%%%%%%%%%%%%%%%%%%%%%%%%%%%%%%%%%%%%%%%%%%%%%%%

\section{Conclusions}
\label{secc:main_conc}

In this paper, we made use of 2D photometric S\'ersic fits catalogs \citep{Meert+2015, Meert+2016} for galaxies in the SDSS DR7 to obtain a volume limited sample of galaxies $\ms\approx 10^9 \msun$ and with a central resolution at least of $\approx 1$ kpc in radius. Our galaxy sample consists of galaxies that have $m_{r}$ < 17.77, $b/a$ > 0.5, 
and a redshift range of 0.005 $< z <$ 0.067,  \citep{Meert+2013, Meert+2015}. From the photometric catalog, we calculated the mass profile, half-mass radius $R_{e,\ast}$, the average mass density at 1 kpc $\Sigma_{1}$, and the colour gradients (CGs) based on their S\'ersic parameters for the $r$, $g$, and $i$ bands. CGs were obtained by calculating the surface brightness and luminosity profiles with the S\'ersic parameters given in the photometric catalog and then deriving the colour profile.
To obtain \ms\, we derived five stellar masses and corresponding mass profiles based on five different mass-to-light ratios and computed the geometric mean of the five determinations, following the methodology described in \citet{Rodriguez-Puebla+2020} (see their Appendix A, eq. 34). In this paper, we focused on central galaxies by using the \cite{Yang+2012} catalog by assuming that the most massive galaxies, in 
%Joel added
stellar mass terms, are the central ones.

In this study, we separated our sample of galaxies into six sub-samples based on their distance from the star-formation-main-sequence, SFMS, relationship, $\Delta_{\rm MS} = \log{\rm SFR}-
\langle \log {\rm SFR}\rangle$. 
Our six sub-samples were defined as follows: Highly star-forming (HSF, 
$\Delta_{\rm MS}>0.25$), Upper main sequence (UMS, $0<\Delta_{\rm MS}
<0.25$), Lower main sequence (LMS, $-0.25<\Delta_{\rm MS}<0$), 
Bottom of the main sequence (BMS, $-0.45<\Delta_{\rm MS}<-0.25$), Green 
Valley (GV, $-1<\Delta_{\rm MS}<-0.45$), and Quiescent (Q, 
$\Delta_{\rm MS}<-1$) galaxies. We provide the details of the subsample 
definitions in Table \ref{tab:sfrs_sub-samples}. Our methods were 
similar to those described in \citet{Fang+2018} and \citet{Rodriguez-Puebla+2020a}. We use the SFRs from \citet{Salim+2018}.

The main goal of this work was to determine whether a relationship between SFR and galaxy structure exists. In particular, we study whether there is a  relationship between SFR and $\re$ as implied by the Kennicutt-Schmidt law. 
 
Our main results are:

\begin{itemize}
\item There is a characteristic mass of $M_c = 2\times10^{10}M_{\odot}$ beyond which the mean of the SFMS bends. At stellar masses below $M_c$, the star-forming main sequence (SFMS) adheres to a power-law relation, where the star formation rate (SFR) scales as ${\rm SFR} \propto M_\ast^{0.85}$. However, at higher masses, the SFMS flattens out, with ${\rm SFR}(\ms) \sim {\rm const.}$. 

\item When converted to a dark matter halo mass, $M_c$ is consistent with the halo mass scale of $M_{\rm vir}\sim 10^{11.8}$ where the formation of virial shocks occur \citep[see also][]{Daddi+2022}.

\item SFMS galaxies exhibit larger radii, both in terms of half-mass and half-light, compared to GV and quiescent galaxies. On average, they possess a lower central mass density ($\Sigma_{1}$) and a lower Sersic index ($n$) in comparison to GV and quiescent galaxies. Additionally, SFMS galaxies tend to have more negative colour gradients (CGs) ($\nabla_{g-i} < 0$, indicating redder centers) in general. However, we found that the continuous relationship between SFR, size, and structural properties ($\Sigma_{1}, n$, and $\nabla_{g-i}$) at a fixed stellar mass is non-monotonic. 

\item The sizes and structural parameters of the galaxies exhibit a notable change in behaviour across the SFMS at the characteristic mass of $M_c = 2 \times 10^{10} M_{\odot}$.
\end{itemize}

For galaxies below $M_* < M_c$ we found:
\begin{itemize}
\item Galaxies above the SFMS, in particular, HSF galaxies ($\Delta_{\rm MS}> \sigma_{\rm MS}$) typically exhibit the smallest sizes among all SFMS sub-samples, considering both half-light and half-mass radii. Regarding their $\Sigma_{1}$, they have the highest central density and tend to have the least negative CGs (bluer centers) compared to other SF sub-samples.

\item Moreover, lower-mass HSF and UMS galaxies exhibit sizes and central densities, $\Sigma_1$, that are comparable to those of GV and quiescent galaxies. Given this resemblance, we posit that galaxies above the SFMS might serve as the progenitors for at least a fraction of the GV and quiescent galaxies. 

\item At masses below $M_* < M_c$, galaxies at the BMS, $-0.45 < \Delta_{\rm MS} <- \sigma_{\rm MS}$,  exhibit more disk-like characteristics. They have the lowest central density and tend to show negative CGs (indicating redder centers) compared to other SF sub-samples. Given that GV and quiescent galaxies are smaller and centrally concentrated with flatter CGs, the above results are counterintuitive, as it is expected that BMS galaxies would be natural candidates for progenitors of GV galaxies and/or quiescent ones. 

\item Our results strongly suggest that low-mass galaxies ($M_* < M_c$) can undergo quenching through two distinct pathways. The first pathway involves galaxies above the SFMS, where we speculate that the transition from SFMS to a more quiescent state is relatively rapid. These galaxies experience a morpho-structural change before transitioning out of the SFMS phase. The second pathway involves galaxies with $-0.45 < \Delta_{\rm MS} < - \sigma_{\rm MS}$ (BMS galaxies), where the transition from the SFMS phase to a quiescent state might occur over a longer timescale. In this case, the departure from the SFMS is not related to a change in morphology but to a process that does not involve an increase in central density.
\end{itemize}

At higher masses ($M_* > M_c$) we found:
\begin{itemize}
\item  BMS galaxies exhibit greater compactness and 
%Joel are 
more centrally concentrated light profiles with the least red centers. Structurally, these massive BMS galaxies are closer to GV galaxies. Conversely, massive HSF galaxies are less centrally concentrated compared to BMS and GV galaxies. In this scenario, BMS galaxies align with the anticipated outcome of being the most likely candidates for progenitors of GV galaxies and/or quiescent ones. However, it is evident that structural transformations are already occurring within the SFMS for high-mass galaxies.
\end{itemize}

We also studied which among the structural properties is the most relevant in inferring $\Delta_{\rm MS}$ by using the Generalised Additive Models (GAM \citealp{HastieTibshirani1990}). We found

\begin{itemize}
    
\item For low mass galaxies, ($M_* < M_c$), CGs and $\Sigma_1$ are the properties ranked on the top of importance respectively for HSF and UMS. For LMS and BMS \re\ was ranked on the top, contrary to previous studies. 

\item For high mass galaxies, ($M_* > M_c$), $\re$ was ranked on the top of importance for HSF and UMS  galaxies while for LMS and BMS galaxies, $n_r$ and CGs were ranked on the top, respectively. 

\end{itemize}

Additionally, we developed an empirical model based on the global Kennicutt-Schmidt (KS) law to understand the relation between the SFR and size for a given stellar mass. Our results show that

\begin{itemize}

\item For galaxies below $M_c$, the size of galaxies contributes to the scatter around the SFMS in accordance with the KS law. However, at higher masses, the relationship between SFR and size disappears.

%Joel changed ${\rm SFR} \propto R_e^{0.28}$ to ${\rm SFR} \propto R_e^{-0.28}$
\item The KS law predicts that the expected relation between SFR and size, for a given 
%Joel added
stellar mass and assuming a constant gas mass, is ${\rm SFR} \propto R_e^{-0.28}$. This is consistent with the observed trends from our SDSS sample, see Figure \ref{fig:re_MS_mass_bins}, \ref{fig:sersic_ms_SFR_Re}, and \ref{fig:grad_color_ms_SFR_Re}.

\item We inverted the global KS law by considering the scatter around this relation and combined it with an empirical model that computes molecular masses from SFRs and \ms. Our analysis reveals that compact galaxies exhibit higher H$_2$-to-HI mass ratios for a given stellar mass (Fig. \ref{fig:H2_to_HI_ratio_predicted_as_a_func_of_Re} and Eq. \ref{eq:fh2_sfr_compact}). Furthermore, we observed a monotonic correlation between H$2$-to-HI mass ratios and $\Delta{\rm MS}$. These findings suggest that any interplay between SFR and radius for a given stellar mass is not primary but a consequence of the relationship between radius and H$_2$-to-HI mass ratios. 
\end{itemize}

Similar to previous studies \citep[e.g.,][]{Schawinski+2014, Woo+2019, Yesuf+2021}, our findings emphasize that quenching is not a straightforward transition of SMFS galaxies evolving through the GV and then ultimately reaching the quiescent region. The evidence presented here suggests that different quenching mechanisms, operating on different time scales, may have influenced the progenitors of GV and quiescent galaxies. This paper marks the beginning of a series that delves into the question of which galaxy properties are the primary predictors for the position relative to the SFMS. In \citet{Kakos+2024}, we will examine the two-point correlation function and the stellar-to-halo mass relations as a function of the distance to the SFMS.

\section*{Acknowledgements}

ARP acknowledges financial support from CONACyT through ``Ciencia Basica'' grant 285721. ARP and VAR CONAHCyT ``Ciencia de Frontera'' grant G-543, and DGAPA-PAPIIT IN106823 and IN106124.

%%%%%%%%%%%%%%%%%%%%%%%%%%%%%%%%%%%%%%%%%%%%%%%%%%
\section*{Data Availability}

The data underlying this article will be shared on reasonable request to the corresponding author.



\bibliographystyle{mnras}
\bibliography{Bibliography} % if your bibtex file is called example.bib

%\begin{thebibliography}{99}
%\end{thebibliography}

%%%%%%%%%%%%%%%%%%%%%%%%%%%%%%%%%%%%%%%%%%%%%%%%%%

%%%%%%%%%%%%%%%%% APPENDICES %%%%%%%%%%%%%%%%%%%%%

\appendix

\section{1D shape functions and predictions}
\label{app:EBM_statistics}

\begin{figure*}
\centering
\includegraphics[width=0.25\textwidth]{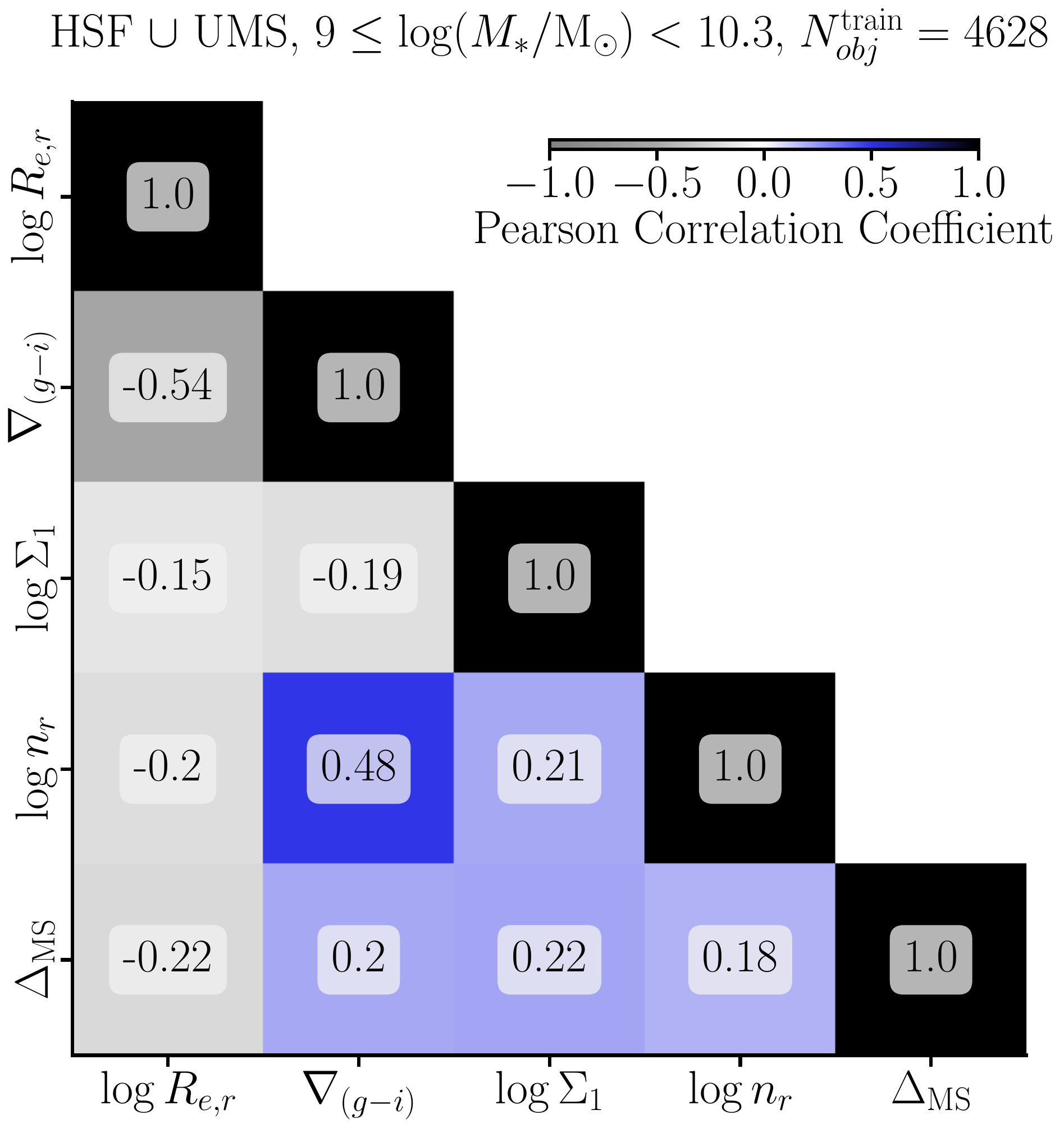}
\hspace{3cm}
\includegraphics[width=0.25\textwidth]{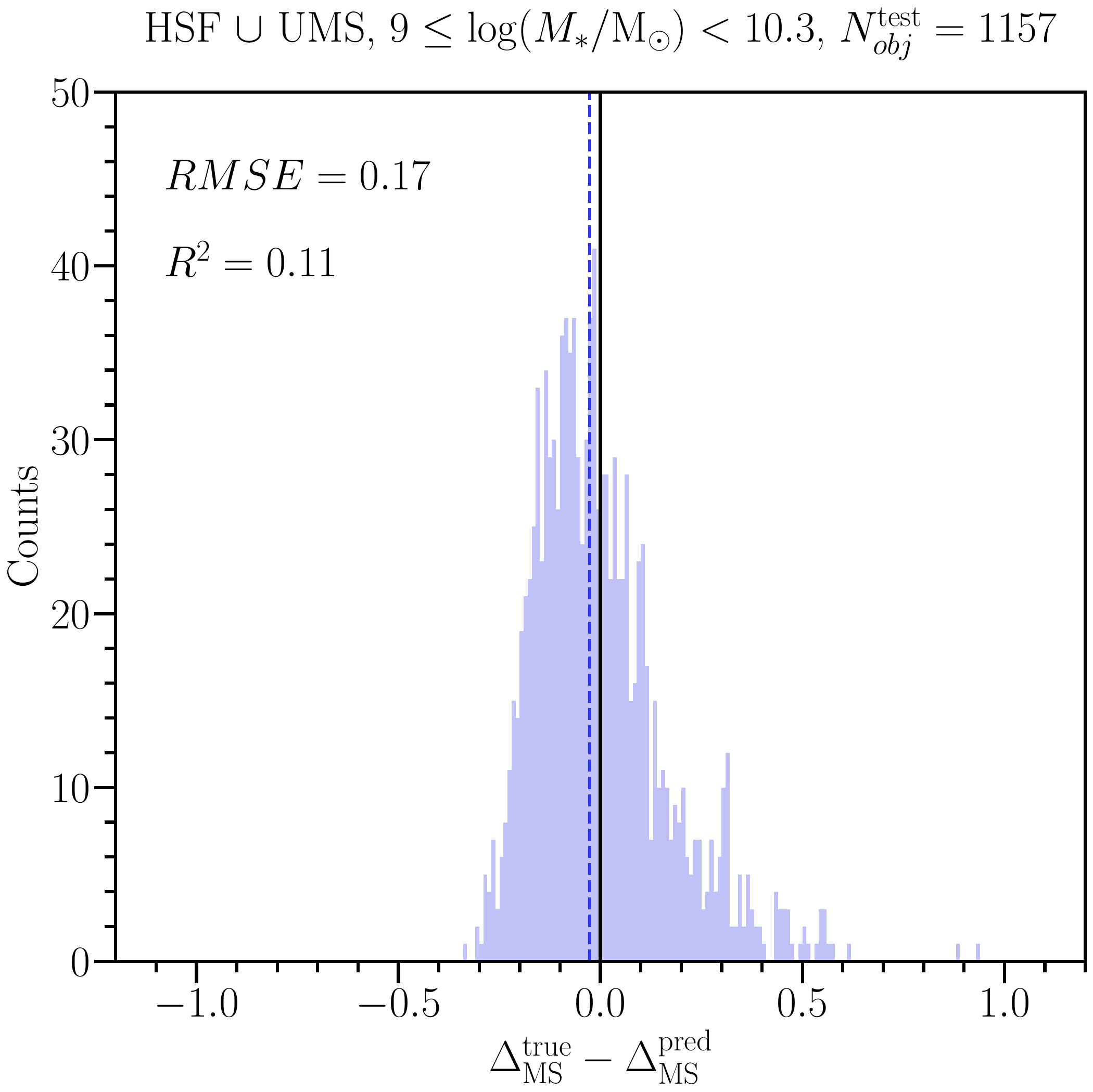}
\\
\includegraphics[width=0.22\textwidth]{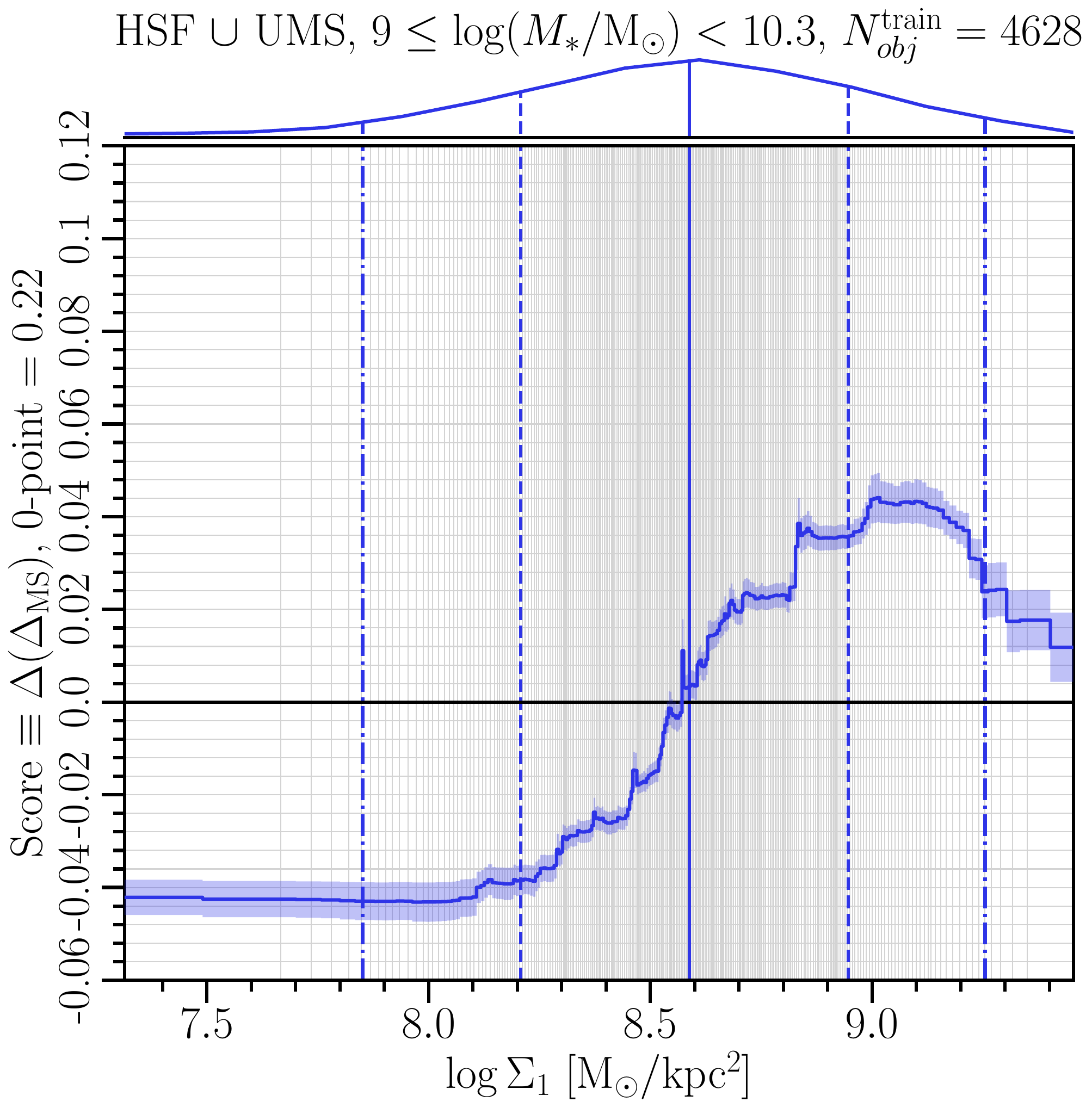}
\hfill
\includegraphics[width=0.22\textwidth]{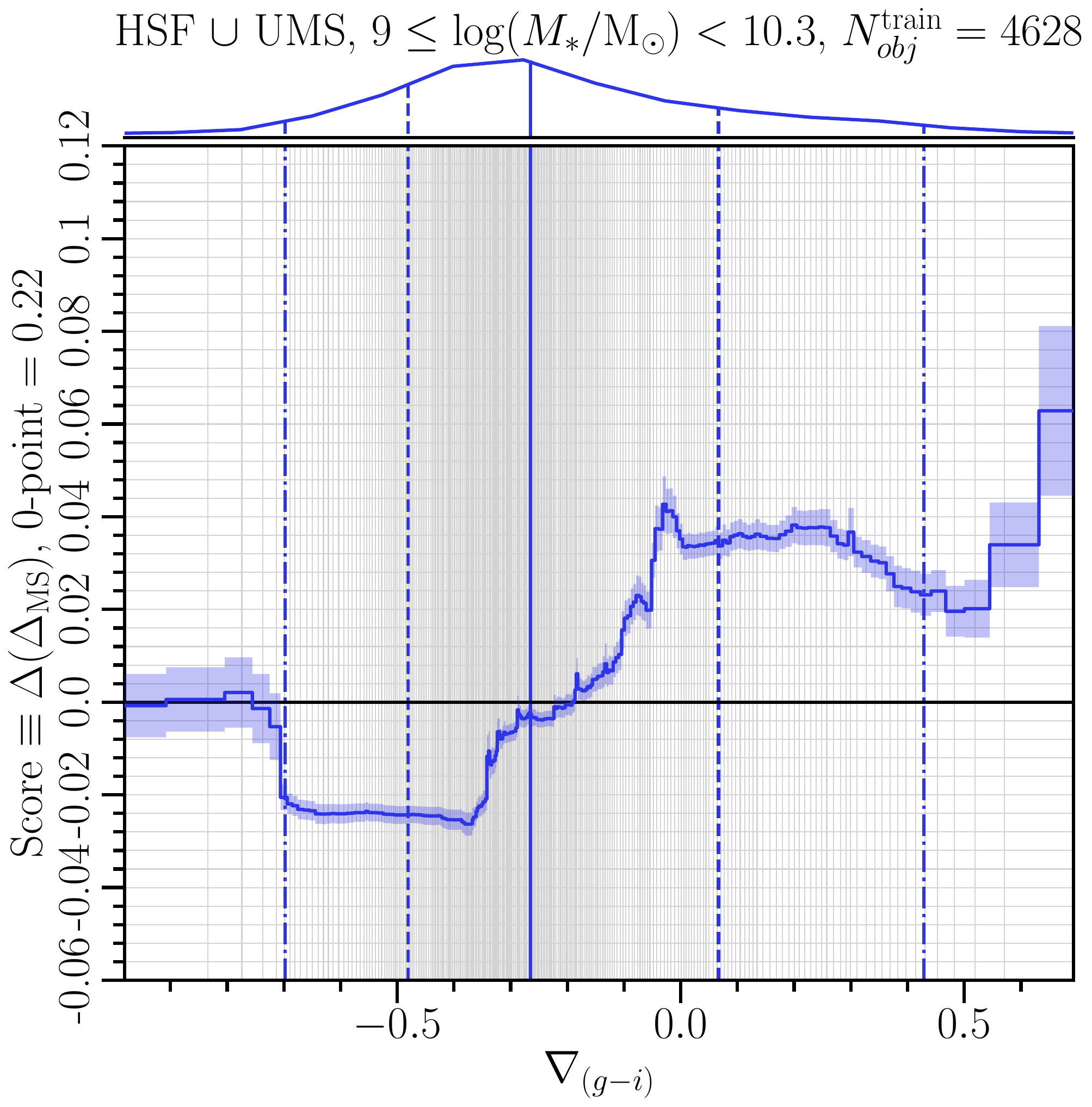}
\hfill
\includegraphics[width=0.22\textwidth]{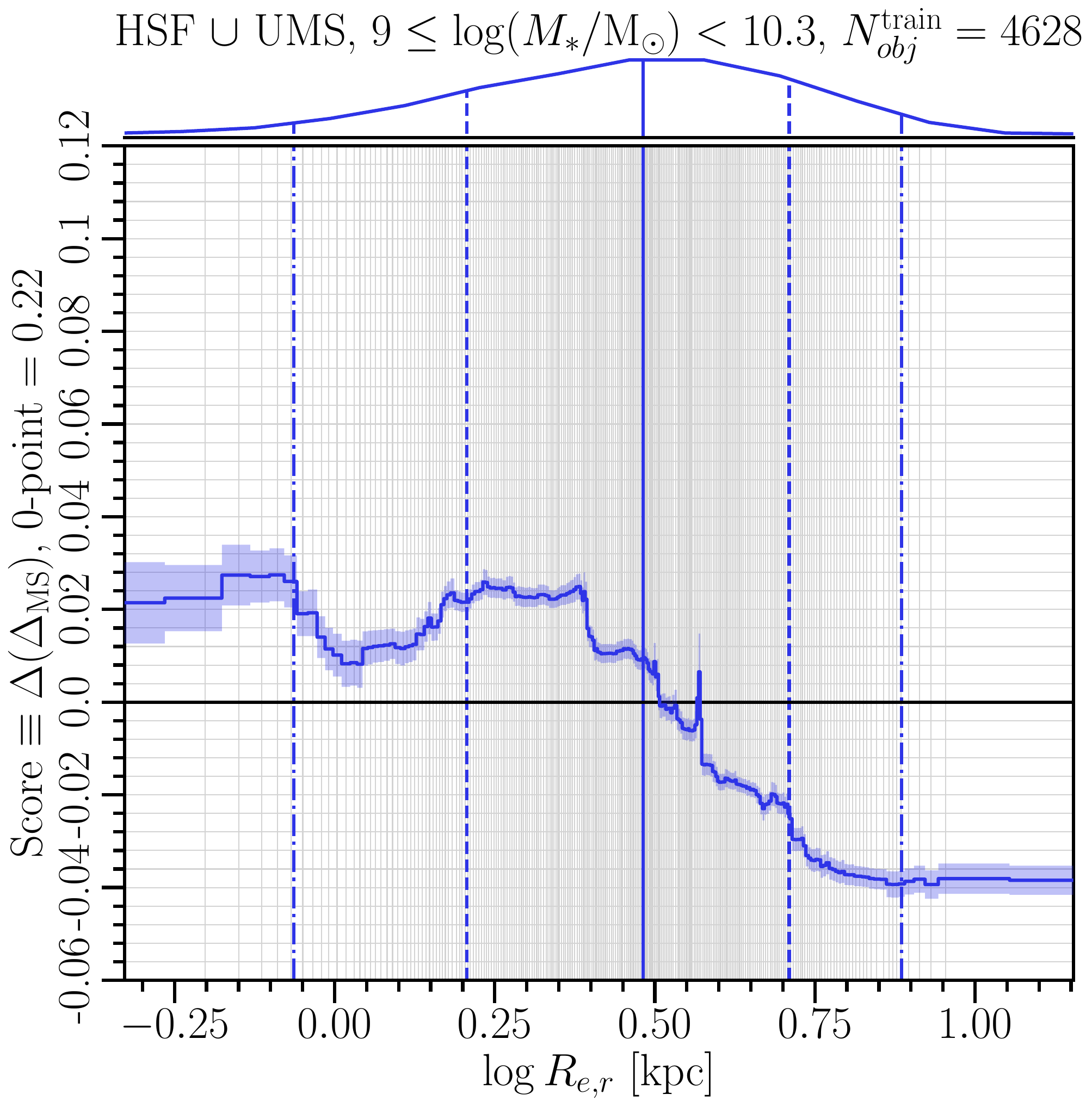}
\hfill
\includegraphics[width=0.22\textwidth]{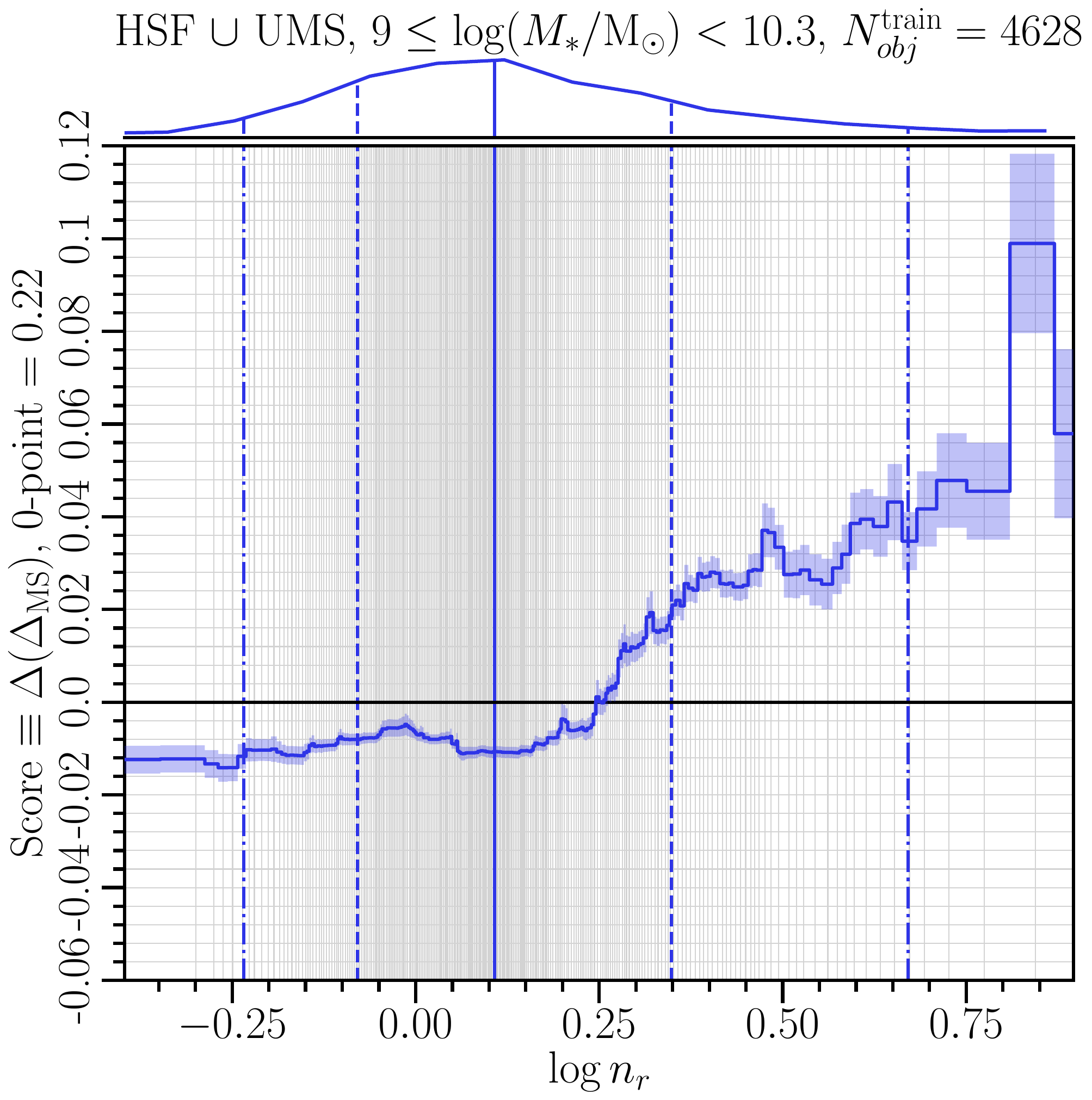}
\hfill

\caption{Correlation matrix of the training sample (upper-left panel), residual distribution between the predicted and actual values of $\Delta_\mathrm{MS}$ for the test sample (upper-right panel), and 1D shape functions (lower panels) sorted for importance (from the left to the right) for HSF and UMS galaxies (i.e. SF galaxies above the SFMS) with $9\leq\log(M_*/\mathrm{M_\odot})<10.3$. The correlation matrix cells are coloured as a function of the Pearson Correlation Coefficient of each pair of features and report its value (rounded up to two decimal digits). For each 1D shape function of $\log\Sigma_1$, $\nabla_{(g-i)}$, $\log R_e$, and $\log n_r$, we highlight the median (vertical thick solid line), and the limits for the $68\%$ (vertical thick dashed lines), and $95\%$ (vertical  thick dash-dotted lines) ranges of the distribution, the latter displayed on the top of each panel. The vertical grey thin lines mark the edges of the equal-density bins used to evaluate the 1D shape functions. For all the 1D shape functions, the shaded region represents the pseudo-errors from the bagging procedure.
The horizontal black thick line tracing the score equal to 0 corresponds to the mean distance from the MS, $\Delta_\mathrm{MS}$, that is  $\simeq 0.22 \, \mathrm{dex}$.
The y-axis range is shared among all the plots to facilitate the comparison of the relative contribution of each feature. The residual distribution plot reports the values of $RMSE$ and $R^2$. The vertical solid line corresponds to 0, while the coloured dashed line is tracing the median of the distribution.}
\label{fig:hsf_ums_9_10.3}
\end{figure*}

\begin{figure*}
\centering
\includegraphics[width=0.25\textwidth]{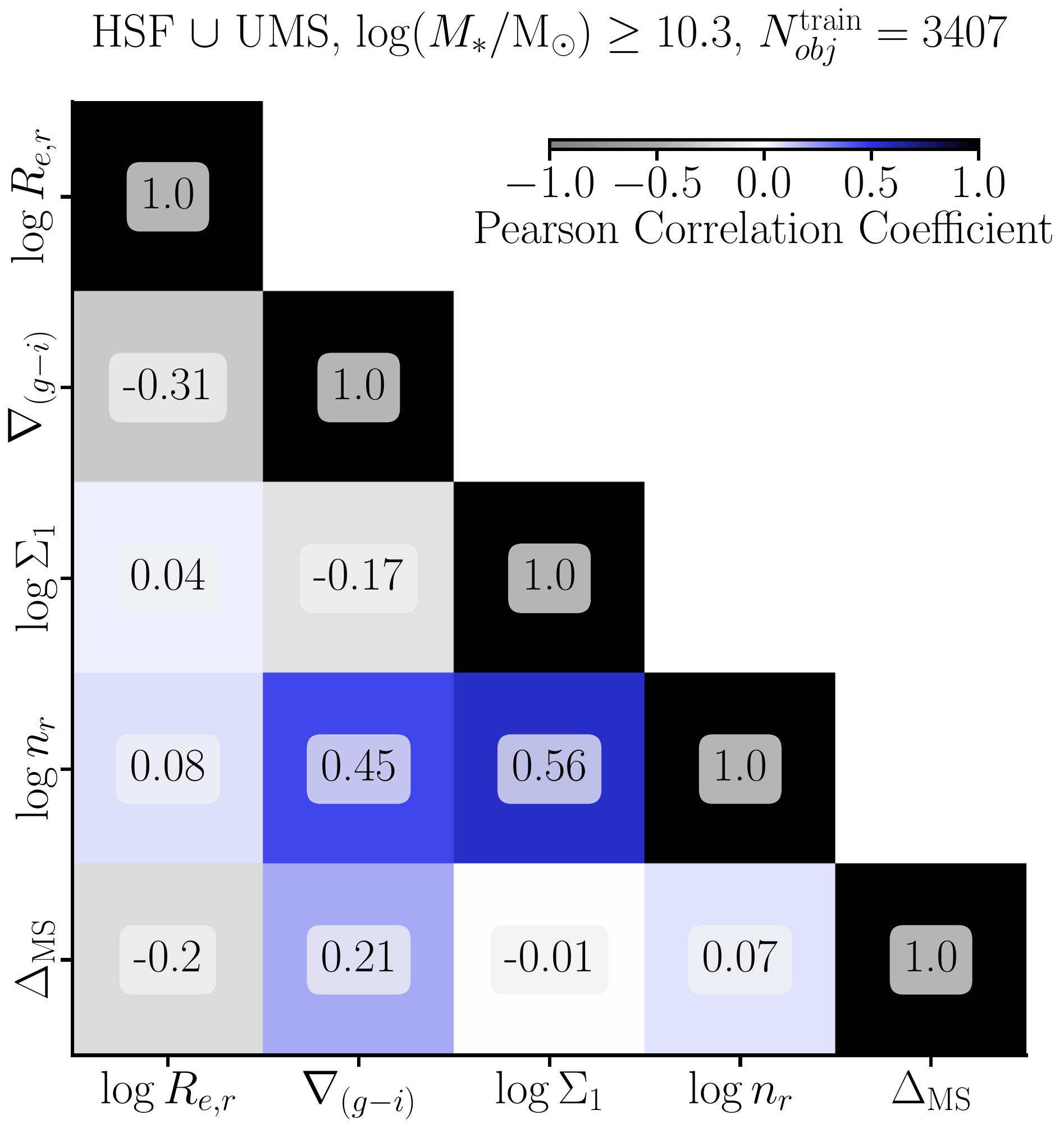}
\hspace{3cm}
\includegraphics[width=0.25\textwidth]{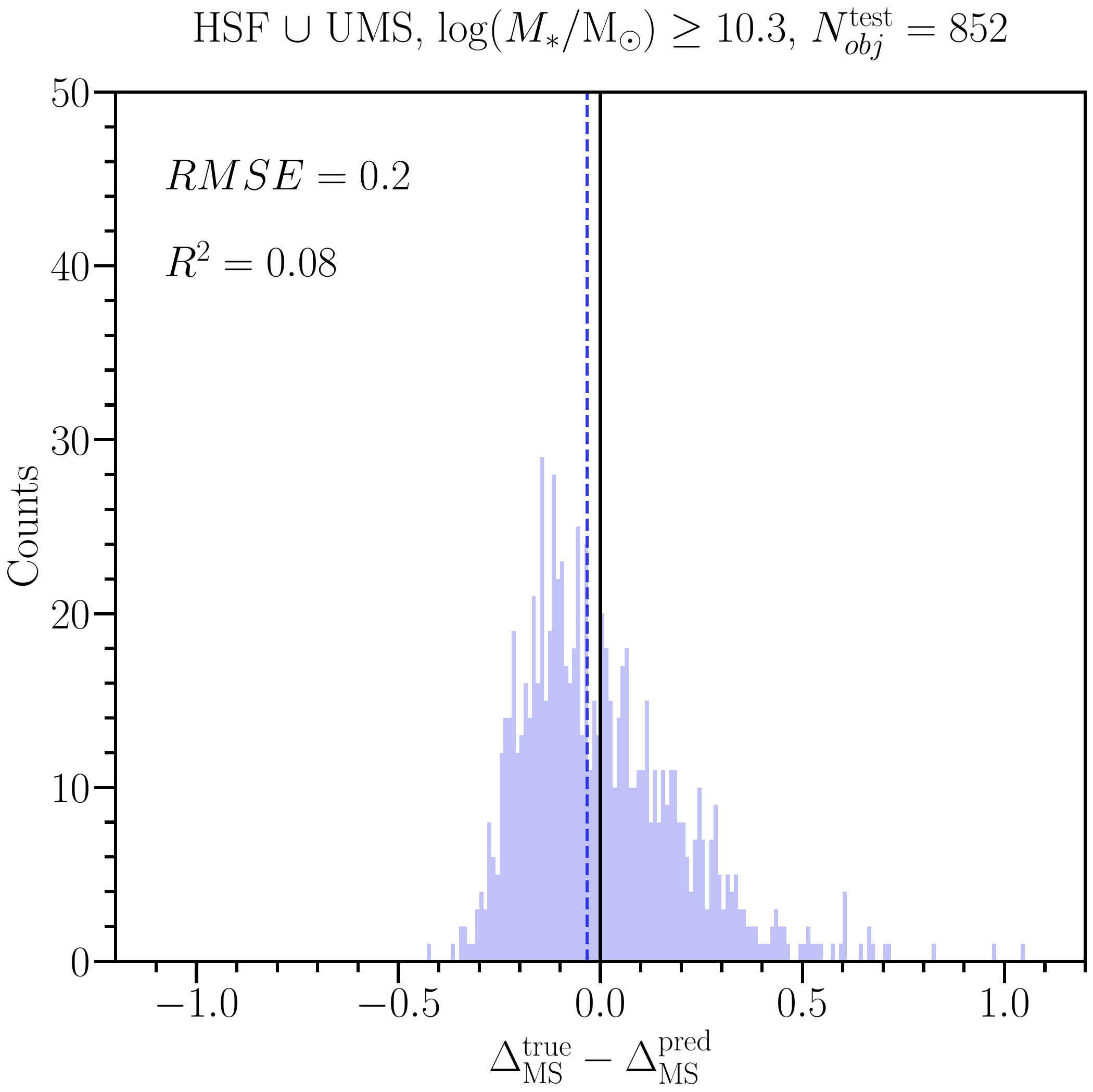}
\\
\includegraphics[width=0.22\textwidth]{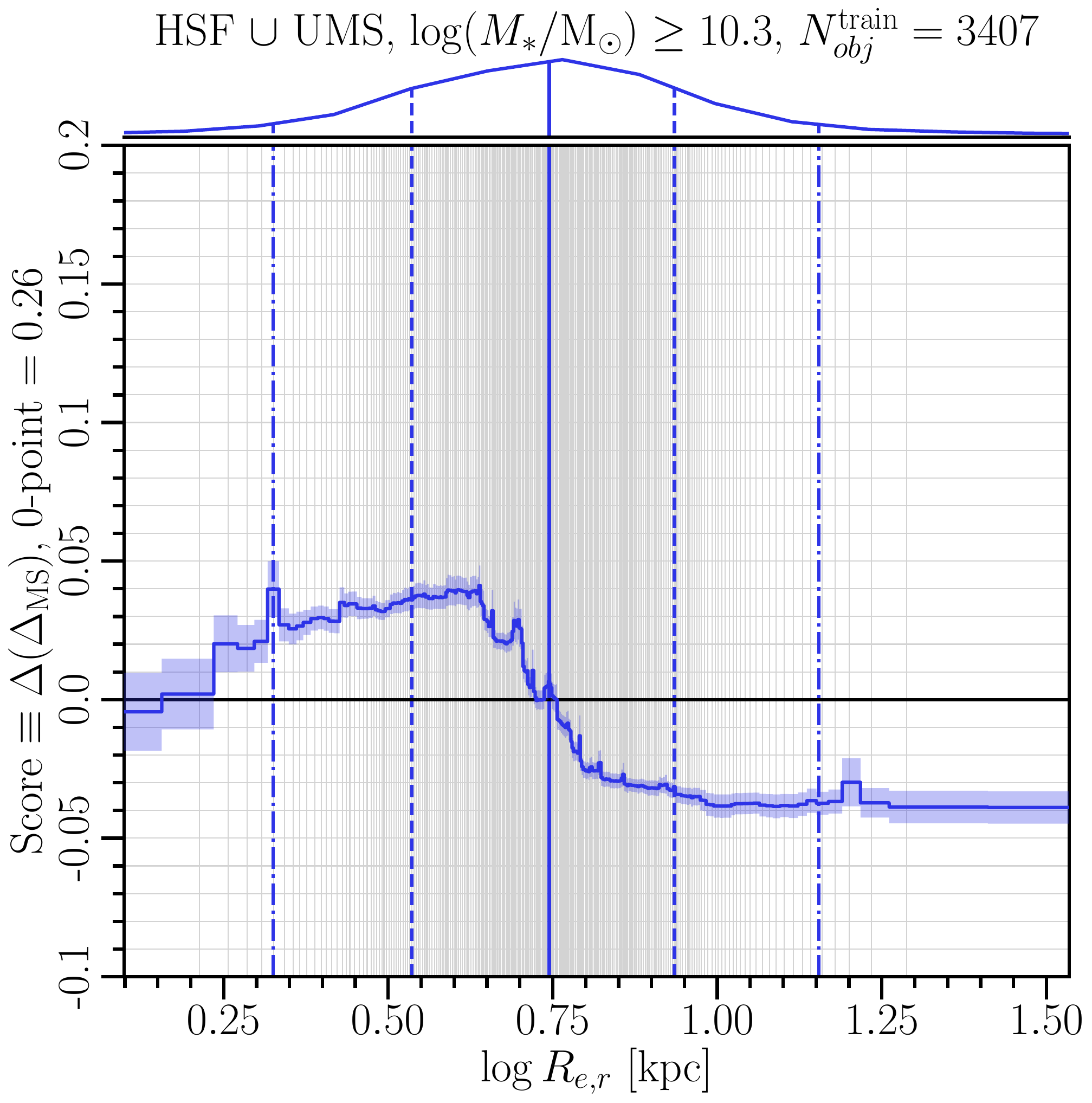}
\hfill
\includegraphics[width=0.22\textwidth]{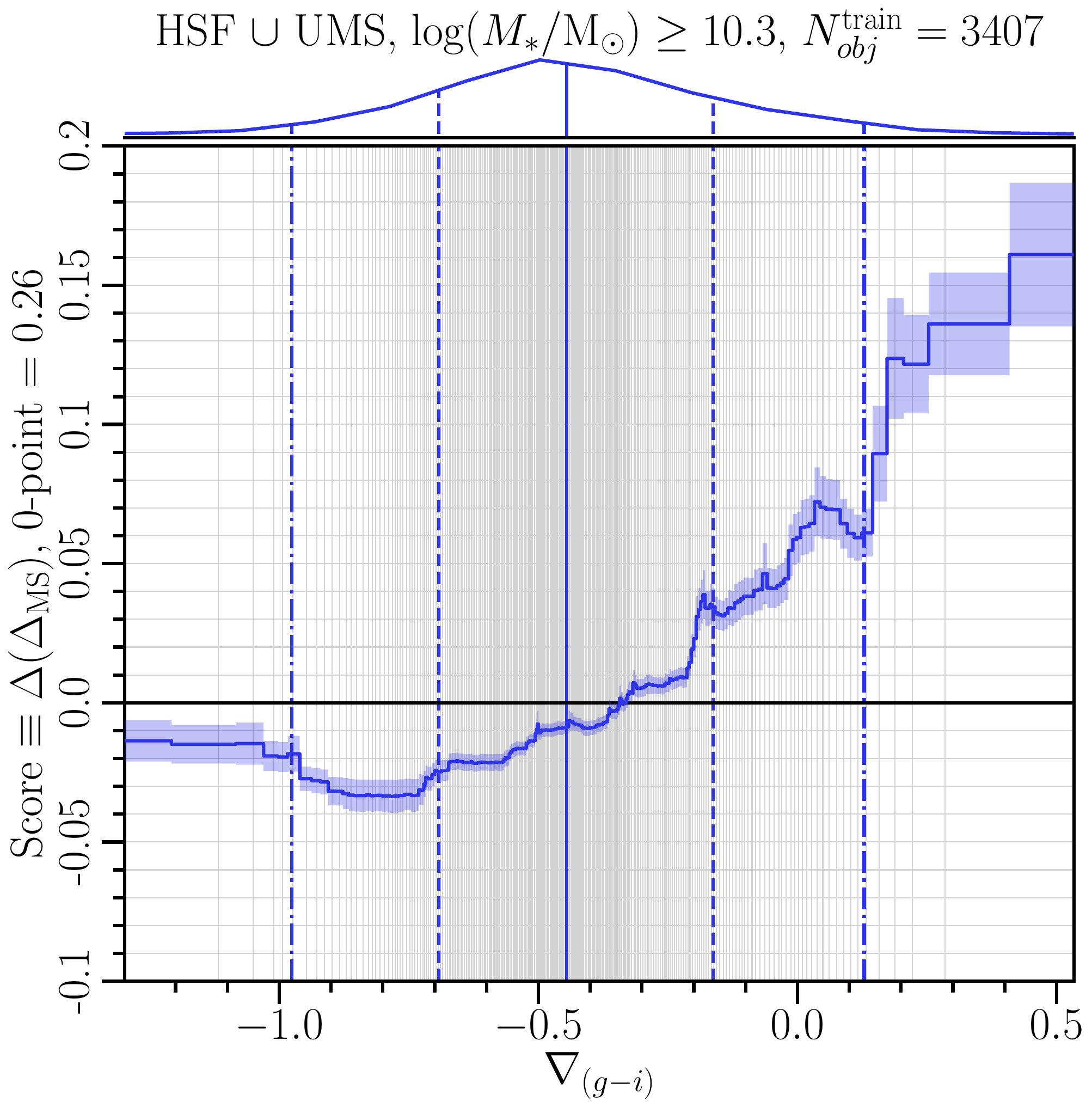}
\hfill
\includegraphics[width=0.22\textwidth]{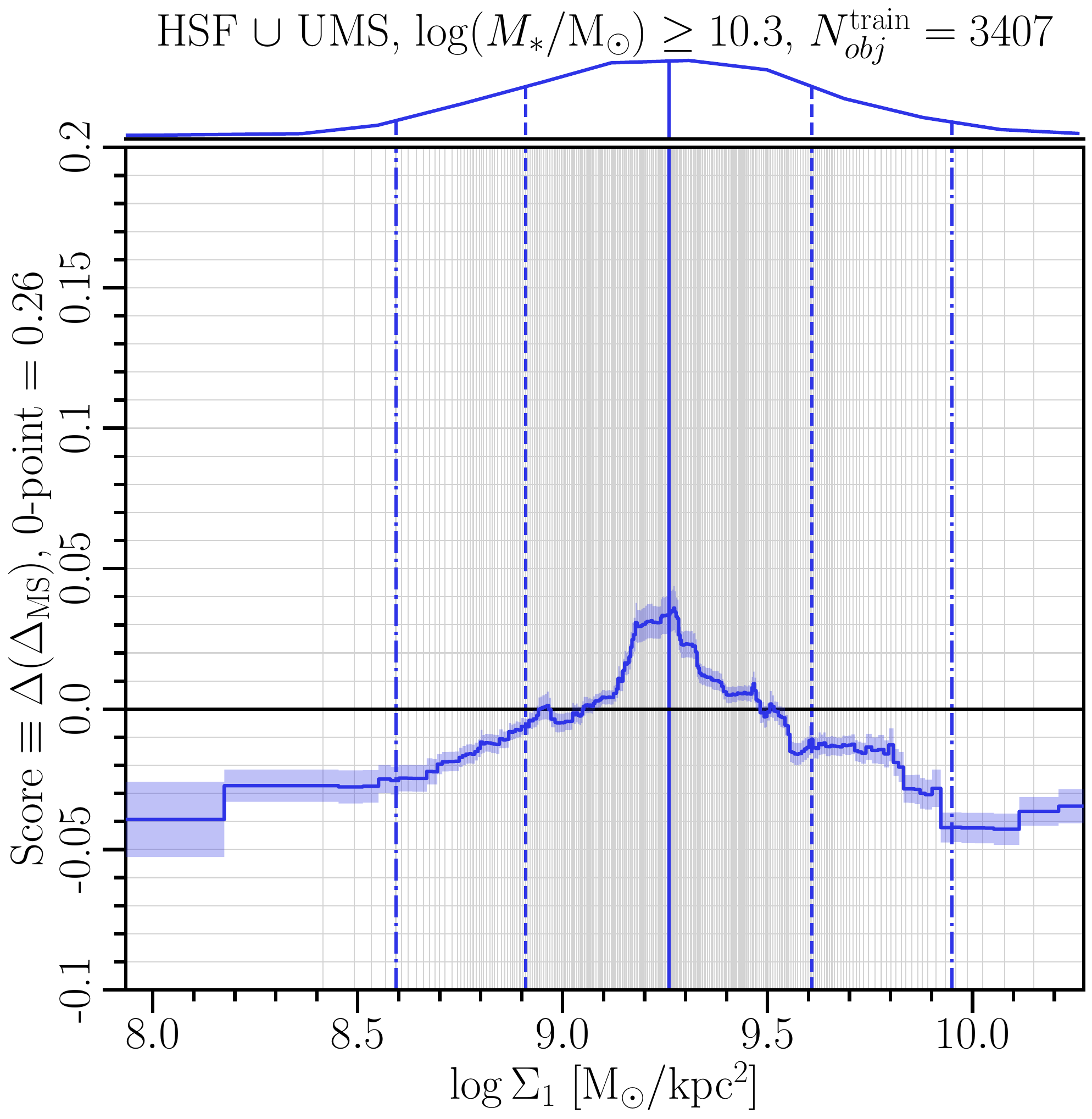}
\hfill
\includegraphics[width=0.22\textwidth]{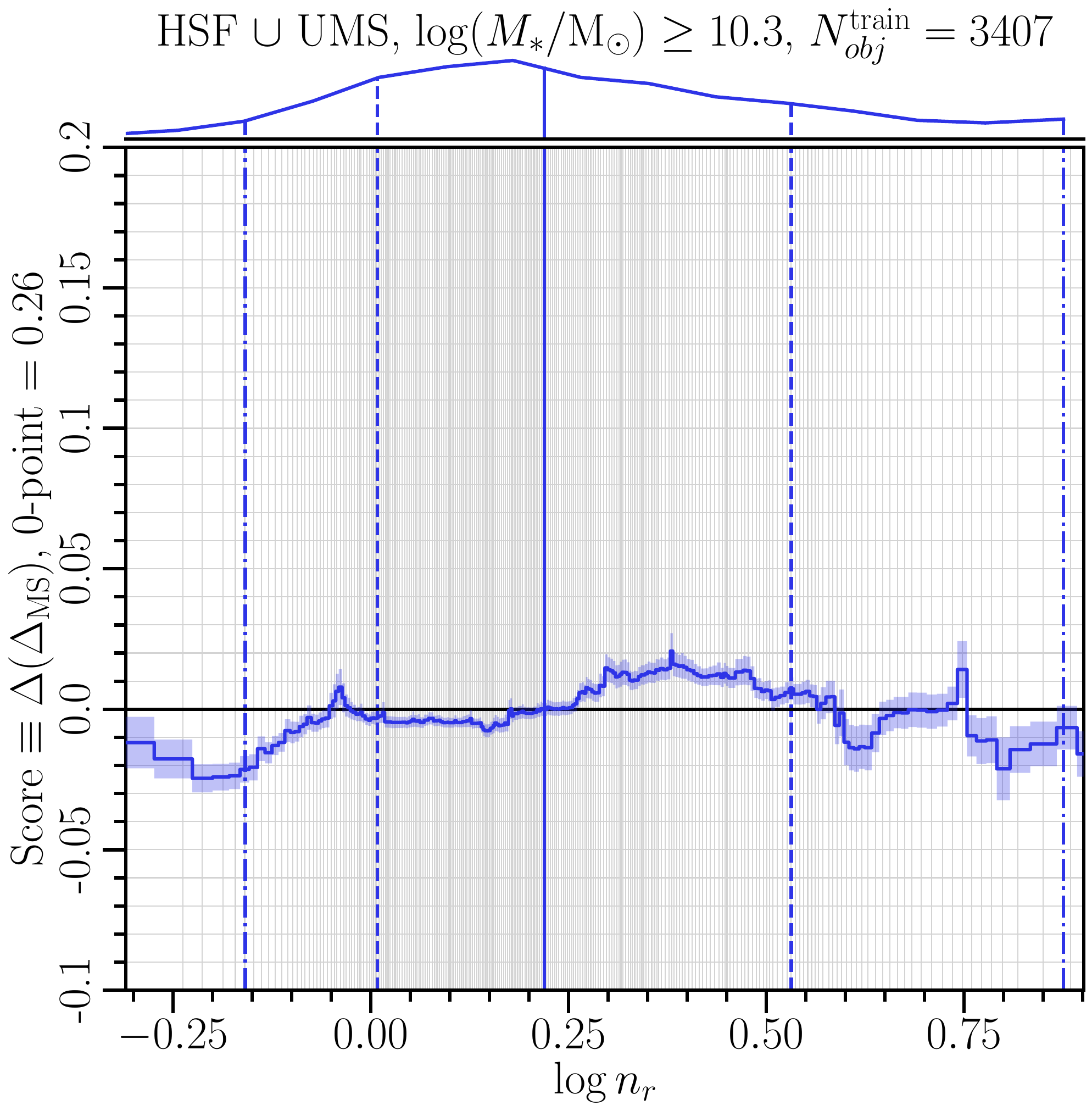}
\caption{Same as \autoref{fig:hsf_ums_9_10.3}, but for HSF and UMS galaxies (i.e. SF galaxies above the SFMS) with $\log(M_*/\mathrm{M_\odot})\geq10.3$. The score equal to 0 corresponds to $\simeq 0.26\, \mathrm{dex}$.}
\label{fig:hsf_ums_10.3_15}
\end{figure*}

\begin{figure*}
\centering
\includegraphics[width=0.25\textwidth]{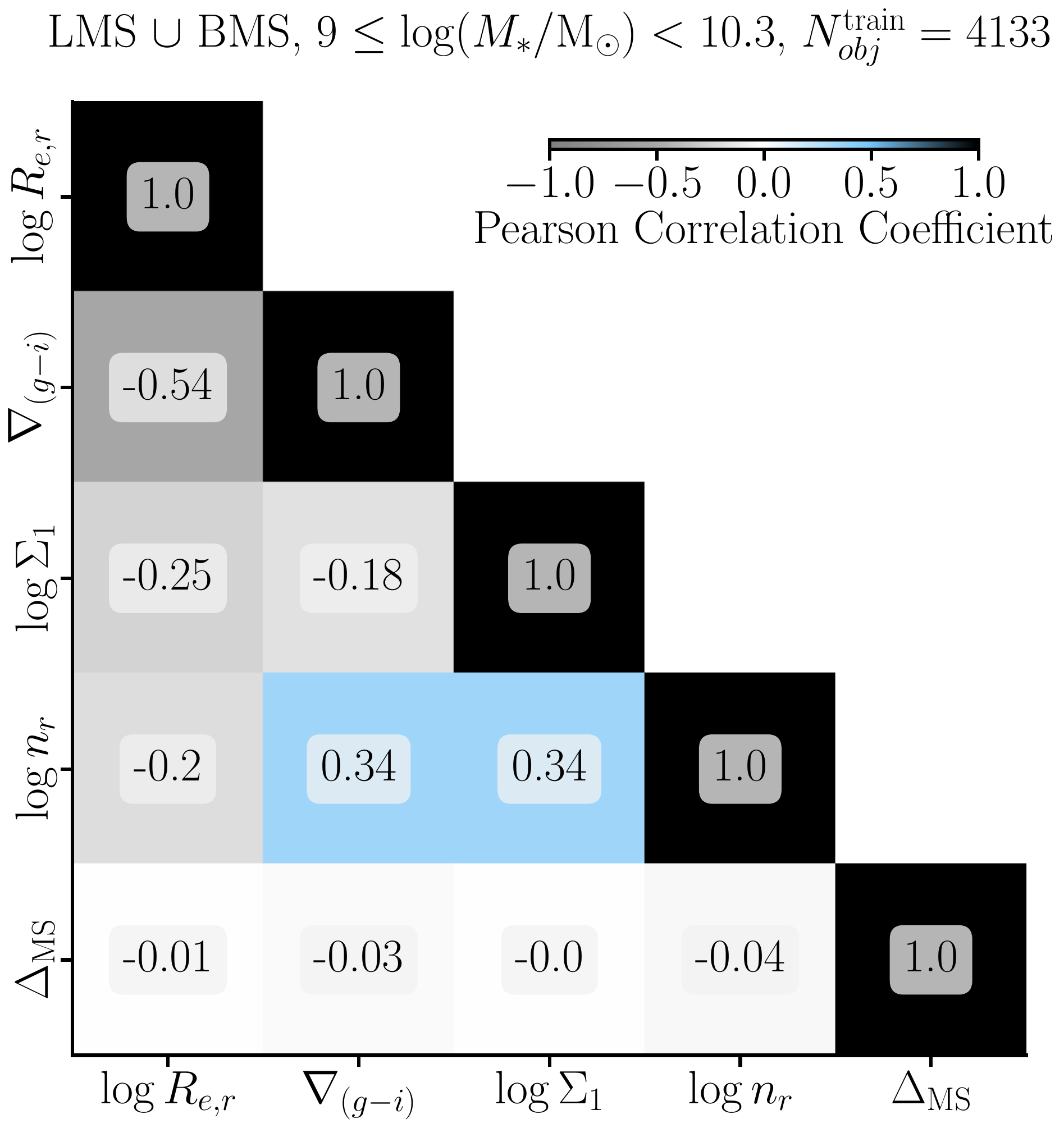}
\hspace{3cm}
\includegraphics[width=0.25\textwidth]{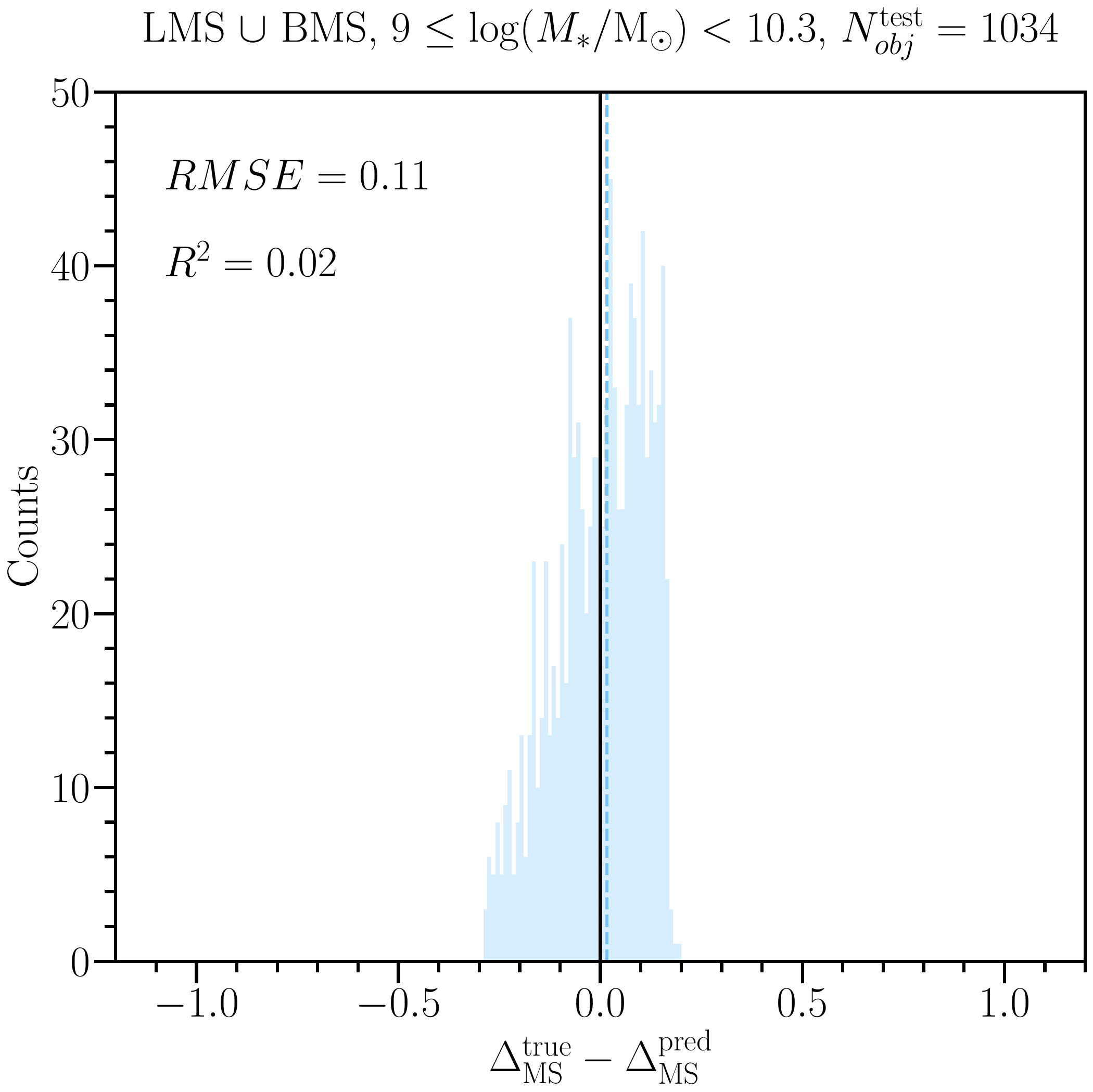}
\\
\includegraphics[width=0.22\textwidth]{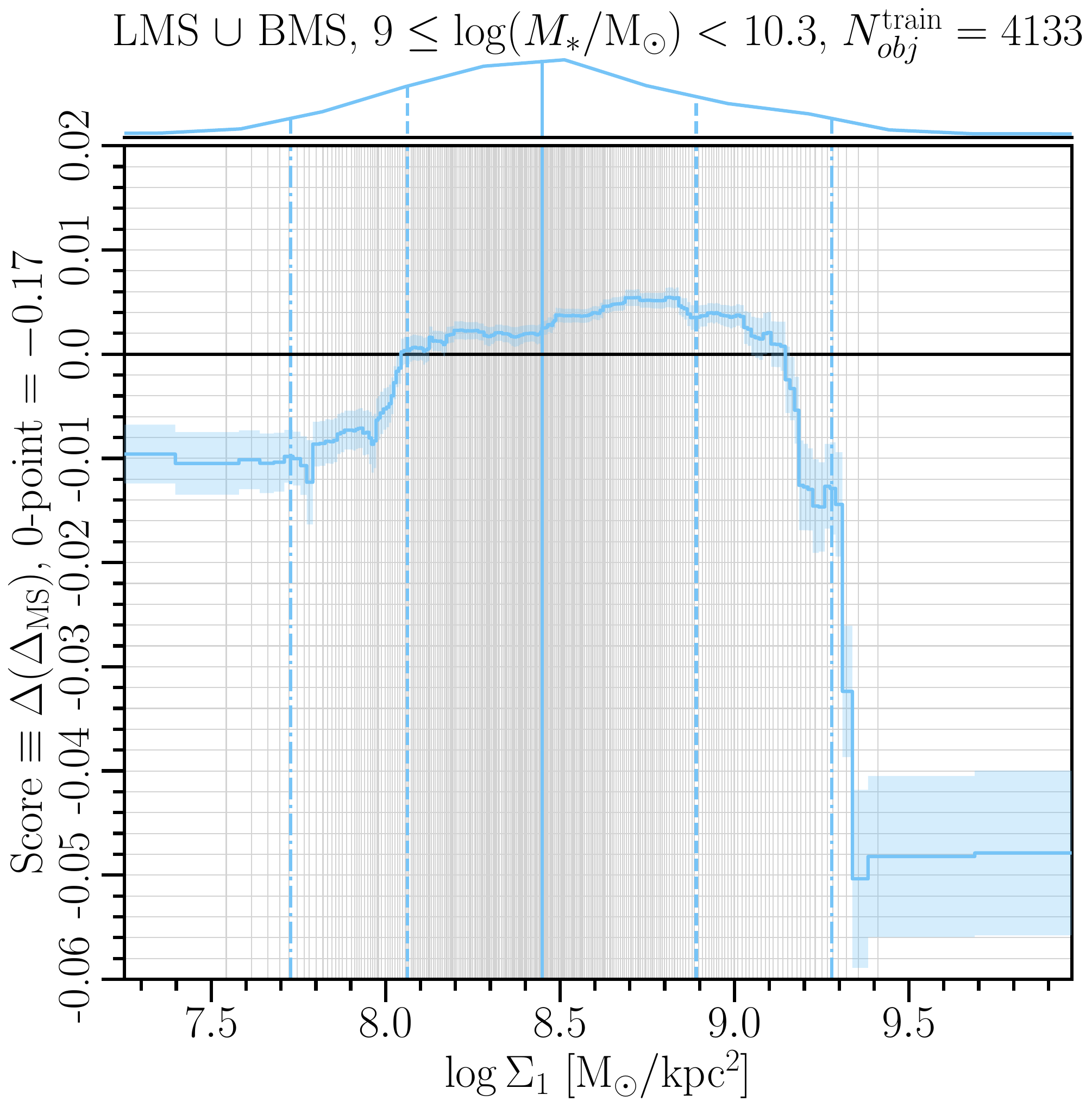}
\hfill
\includegraphics[width=0.22\textwidth]{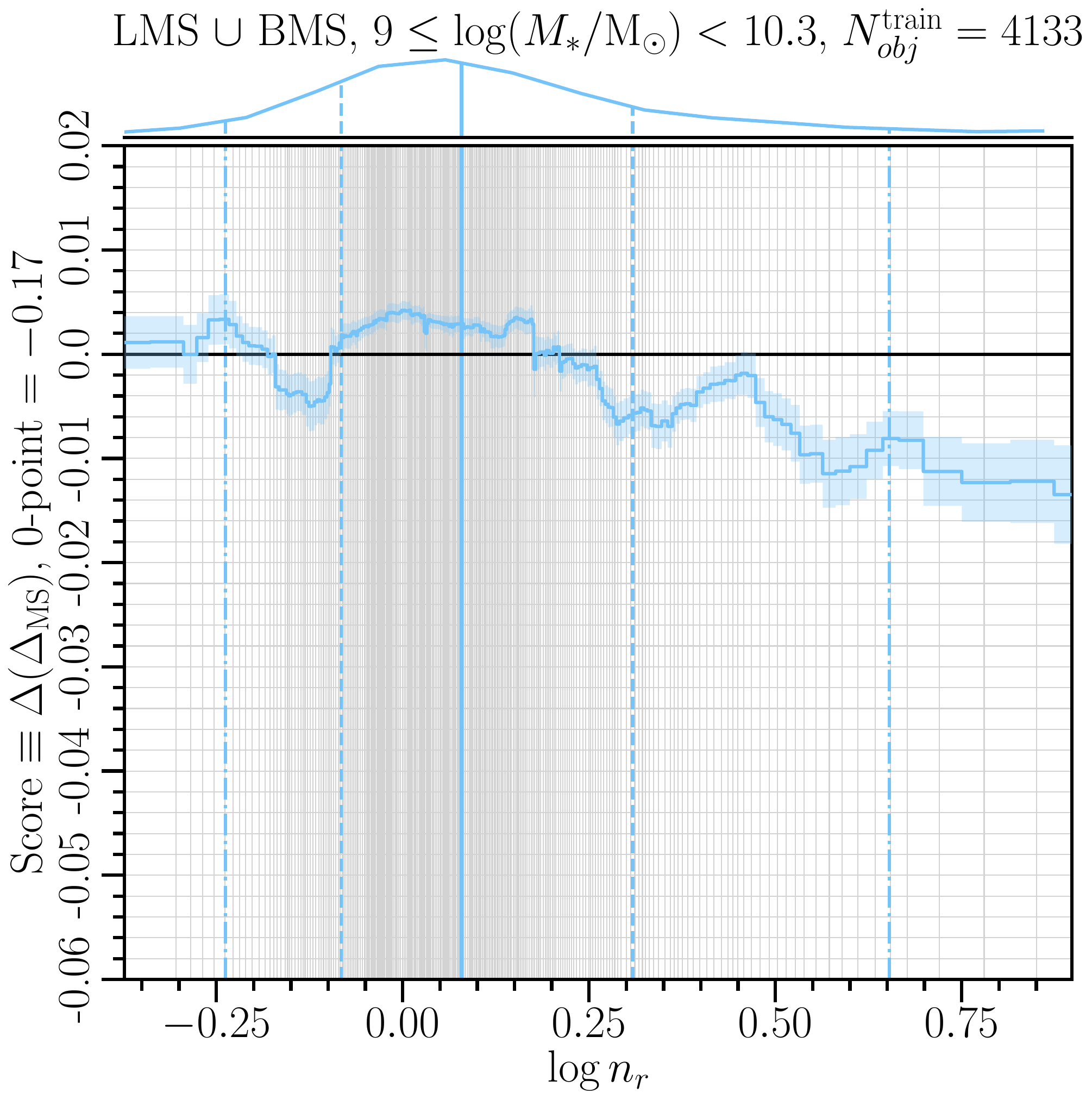}
\hfill
\includegraphics[width=0.22\textwidth]{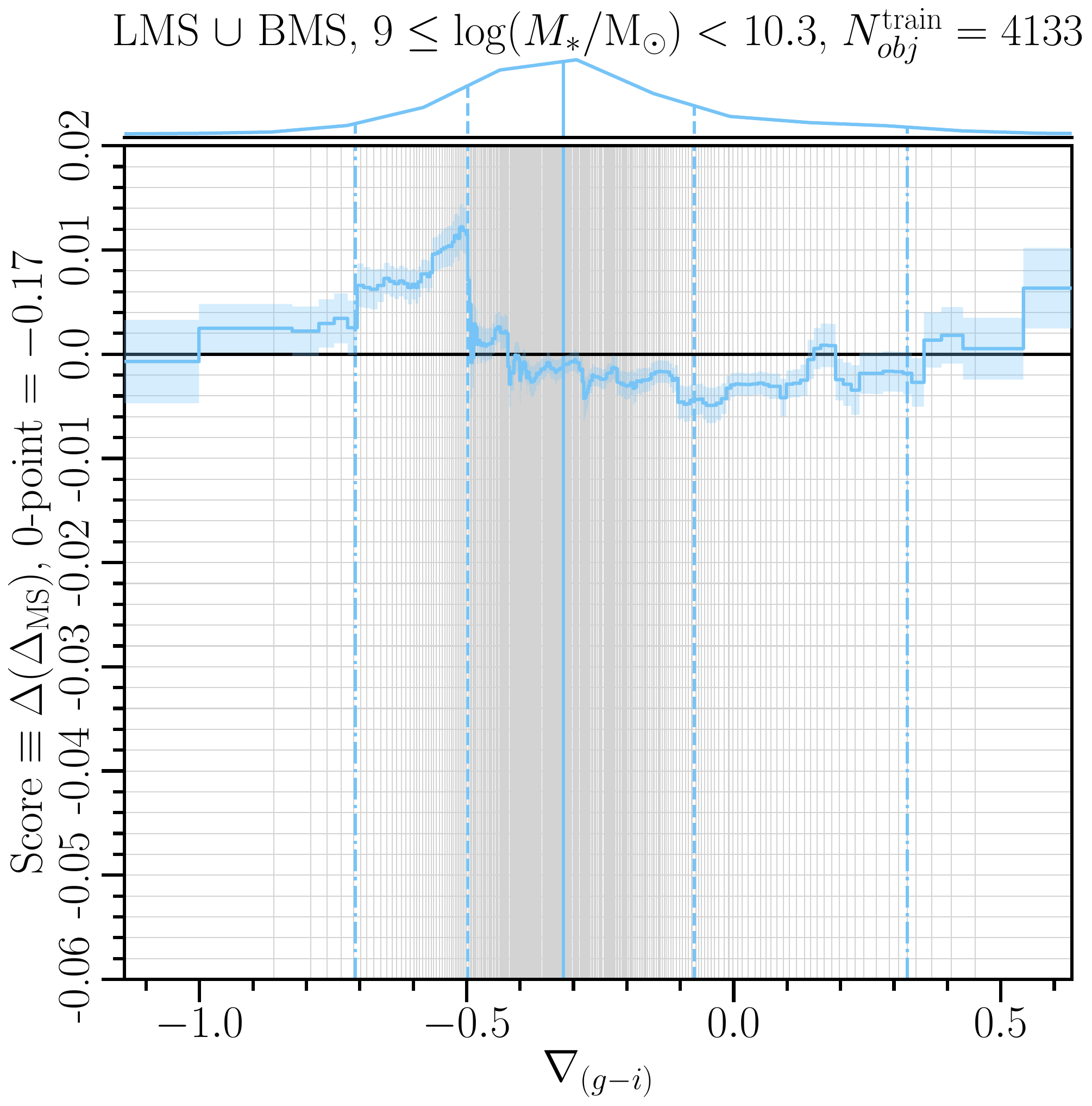}
\hfill
\includegraphics[width=0.22\textwidth]{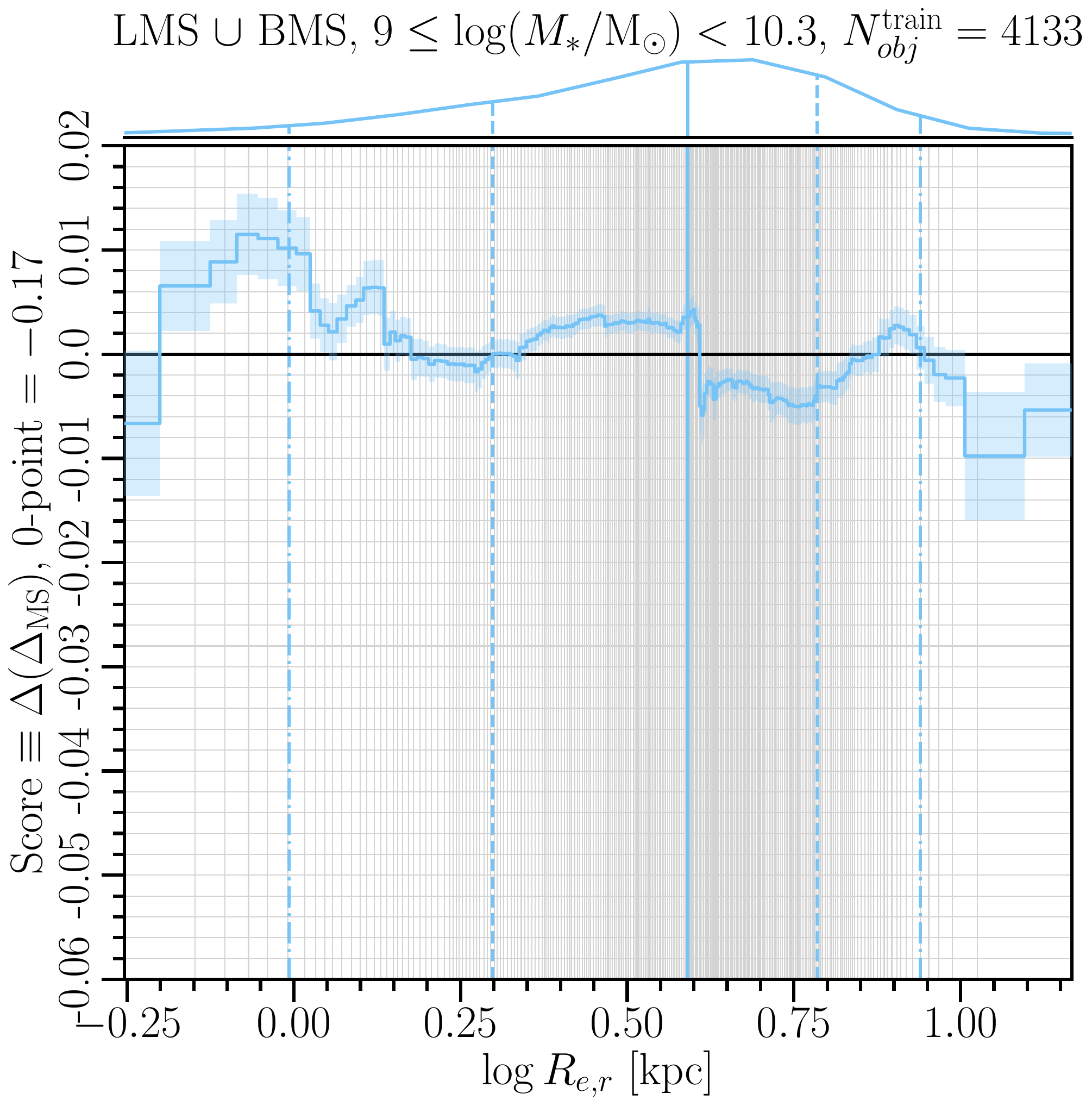}
\caption{Same as \autoref{fig:hsf_ums_9_10.3}, but for LMS and BMS galaxies (i.e. SF galaxies below the SFMS) with $9\leq\log(M_*/\mathrm{M_\odot})<10.3$. The score equal to 0 corresponds to $\simeq -0.17\, \mathrm{dex}$.}
\label{fig:lms_bms_9_10.3}
\end{figure*}

\begin{figure*}
\centering
\includegraphics[width=0.25\textwidth]{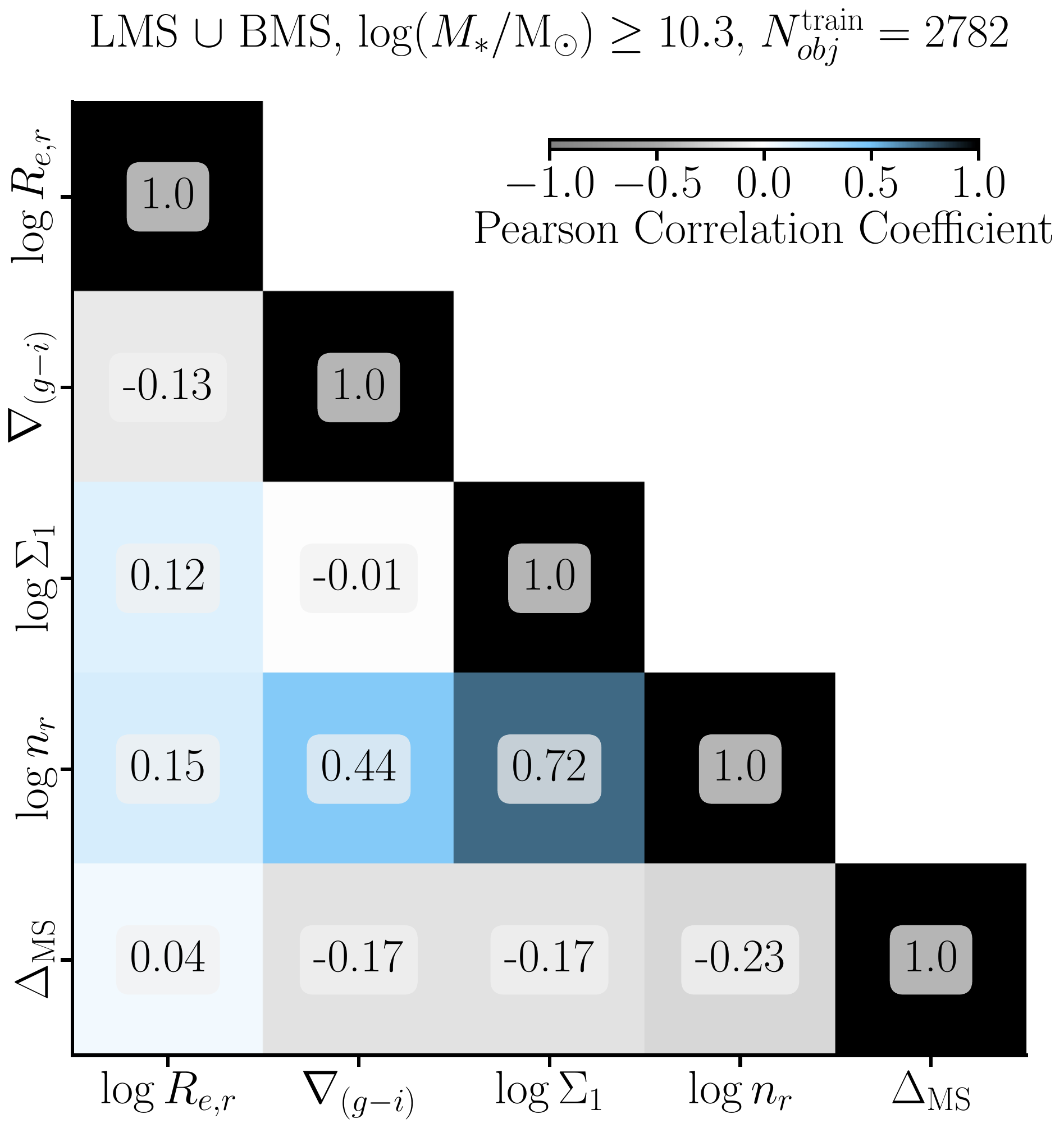}
\hspace{3cm}
\includegraphics[width=0.25\textwidth]{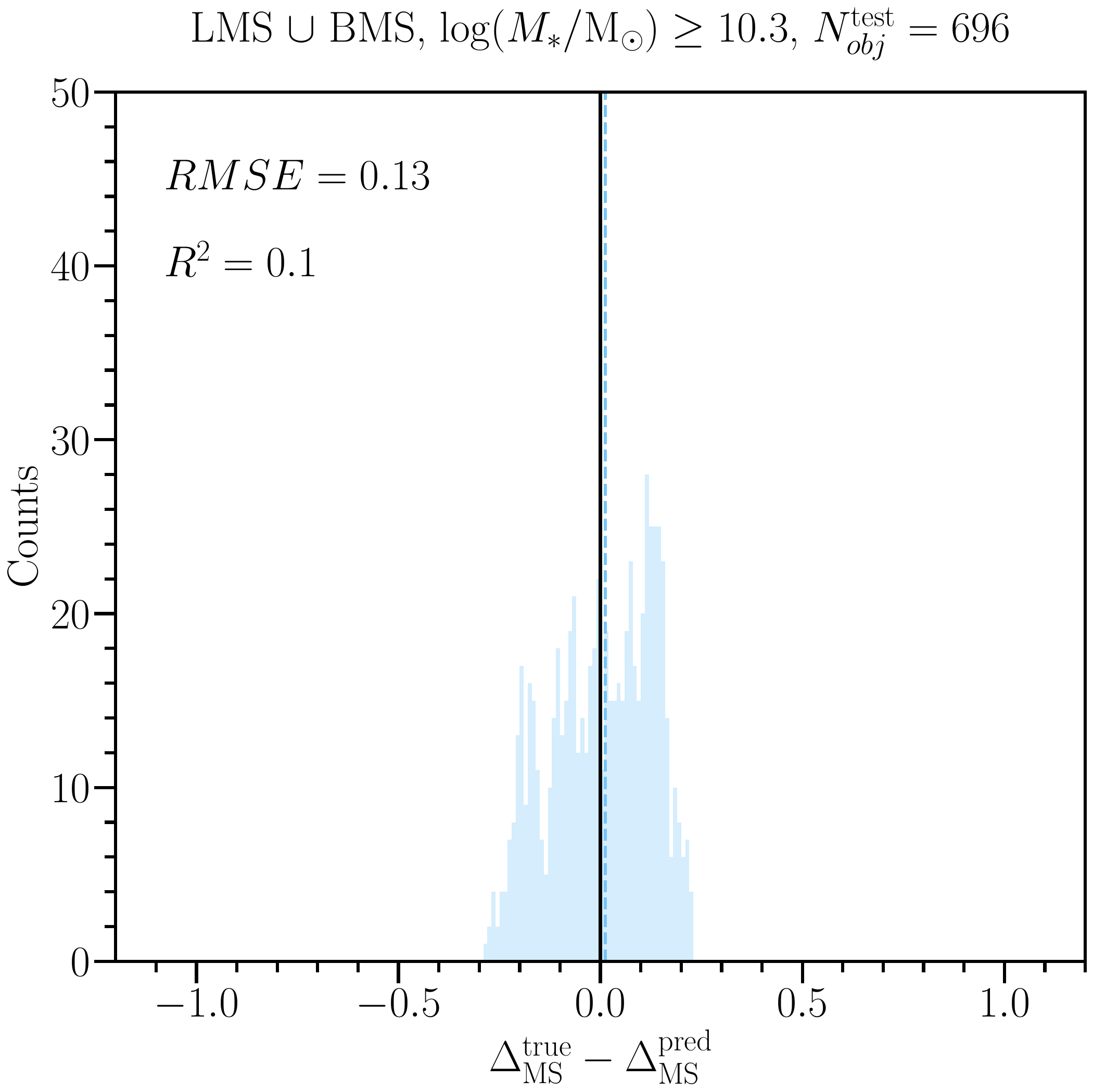}
\\
\includegraphics[width=0.22\textwidth]{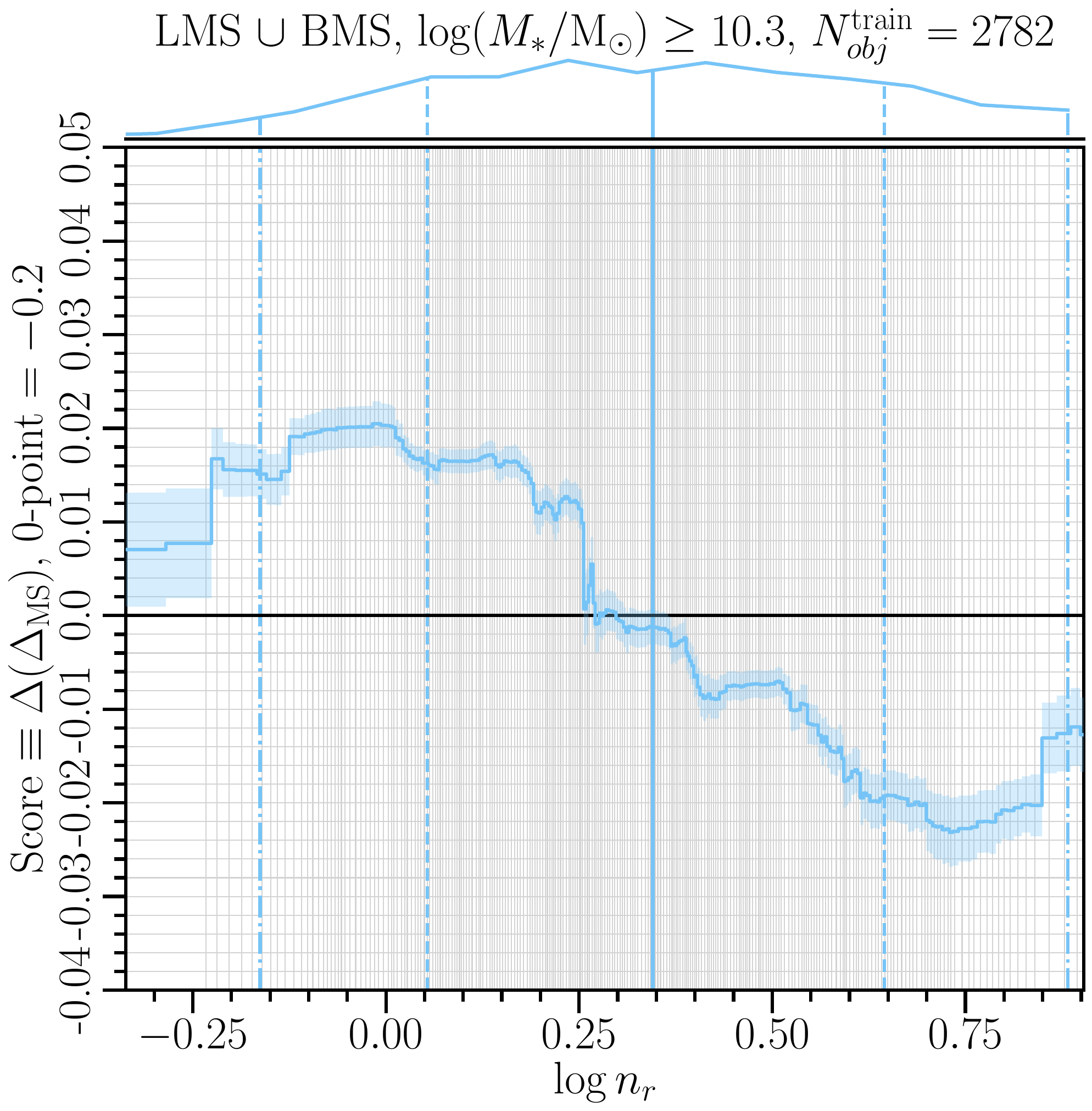}
\hfill
\includegraphics[width=0.22\textwidth]{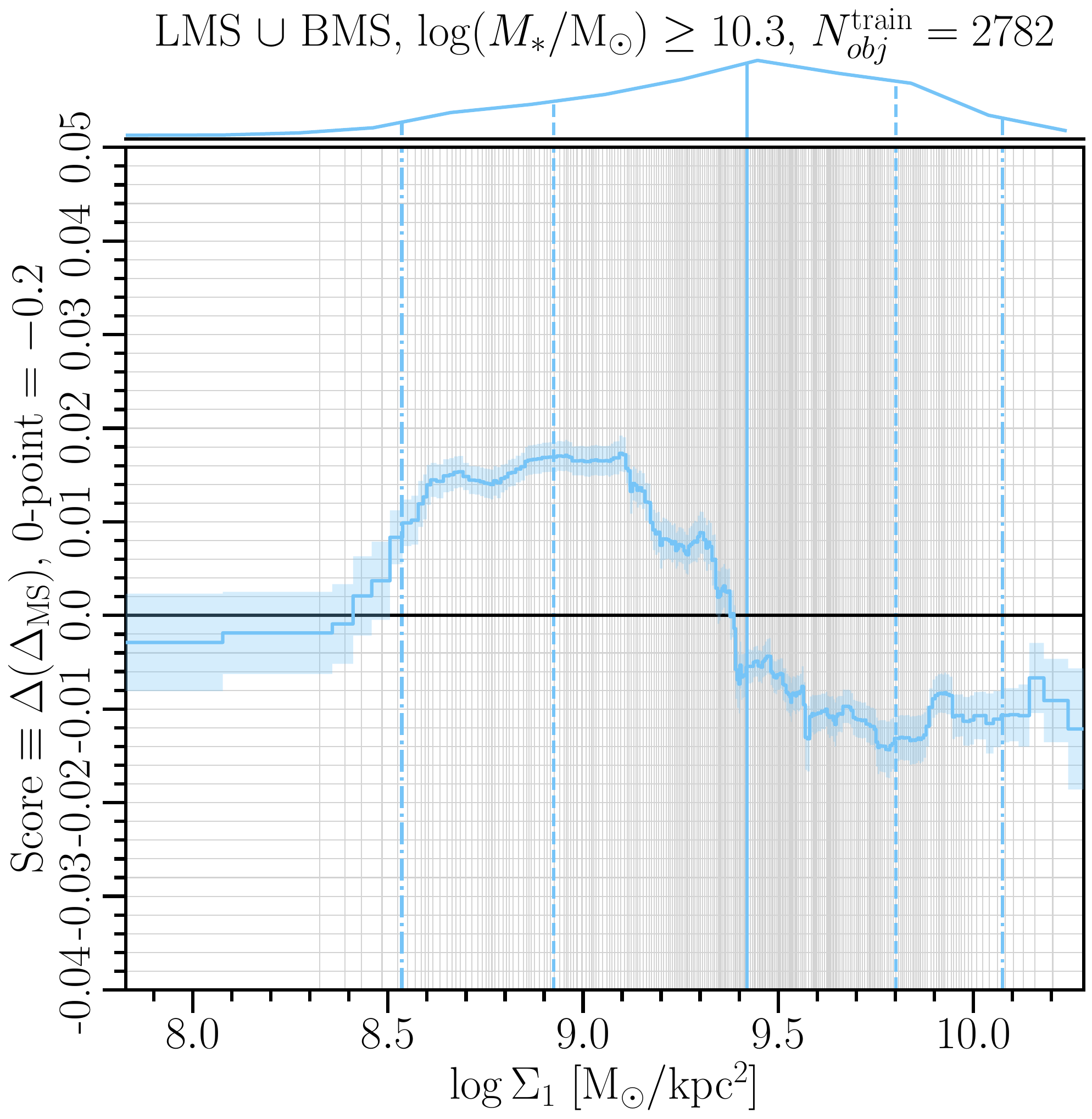}
\hfill
\includegraphics[width=0.22\textwidth]{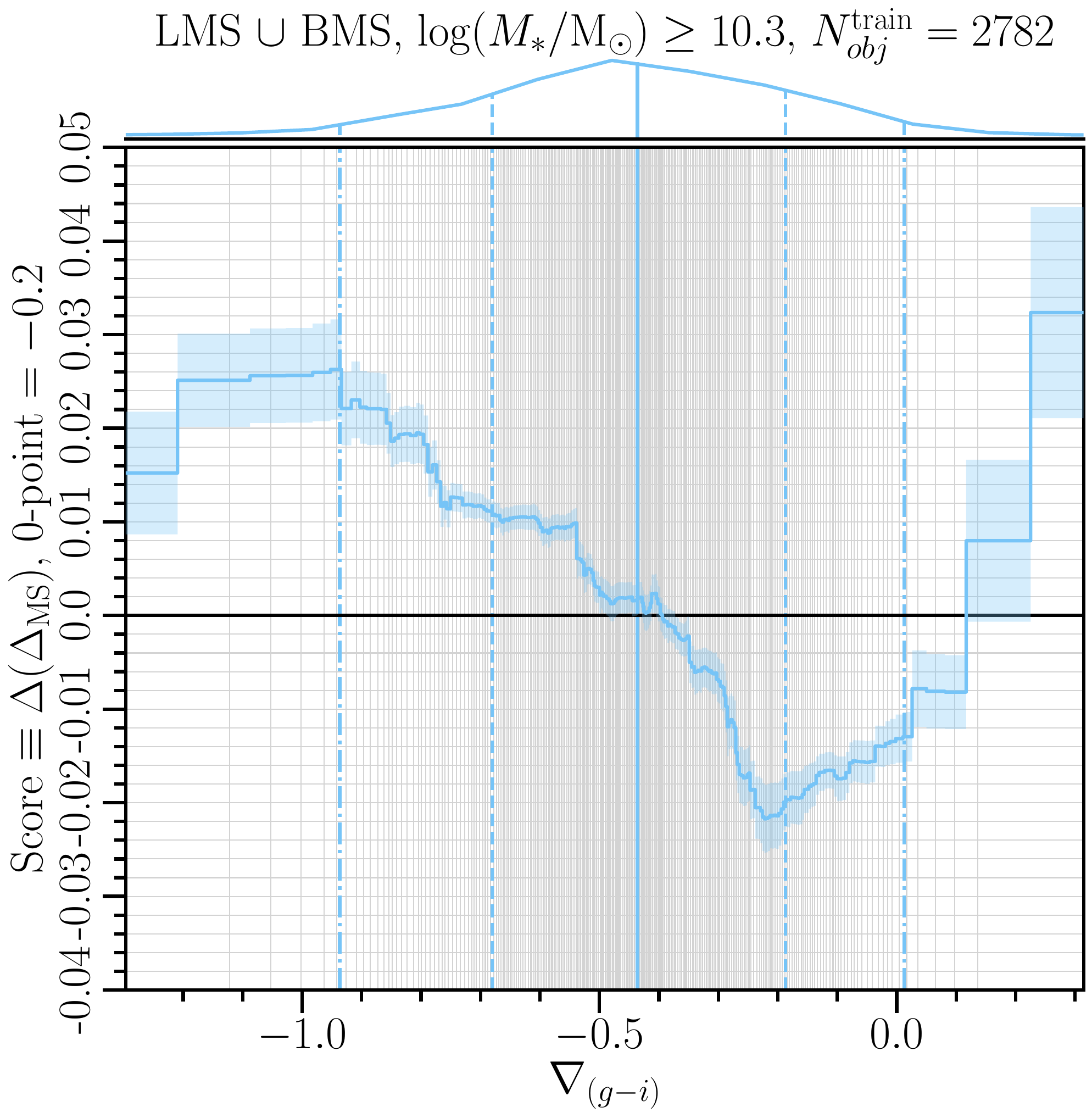}
\hfill
\includegraphics[width=0.22\textwidth]{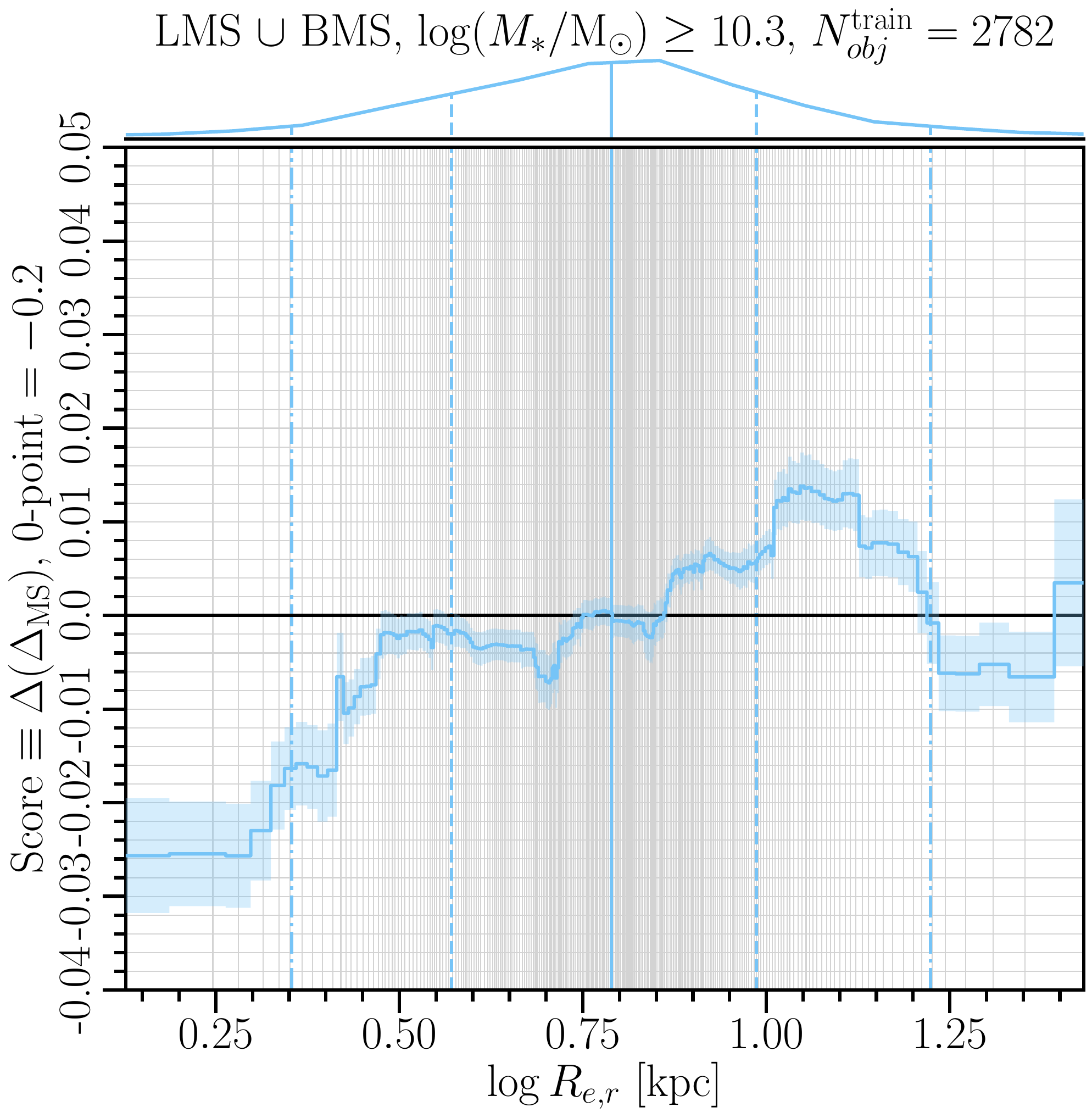}
\caption{Same as \autoref{fig:hsf_ums_9_10.3}, but for LMS and BMS galaxies (i.e. SF galaxies below the SFMS) with $\log(M_*/\mathrm{M_\odot})\geq10.3$. The score equal to 0 corresponds to $\simeq -0.2\, \mathrm{dex}$.}
\label{fig:lms_bms_10.3_15}
\end{figure*}

% \carlo{Adding a brief description.}
In \autoref{sec:ga2m} we presented an analysis made by exploiting EBM to support the findings described in \autoref{secc:Results}.
By training each model for the SF galaxies above the SFMS (i.e. UMS and HSF) and below (i.e. of LMS and BMS, see \autoref{secc:SF_MS} and \autoref{tab:sfrs_sub-samples}), each subsample split further into two stellar mass bins, i.e. $9\leq\log(M_*/\mathrm{M_\odot})<10.3$ and $\log(M_*/\mathrm{M_\odot})\geq10.3$, we obtained the relative importance that single features have in inferring the distance from the Main Sequence of galaxies, $\Delta_\mathrm{MS}$. \autoref{fig:gal_prop_importance} shows for the aforementioned galaxy subsamples in the two stellar mass bins the rankings of importance of the features  obtained by computing the absolute mean scores in the 1D shape functions. In Figures \ref{fig:hsf_ums_9_10.3}-\ref{fig:lms_bms_10.3_15}, we illustrate for each subset the correlation matrix among all the features, also including  $\Delta_\mathrm{MS}$, and the 1D shape functions.

The aim of the application of EBM, in the case presented in this manuscript, is not to train a set of models on our sample and make predictions on other galaxy samples, but to exploit the high intelligibility of the method to confirm the results presented in \autoref{secc:Results}. However, by the very nature of the method, each GAM has been trained on a \emph{training sample} ($80\%$ of each galaxy subsample), and tested on a \emph{test sample} (the remaining $20\%$ of each subset). 
We provide here two metrics to quantify the goodnes of the fit: the root mean squared error $RMSE$, and the coefficient of determination $R^2$. The $RMSE$ provides a measurement of how predicted values differ from the observed values:
\begin{equation}
    RMSE=\sqrt{\frac{1}{N} \sum_{i=1}^N (\mathcal{T}_i - \mathcal{P}_i)^2},
\end{equation}
where $\mathcal{T}_i$ and $\mathcal{P}_i$ are the true (or observed) and predicted values for the $i$-th object, respectively, while $N$ is the size of the sample, in this case, the test sample.
Specifically, the closer to 0, the lower the discrepancy.
Concerning $R^2$, this value informs us about the dispersion of the data and the goodness of the model:
\begin{equation}
    R^2=1-\frac{\sum_{i=1}^N (\mathcal{T}_i - \mathcal{P}_i)^2}{\sum_{i=1}^N (\mathcal{T}_i - \bar{\mathcal{T}})^2},
\end{equation}
where $\bar{\mathcal{T}}$ is the mean of all the $\mathcal{T}_i$ with $i=\{1,...,N\}$. In this case, the closer to 1, the better the model describes the dataset.
Hence, the $RMSE$ provides a measurement of how the true and the predicted values differ, while $R^2$ gives an estimate of how the features comprehensively explain the variation in the predicted quantity.

The upper-right panels of Figures \ref{fig:hsf_ums_9_10.3}-\ref{fig:lms_bms_10.3_15} show the residual histograms between the nominal $\Delta_\mathrm{MS}$ values in the test samples and their values predicted by the model. In each plot, together with the size of the corresponding test sample, the $RMSE$ and $R^2$ values are reported.
In the case of the HSF and UMS galaxies, we found for both stellar mass bins, two Gaussian-like distributions, slightly skewed towards the predicted values. The $RMSE$ are $\sim0.2$, while the $R^2\sim0.1$. For the LMS and BMS systems, instead, the $RMSE$ reduces to $\sim 0.1$ and the $R^2$ is around $0.02$ at $9\leq\log(M_*/\mathrm{M_\odot})<10.3$, and lightly improved in the high-mass tail, with a value of $\sim0.1$. These last two distributions, however, do not resemble a Gaussian and are more noisy than the previous two.
A possible explanation for the low $R^2$ values for the 4 subsets can be found in the nature of the galaxies samples and their large scatter with respect to the $\Delta_\mathrm{MS}$.

In conclusion, we stress here again that the scope of applying EBM was to give support to the results presented in \autoref{secc:Results} thanks to the high information level of the method. Indeed, the 1D shape functions presented in Figures \ref{fig:hsf_ums_9_10.3}-\ref{fig:lms_bms_10.3_15} confirm all the previous findings, in particular when we focus on the regions containing $68\%$ of each subsample (delimited by dashed lines).

% \begin{table}
%  \caption{...}
%  \label{tab:ga2m}
%  \begin{tabular}{lccc}
%   \hline
%   Subsample & Training ($80\%$) / test ($20\%$) & $RMSE$ & $R^2$\\
%   \hline
  
%   HSF [$9;10.3$)       &    1614 / 404 & 0.16 & 0.04 \\
%   HSF [$10.3;+\infty$) &    1500 / 375 & 0.17 & 0.07 \\

%   UMS [$9;10.3$)       &    3088 / 772 & 0.07 & 0.03 \\
%   UMS [$10.3;+\infty$) &    1956 / 489 & 0.07 & $\lesssim 0.01$\\

%   LMS [$9;10.3$)       &    3092 / 773 & 0.07 & $\lesssim 0.01$\\
%   LMS [$10.3;+\infty$) &    1840 / 460 & 0.07 & 0.02\\

%   BMS [$9;10.3$)       &    1111 / 278 & 0.06 & $\lesssim 0.01$\\
%   BMS [$10.3;+\infty$) & \,\,988 / 247 & 0.06 & 0.02\\

%   \hline
%  \end{tabular}
% \end{table}

% \section{Some extra material}

% If you want to present additional material which would interrupt the flow of the main paper,
% it can be placed in an Appendix which appears after the list of references.

%%%%%%%%%%%%%%%%%%%%%%%%%%%%%%%%%%%%%%%%%%%%%%%%%%

\section{The inverted global KS law and new fits for Molecular Hydrogen masses}

\label{app:fits_to_KSlaws}

In this section, we provide a detailed description of our fits to the inverted Kennicutt-Schmidt (KS) law model and Equation (\ref{eq:M_h2}). To accomplish this, we utilize two samples. The first sample is based on the compilation by Callete et al. (in prep.), heavily relying on the xCOLDGASS sample \citep{Saintonge+2017}. In Callete et al., we chose to apply the conversion factor between CO luminosity and H$_2$ masses proposed by \citet{Accurso+2017}. Additionally, we cross-matched the Callete et al. sample with those from \citet{Meert+2015,Meert+2016} to obtain 2D photometric Sérsic fits for the $g$, $r$, and $i$ bands. The second sample is described by \citet{de_los_Reyes_Kennicutt2019}, who recently revisited the global KS law for spiral galaxies. We corrected their H$_2$ masses to be consistent with the \citet{Accurso+2017} correction factors and changed their IMF to ours for consistency. 

Our strategy for constraining the best-fit models for the inverted global KS law and Equation (\ref{eq:M_h2}) is as follows: we utilize the compilation by Callete et al. and estimate $\Sigma_{\rm HI}$ and $\Sigma_{\rm H_2}$ such that the relation $\Sigma_{\rm gas}$ as a function of $\Sigma_{\rm SFR}$ aligns with the measurements from \citet{de_los_Reyes_Kennicutt2019}. Subsequently, we fit the resulting inverted global KS law using only the Callete et al. sample. We chose the Callete et al. sample over the \citet{de_los_Reyes_Kennicutt2019} sample because the latter requires additional homogenization efforts with our data (e.g., photometry, morphology, etc.), which goes beyond the scope of this paper. Furthermore, the Callete et al. sample is already consistent with the stellar masses, SFRs, and half-light radius in our SDSS catalog. Below, we provide a more detailed description of additional calculations performed for the Callete et al. sample. 

As outlined in the main manuscript, the global KS law is typically derived by computing the gas mass and SFR within a star-forming region with a radius that encloses approximately 95\%\ of the H$\alpha$ flux of the galaxy. This radius is denoted as $R_{\rm SF}$. According to \citet{de_los_Reyes_Kennicutt2019}, this star-forming radius is smaller than the \citet{Holmberg1950} radius ($R_{25}$, the $B$-band 25 mag arcsec$^{-2}$ isophote). Specifically, $R_{\rm SF} = R_{25} / 1.83$. To adhere closely to this definition, we employ the model by \citet{Salim+2023} for the isophotal radius at 25 mag arcsec$^{-2}$ in the $r$-band, given by their Equation 6, which depends on stellar mass and the effective radius.

The relation provided by \citet{Salim+2023} includes a scatter of 0.05 dex, which we incorporate as log-normal random scatter to our sizes. Assuming the $B$ and SDSS $g$-bands are very similar for SFMS galaxies, we utilize the wavelength dependence of the effective radius with stellar mass for the SFMS from \citet{Kawinwanichakij+2021} to convert the SDSS $r$-band into the $g$-band. We adopt the mean value reported by \citet{de_los_Reyes_Kennicutt2019} to convert the isophote into a star-forming radius, i.e., $R_{25}/R_{\rm SF} = 1.83$. This value is then used to transform our SDSS $R_{25}$ into $R_{\rm SF}$ radii, incorporating a random scatter of 0.08 dex. 

%%%%%%%%%%%%%%%%%%%%%%

%%%%%%%%%%%%%%%%%%%%%%%%%%%%%%%%%%%%%%%%%%%%%%%%%%%%%%%%%%%%%%
\begin{figure*}
    \vspace{2pt}
    \includegraphics[height=2.3in,width=2.8in]{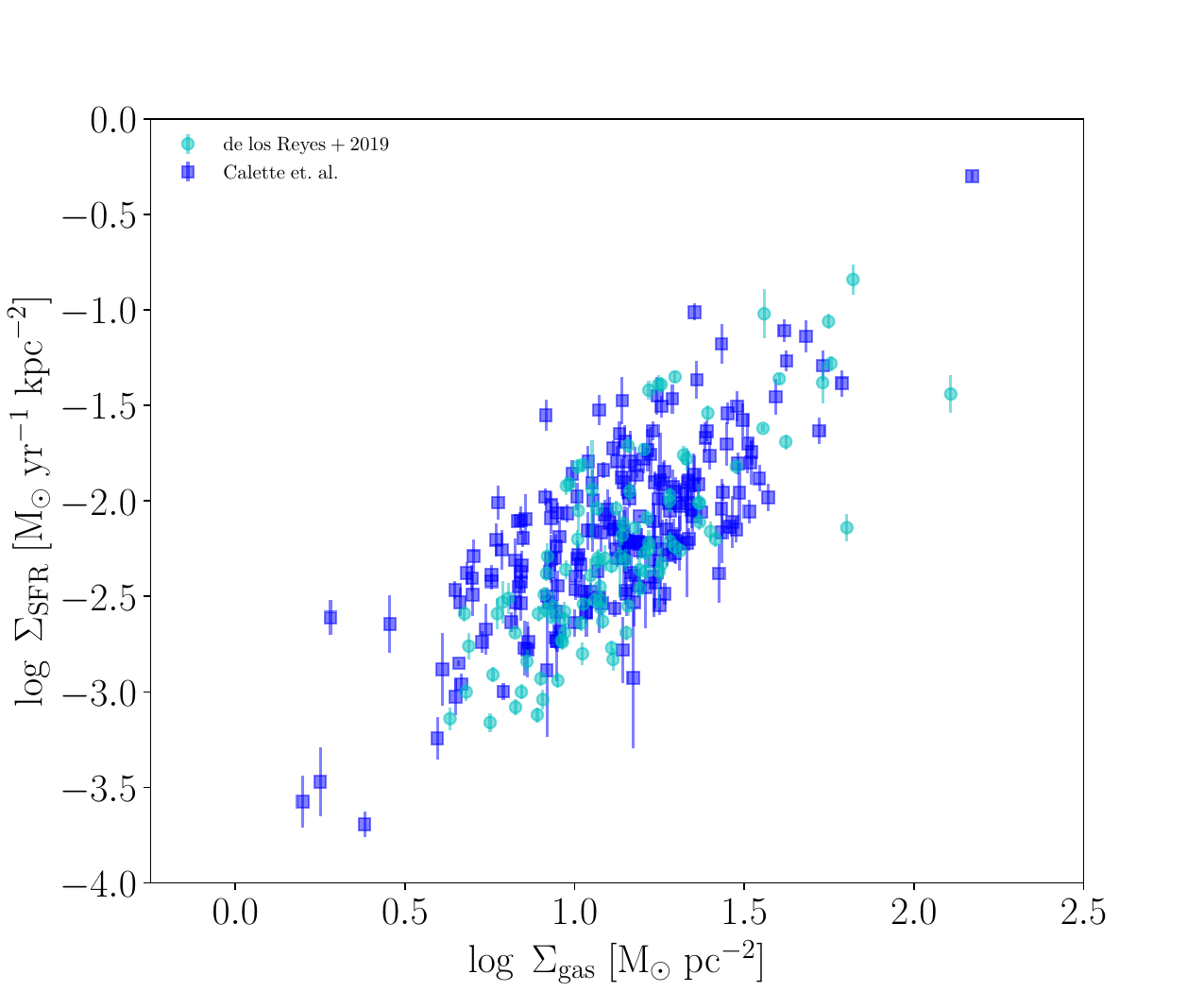}
    \includegraphics[height=2.3in,width=2.8in]{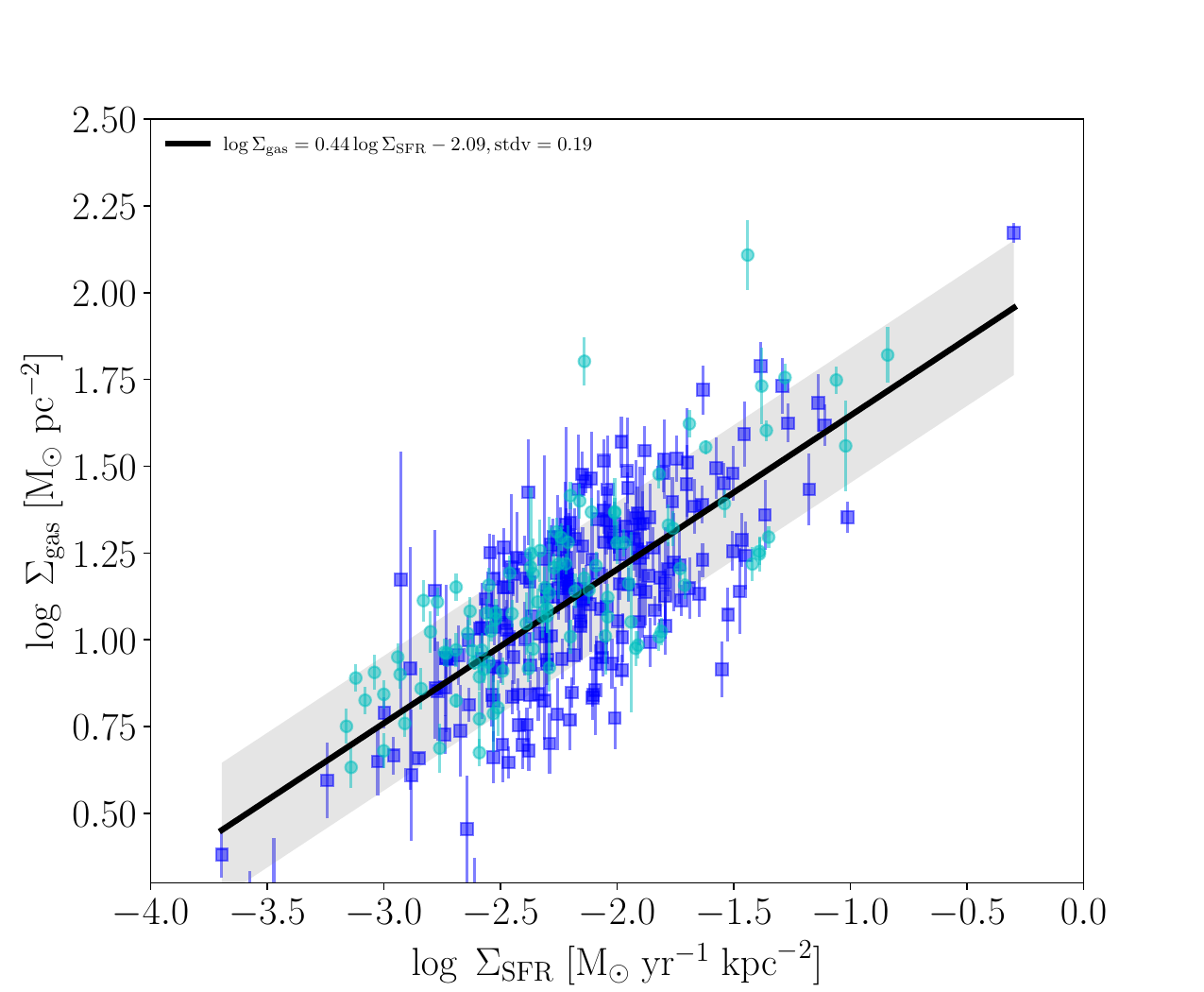}
    \caption{{\bf: Left Panel:} Kennicutt-Schmidt law for our sample (see the text for detials, blue filled squares) in comparison to the \citet{de_los_Reyes_Kennicutt2019} sample (cyan filled squares). Overall, we observe a strong agreement between both datasets. Therefore our sample accurately captures the structure of the global KS law. {\bf: Right Panel:} Inverted global KS law. As described in the text, our main conclusions remain unbiased when including or excluding the data from \citet{de_los_Reyes_Kennicutt2019} by fitting to a power-law model.} 
\label{fig:KS_gas}    
\end{figure*}
%%%%%%%%%%%%%%%%%%%%%%%%%%%%%%%%%%%%%%%%%%%%%%%%%%%%%%%%%%%%%%

%%%%%%%%%%%%%%%%%%%%%%%%%%%%%%%%%%%%%%%%%%%%%%%%%%%%%%%%%%%%%%
\begin{figure}
%    \vspace{2pt}
    \centering
    \includegraphics[height=2.3in,width=2.8in]{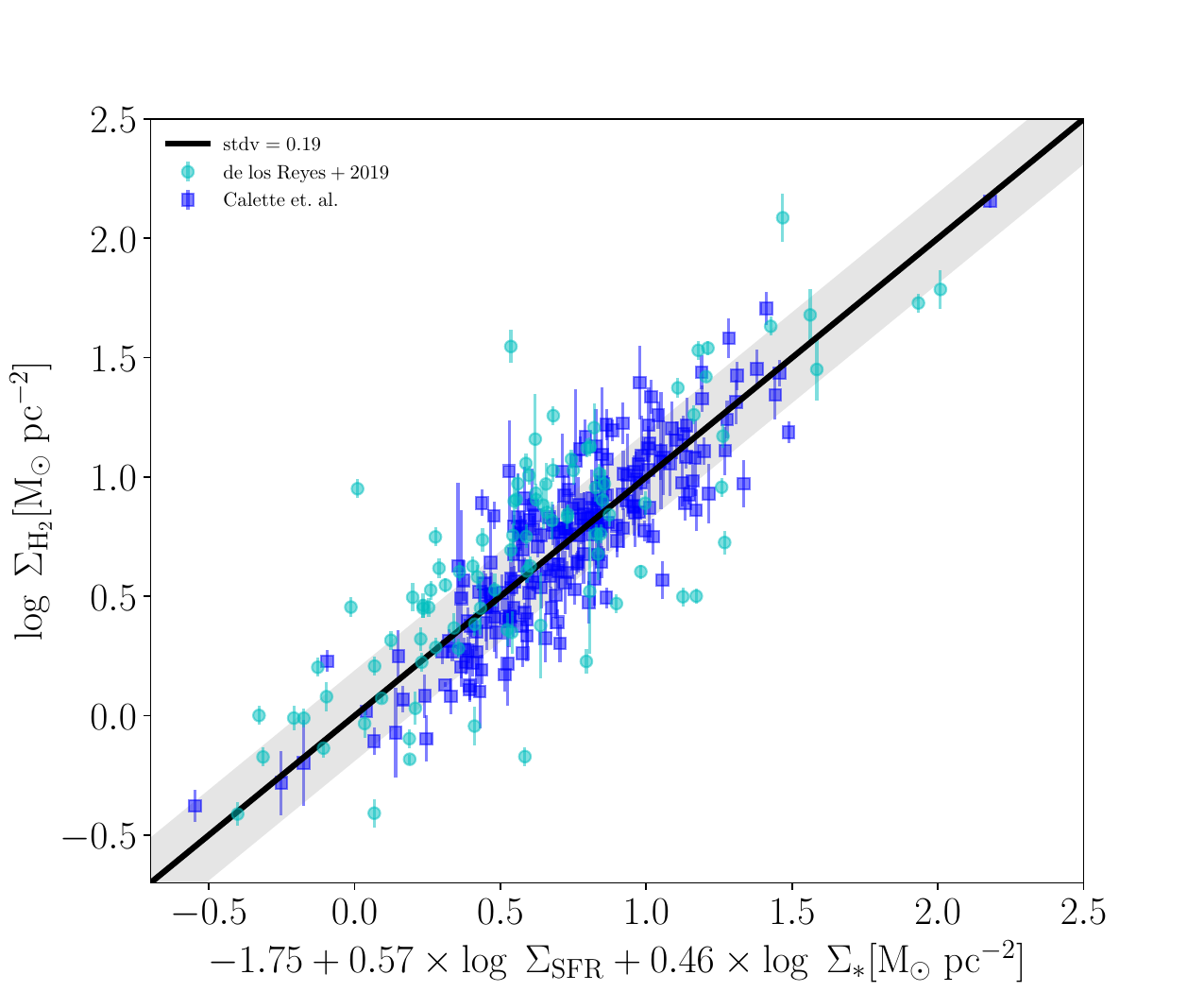}
    \caption{ Best-fit model for our sample (blue filled squares), obtained by combining $\Sigma_{\rm SFR}$ and $\Sigma_{\ast}$ to derive $\Sigma_{\rm H_2}$ as detail in the text. The predicted $\Sigma_{\rm H_2}$ values for the \citet{de_los_Reyes_Kennicutt2019} sample is consistent with our best-fit model but exhibit a larger scatter.} 
\label{fig:sfr_ms_h2}    
\end{figure}
%%%%%%%%%%%%%%%%%%%%%%%%%%%%%%%%%%%%%%%%%%%%%%%%%%%%%%%%%%%%%%

Having defined the star-forming region for each galaxy in our sample (Calette et al. in prep.), we re-estimated the HI gas mass within $R_{\rm SF}$ by assuming an exponential disc profile for the total gas mass (i.e., $M_{\rm gas} = M_{\rm H_2} + M_{\rm HI}$). We assumed that at large distances, the gas profiles of galaxies follow the HI size-mass relation, defined as the radius at which the average HI surface density profile reaches a value of $\Sigma_{\rm HI,1} = $1 \msun\ pc$^2$ \citep{Broeils+1997}. We compute for each galaxy HI radii, $R_{\rm HI}$, utilizing the HI size-mass relation from \citet{Wang+2016} and by adding a random scatter of 0.06 dex. Furthermore, we assumed that at distances of $R_{\rm SF}$, the HI profiles are well-characterized by a disk profile. The final step is to compute the scale-length radius of the HI disk by solving the following fundamental equation
\begin{equation}
    \Sigma_{\rm HI,1} - \frac{M_{\rm HI}}{2\pi R_{d, \rm HI}^2}
    \exp\left(-\frac{R_{\rm HI}}{R_{d, \rm HI}}\right) = 0,
\end{equation}
where $R_{d, \rm HI}$ is the scale-length radius, which is the unknown variable in the above equation. Having obtained the scale-length radius we can then obtain the HI mass at any radii, in particular at $R_{\rm SF}$. Consequently, we define the following quantities for the analysis of the Calette et al. (in prep) data: $\Sigma_{\rm SFR} = 0.95 \times {\rm SFR}/\pi R_{\rm SF}^2$, $\Sigma_{\ast} = M_\ast/\pi R_{\rm SF}^2$, $\Sigma_{\rm gas} = M_{\rm gas}/\pi R_{\rm SF}^2$, and $\Sigma_{\rm H_2} = M_{\rm H_2}/\pi R_{\rm SF}^2$. In these definitions, ${\rm SFR}$, $\ms$, and $M_{\rm H_2}$ represent the global SFRs and stellar masses from the sample, while $M_{\rm gas}$ is the gas mass within the star formation radius. Notice that in all the analyses described below, we only include galaxies with detections in both HI and H$_2$ gas masses. 

The left panel of Figure \ref{fig:KS_gas} displays the Kennicutt-Schmidt (KS) law for our sample (blue filled squares) in comparison to the \citet{de_los_Reyes_Kennicutt2019} sample (cyan filled squares). The distribution structure of our data closely mirrors that of \citet{de_los_Reyes_Kennicutt2019}, which is reassuring given the meticulous steps taken to obtain our data. By definition, the global KS law is expressed as:
\begin{equation}
    \Sigma_{\rm gas} = \alpha_{\rm KS} \Sigma^{\eta}_{\rm SFR}.
\end{equation}
The best-fitting parameters obtained from our sample are $n = 1.28 \pm 0.01$ and $\log A_{\rm KS} = -3.6 \pm 0.02$, with a dispersion of 0.31 dex. Combining information from both datasets, we derive the following parameters: $n = 1.4 \pm 0.01$ and $\log A_{\rm KS} = -3.79 \pm 0.01$, with a dispersion of 0.32 dex. Overall, we observe a strong agreement between both datasets, affirming that our sample accurately captures the structure of the global KS law.

The right panel of Figure \ref{fig:KS_gas} illustrates the inverted global KS law, expressed as
\begin{equation}
    \Sigma_{\rm gas} = \alpha_{\rm KS} \Sigma^{\eta}_{\rm SFR}.
\end{equation}
For our sample, we find $n = 0.44 \pm 0.02$ and $\log \alpha_{\rm KS} = 2.09 \pm 0.04$, with a dispersion of 0.19 dex. For completeness, we performed fits on the combined dataset, incorporating the data from \citet{de_los_Reyes_Kennicutt2019}. The resulting parameters are $n = 0.43 \pm 0.02$ and $\log \alpha_{\rm KS} = 2.08 \pm 0.03$. Importantly, we observe strong consistency between these parameters and the ones obtained from the previous fits. Therefore, whether including or excluding the data from \citet{de_los_Reyes_Kennicutt2019}, our main conclusions remain unbiased.

In Figure \ref{fig:sfr_ms_h2}, we present the best-fit model for our sample (blue filled squares), obtained by combining $\Sigma_{\rm SFR}$ and $\Sigma_{\ast}$ to derive $\Sigma_{\rm H_2}$. The best-fit model is expressed as
\begin{equation}
    \Sigma_{\rm H_2}({\rm SFR},M_{\ast}) = 10^{-1.75\pm0.03} \times \Sigma_{\rm SFR}^{0.57\pm0.03} \times \Sigma_{\ast}^{0.46\pm0.05}.
    \label{eq:App_M_h2}
\end{equation}
with a dispersion of 0.19 dex. Additionally, we compare the predicted $\Sigma_{\rm H_2}$ values for the \citet{de_los_Reyes_Kennicutt2019} sample, which align with our best-fit model but exhibit a larger scatter.

%%%%%%%%%%%%%%%%%%%%%%%%%%%%%%%%%%%%%%%%%%%%%%%%%%%%%%%%%%%%%%
\begin{figure}
    \vspace{2pt}
    \includegraphics[height=5.8in,width=3.6in]{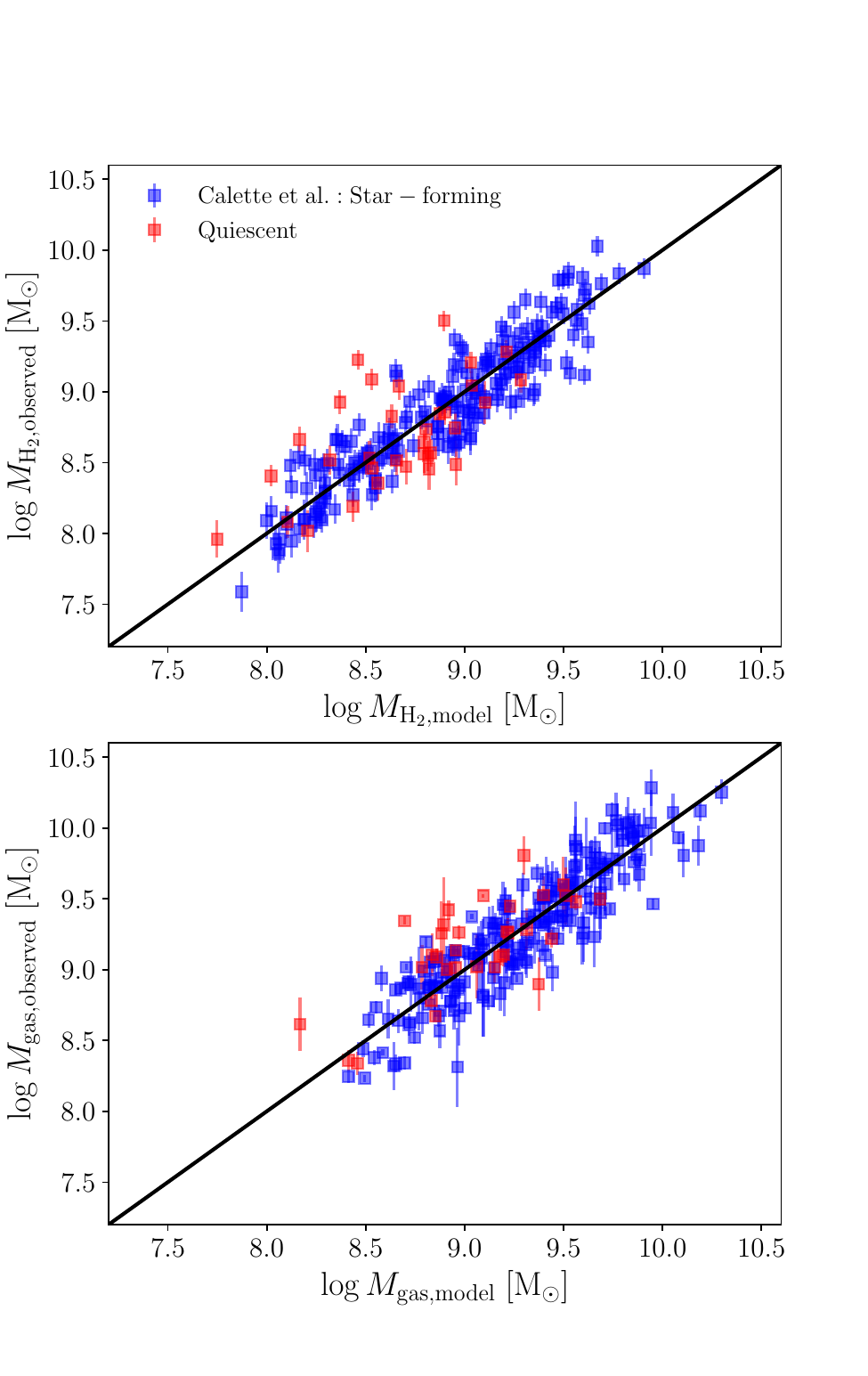}
    \caption{Comparison of the inferred H$_{2}$ (top panel) and total cold gas (bottom panel) masses with observations from the Calette et al. compilation (in prep). Only H$_2$ and HI detections were considered. The solid red line represents the one-to-one relation.}
\label{fig:model-vs-observations-gas}    
\end{figure}
%%%%%%%%%%%%%%%%%%%%%%%%%%%%%%%%%%%%%%%%%%%%%%%%%%%%%%%%%%%%%%

%%%%%%%%%%%%%%%%%%%%%%%%%%%%%%%%%%%%%%%%%%%%%%%%%%%%%%%%%%%%%%
\begin{figure}
    \includegraphics[height=3.3in,width=3.6in]{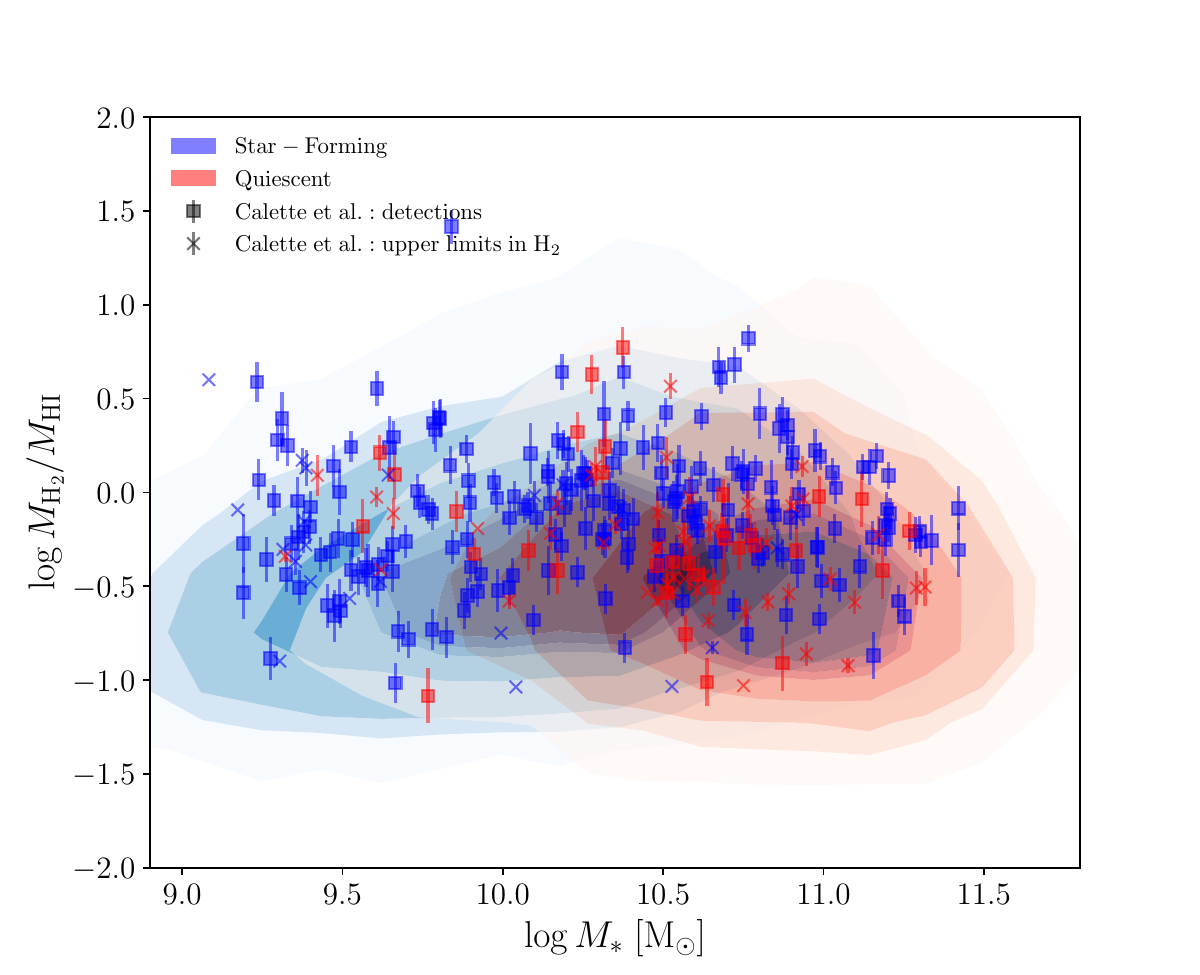}
    \caption{Predicted H$_{2}$/HI ratios as a function of \ms\ shown as blue and light red contours respectively for SFMS and quiescent (including green valley) galaxies. The contours show 5, 16, 0.25, 5, 75, 84 and 95 $\%$ of the distribution. Blue (red) filled squares show the SFMS (quiescent) compilation from Calette et al. \citep[in prep., see also,][]{Calette+2018,Calette+2021}. The skeletal symbols show upper limits in H$_2$ but detections in HI. Notice how our model is consistent with the distribution of observed points.}
\label{fig:H2_to_HI_ratio_predicted_iso}    
\end{figure}
%%%%%%%%%%%%%%%%%%%%%%%%%%%%%%%%%%%%%%%%%%%%%%%%%%%%%%%%%%%%%%

%%%%%%%%%%%%%%%%%%%%%%%%%%%%%%%%%%%%%%%%%%%%%%%%%%%%%%%%%%%%%%
\begin{figure}
    \includegraphics[height=2.8in,width=3.6in]{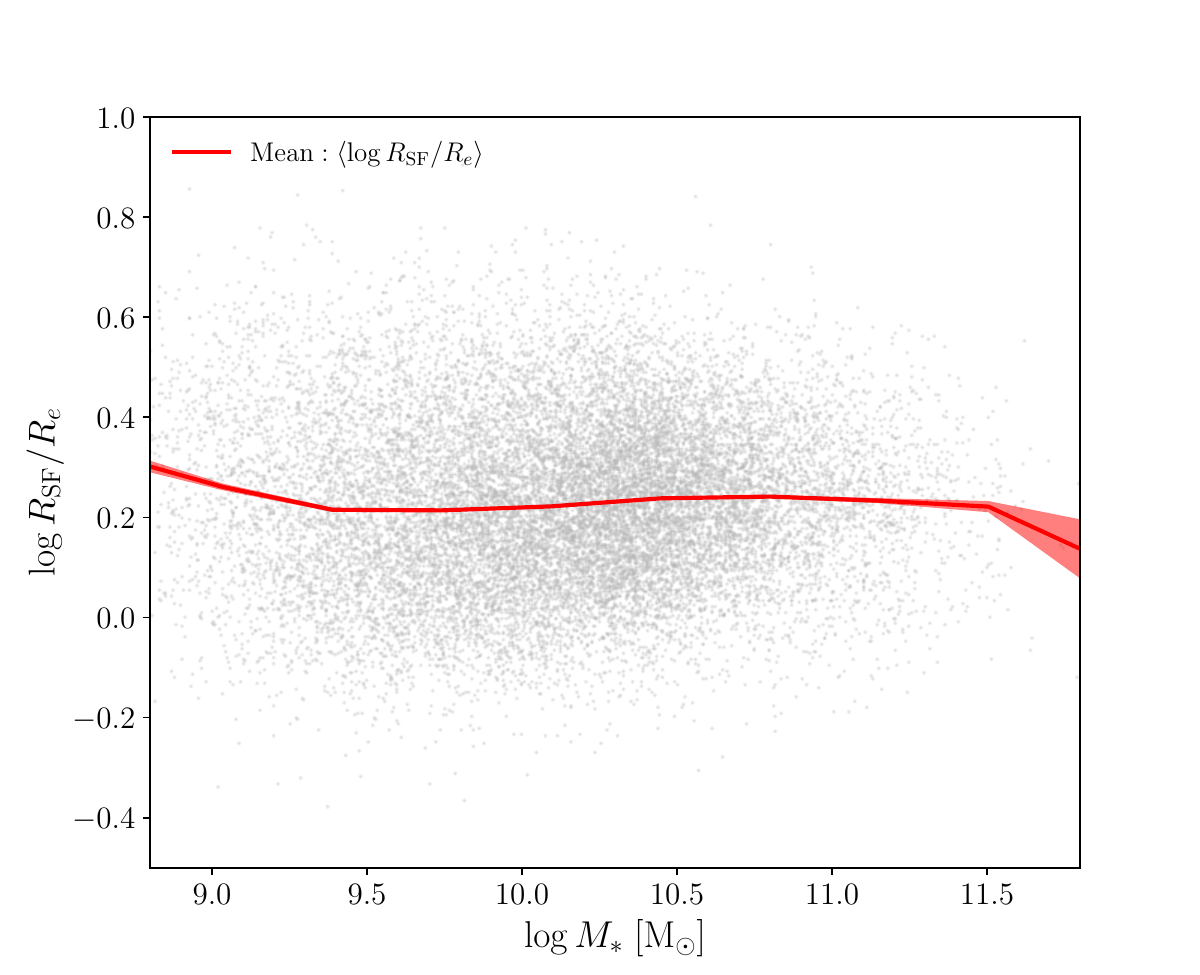}
    \caption{Ratio between the star forming radius, $R_{\rm SF}$ and the half-light radius, $R_e$, as a function of stellar mass. The ratio depends little with mass but, on average, this is approximately $\langle \log R_{\rm SF} / R_e \rangle \sim 0.23$.}
\label{fig:R_halpatoRe_vs_Ms}    
\end{figure}
%%%%%%%%%%%%%%%%%%%%%%%%%%%%%%%%%%%%%%%%%%%%%%%%%%%%%%%%%%%%%%

The predicted $M_{\rm gas}$ from our inverted global KS law versus the measured one, using the data from the Calette et al. (in prep) compilation, is shown in the bottom panel of Figure \ref{fig:model-vs-observations-gas}. The prediction nicely recovers the gas mass for the majority of the galaxies. 
Similarly, using the Calette et al. compilation, in the upper panel of the same figure we show that Eq. (\ref{eq:App_M_h2}) is a good predictor of the observed H$_2$ gas masses not only for star-forming galaxies but for quiescent ones.

Figure \ref{fig:H2_to_HI_ratio_predicted} shows the  H$_{2}$-to-HI mass ratio predicted by our model as a function of stellar mass for our SDSS galaxy sample. The sample of SFMS galaxies is shown by the blue-filled isocontours while GV and quiescent galaxies are shown in light red-filled isocontours. In the same panel, as a reference, we reproduce the Calette et al. compilation divided into star-forming an quiescent galaxies (filled squares represent detections in both H$_2$ and HI gas masses while crosses show non-detection in H$_2$) which is in good agreement with the distribution predicted by our model.  

Finally, in Figure \ref{fig:R_halpatoRe_vs_Ms}, we present the ratio between $R_{\rm SF}$ and $R_e$ as a function of $\ms$ for all our SDSS galaxies. The red line represents the mean $\langle \log R_{\rm SF} / R_e \rangle$, and the shaded area indicates the error around the mean. It is noteworthy that this ratio exhibits little dependence on mass but is larger for low-mass galaxies and smaller for more massive ones. On average, this ratio is approximately $\langle \log R_{\rm SF} / R_e \rangle \sim 0.23$.

% Don't change these lines
\bsp	% typesetting comment
\label{lastpage}
\end{document}